%% file: paper.tex
\documentclass[12pt]{iopart}

\usepackage{array,multirow}
\usepackage{booktabs}
\usepackage{graphicx}
\usepackage{cite}
\usepackage{lscape}
\usepackage{bm}
\usepackage{xspace}
\usepackage{amssymb}
\usepackage{afterpage}
\usepackage{hyperref}

\usepackage{lineno}

\newcommand{\NLDBD}{0$\nu\beta\beta$\xspace}
\newcommand{\XeNLDBDEvt}{a $^{136}$Xe 0$\nu\beta\beta$ event}
\newcommand{\Geant}{\emph{Geant4}}

\newcommand{\evProcesses}{\emph{event processes}}

\begin{document}

\title[Topological background discrimination in the PandaX-III experiment]{Topological background discrimination in the PandaX-III neutrinoless double beta decay experiment}



\author{J~Galan$^{1,2}$, X~Chen$^1$, H~Du$^1$, C~Fu$^1$, K~Giboni$^1$, F~Giuliani$^1$, K~Han$^1$, B~Jiang$^1$, X~Ji$^{1,10}$, H~Lin$^1$, Y~Lin$^1$, J~Liu$^1$, K~Ni$^1$, X~Ren$^1$, S~Wang$^1$, S~Wu$^1$, C~Xie$^1$, Y~Yang$^1$, D~Zhang$^1$, T~Zhang$^1$, L~Zhao$^1$, S Aune$^2$, Y Bedfer$^2$, E Berthoumieux$^2$, D~Calvet$^2$, N~d'Hose$^2$, E~Ferrer-Ribas$^2$, F Kunne$^2$, B Manier$^2$, D~Neyret$^2$, T~Papaevangelou$^2$, L Chen$^3$, S Hu$^3$, S Jian$^3$, P Li$^3$, X~Li$^3$, H Zhang$^3$, M Zhao$^3$, J Zhou$^3$, Y Mao$^4$, H Qiao$^4$, S~Wang$^4$, Y Yuan$^4$, M Wang$^5$, Y Chen$^6$, A N Khan$^6$, N Raper$^6$, J Tang$^6$, W Wang$^6$, C Feng$^7$, C Li$^7$, J Liu$^7$, S Liu$^7$, X Wang$^7$, D~Zhu$^7$, J~F~Castel$^8$, S Cebri\'an$^8$, T Dafni$^8$, I G Irastorza$^8$, G~Luz\'on$^8$, H Mirallas$^{8,1}$, X Sun$^9$, A Tan$^{10}$, W Haxton$^{11}$, Y~Mei$^{11}$, C Kobdaj$^{12}$, Y~Yan$^{12}$}

\address{$^1$ INPAC and School of Physics and Astronomy, Shanghai Jiao Tong University, Shanghai Laboratory for Particle Physics and Cosmology, Shanghai 200240, China }

\address{$^2$ IRFU, CEA, Universit\'e Paris-Saclay, F-91191 Gif-sur-Yvette, France}

\address{$^3$ China Institute of Atomic Energy, Beijing 102413, China }

\address{$^4$ School of Physics, Peking University, Beijing 100871, China}

\address{$^5$ School of Physics and Key Laboratory of Particle Physics and Particle Irradiation (MOE), Shandong University, Jinan 250100, China }
\address{$^6$ School of Physics and Engineering, Sun Yat-Sen University, Guangzhou 510275, China}

\address{$^7$ Department of Modern Physics, University of Science and Technology of China, Hefei 230026, China;
State Key Laboratory of Particle Detection and Electronics, University of Science and Technology of China, Hefei 230026, China}

\address{$^8$ University of Zaragoza, C/P Cerbuna 12 50009, Zaragoza, Spain }

\address{$^9$ College of Physical Science and Technology, Central China Normal University,  Wuhan 430079, China}

\address{$^{10}$ Department of Physics,  University of Maryland, College Park, MD 20742, USA}

\address{$^{11}$ Nuclear Science Division, Lawrence Berkeley National Laboratory, Berkeley, CA 94720, USA }

\address{$^{12}$ Center for Excellence in High Energy Physics and Astrophysics, Suranaree University of Technology, Nakhon Ratchasima 30000, Thailand}

\ead{javier.galan.lacarra@cern.ch}
\ead{ke.han@sjtu.edu.cn}
\ead{jiangbofang@sjtu.edu.cn}


\begin{abstract}
The PandaX-III experiment plans to search for neutrinoless double beta decay (\NLDBD) of $^{136}$Xe in the China JinPing underground Laboratory~(CJPL). The experiment will use a high pressure gaseous Time Projection Chamber~(TPC) to register both the energy and the electron track topology of an event. This article is devoted to demonstrate our particular detector setup capabilities for the identification of \NLDBD and the consequent background reduction.

As software tool we use REST, a framework developed for the reconstruction and simulation of TPC-based detector systems. We study the potential for background reduction by introducing appropriate parameters based on the properties of \NLDBD events. We exploit for the first time not only the energy density of the electron track-ends, but also the electron scattering angles produced by an electron near the end of its trajectory. To implement this, we have added new algorithms for detector signal and track processing inside REST.
Their assessment shows that background can be reduced by about 7 orders of magnitude while keeping \NLDBD efficiency above 20\% for the PandaX-III baseline readout scheme, a 2-dimensional 3\,mm-pitch stripped readout. More generally, we use the potential of REST to handle 2D/3D data to assess the impact on signal-to-background significance at different detector granularities, and to validate the PandaX-III baseline choice.
Finally, we demonstrate the additional potential to discriminate surface background events generated at the readout plane in the absence of $t_o$, by making use of event parameters related with the diffusion of electrons.

\end{abstract}

%
%
%
%
%

\input{introduction.tex}
\input{baseline.tex}
\input{granularity.tex}

\input{fiducial.tex}

\input{conclusions.tex}

\appendix

\input{REST.tex}

\input{MCDataChain.tex}

\input{RawDataChain.tex}

\vspace{0.5cm}
\section*{References}
\bibliography{paper}
\bibliographystyle{unsrt}

\end{document}

%% file: introduction.tex
\section{Introduction}
\label{sc:intro}

The neutrino sector has been one of the most promising research areas for searching for new physics beyond the Standard Model~(SM).
Neutrino oscillations and the inferred non-zero neutrino mass offered a concrete evidence of new physics, as demonstrated by experiments including Super-Kamiokande~\cite{PhysRevLett.81.1562}, SNO~\cite{Ahmad:2002jz}, and etc.
Early in 1937, the possibility for an electrically neutral fermion to be the anti-particle of itself was raised by Majorana~\cite{majorana1937majorana}, and neutrinos are the most promising candidates for the so-called Majorana fermions.
Soon after that, physicists started to use the neutrinoless double beta decay (\NLDBD) to search for the existence of Majorana neutrinos~\cite{Barabash:2011fn}. 
In a SM-allowed two-neutrino double beta decay process, two neutrons in a candidate nucleus decay into two protons simultaneously and emit two electrons together with two electron antineutrinos.
Such a process has been observed in a dozen or so isotopes, including $^{76}$Ge, $^{130}$Te, $^{136}$Xe, and etc (see, for example~\cite{DellOro:2016tmg}).
If neutrinos are Majorana fermions, no electron antineutrinos are released in the double beta decay process.
It has also been argued with the Black Box Theorem that an observation of \NLDBD directly indicates the Majorana nature of neutrinos~\cite{Schechter:1981bd}.
The experimental observation of \NLDBD will also distinctly violate the conservation of the lepton number, and have far-reaching impact beyond the neutrino sector~\cite{avignone2008double}.
After about 80 years of experimental effort, no firm evidence for \NLDBD has been obtained.
Several major experiments are taking data or are under construction and many more smaller-scale efforts are in R\&D phase.
Current status and expected sensitivity can be seen in reference~\cite{Agostini:2017jim} and references within.

The PandaX-III experiment aims to search for \NLDBD of $^{136}$Xe using a high pressure gaseous Time Projection Chamber~(TPC)~\cite{chen2017pandax}.
The experiment will be located in the newly excavated China JinPing underground Laboratory Phase II~(CJPL-II) in Sichuan Province, China~\cite{LI2015576}.
As a calorimeter, PandaX-III measures the energy of two electrons from \NLDBD and constructs an energy spectrum at the Region Of Interest~(ROI) around the Q$_{\beta\beta}$ value of $^{136}$Xe, 2,457.83\,keV\,\cite{PhysRevLett.98.053003}.
The expected energy resolution is better than 3\%~Full-Width-Half-Maximum (FWHM) at the Q$_{\beta\beta}$ value\,\cite{Gonzalez-Diaz:2015oba,Lin:2018mpd}.
PandaX-III follows the standard recipe of a low background experiment, including deep underground lab environment, passive shielding, and careful screening of detector as well as electronics material to minimize the number of dubious background events in the ROI.
The descendants of $^{238}$U and $^{232}$Th, mainly $^{214}$Bi and $^{208}$Tl, contribute most significantly to the background in the ROI.
One of the gamma lines from $^{214}$Bi at 2,448~keV, only 10\,keV below the \NLDBD Q-value, poses a major challenge to event identification with energy alone.
The expected background level at the ROI including a realistic detector response is on the order of $10^{-3}$\,keV$^{-1}$kg$^{-1}$y$^{-1}$ when considering only the event energy.

Additional background suppression can be realized with event tracking information, which is the main topic of this paper.
Different from other detector technologies, a gaseous TPC will record detailed trajectories of the two electrons from a potential \NLDBD event.
PandaX-III will exploit the electron tracking potential using fine-pitched Microbulk Micromegas~\cite{andriamonje2010development} readout planes to reach millimeter level spatial resolution, in comparison with the typical \NLDBD track length of 10\,cm scale at 10\,bar.
The signal identification will exploit the features of the electron tracks topology produced by \NLDBD decays. The two electrons generated by a \NLDBD decay will produce two bright Bragg peaks, occurring on at least 95\% of the decays~\cite{PhysRevD.48.1009}.
On the other hand, gamma background events in the ROI normally have only one Bragg peak. We will review the details on the main characteristics of signal and background events, previously investigated in reference~\cite{PhysRevD.48.1009}, and we will exploit them to perform a discrimination analysis to demonstrate the ultimate background level achievable by the PandaX-III setup.

The work presented in this paper includes an accurate Monte Carlo demonstration of the operation of the PandaX-III detector setup from the point of view of \NLDBD event identification. Previous work, published in reference~\cite{chen2017pandax}, provided a rough estimation of the final background achievable. This paper provides a complete and detailed analysis for PandaX-III including a realistic 2-dimensional detector readout.

PandaX-III adopts the event-oriented software framework REST~\footnote{We include in \ref{sc:REST} necessary technical details on the software used, and the description of the \emph{event data} and \emph{event process} routines used in our analysis. Any keyword starting by \emph{TRest} will do reference to REST and it will be detailed in one of the appendices of this document.} for simulation and data analysis, especially to facilitate the topological analysis. 
With REST and the underlying Geant4 framework, we simulate \NLDBD events and background events with energy deposition and tracking information in the gas medium.
Realistic detector response, including the strip readout scheme, is then added to generate mock data, which are then fed into our data analysis chain for energy and track reconstruction.
We then optimize the efficiency of the discrimination between signal and background by analyzing the topological information of the events.

This paper is organized as follows. In section~\ref{sc:panda3mm} we present the results obtained in terms of signal-to-background enhancement for the PandaX-III baseline readout scheme. In section~\ref{sc:granularity}, we explore the impact of different readout granularities and assess the goodness of choice for the PandaX-III baseline readout. Finally, in section~\ref{sc:fiducial}, we go a step forward in our analysis to demonstrate the capability of our detector to fiducialize in the drift direction of the TPC and reject background from our charge readout plane effectively.

%% file: baseline.tex
\setcounter{footnote}{0}

\section{Background rejection on the PandaX-III baseline readout scheme}
\label{sc:panda3mm}

We present here the results obtained using a realistic detector readout description based on the existing PandaX-III baseline Microbulk Micromegas design~\cite{Galan_2016}. For rigourosity, we provide in~\ref{sc:Montecarlo} details on how we generate our Monte Carlo \emph{event data} including any necessary details such as physics considerations and detailed detector readout description for PandaX-III, based on a 2-dimensional charge readout, which represents a major milestone on the event reconstruction and data analysis of the experiment. In section~\ref{sc:topoParams}, we define the parameters, extracted during the \emph{event data} processing, that will be used for event pattern recognition, leading to our final background discrimination results reported in section~\ref{sc:results}.
\subsection{Topological parameters used for pattern recognition}\label{sc:topoParams}

This section does not focus on the nature, or the source, of background events, but mostly on the type of background events that might mimic a \NLDBD in the ROI at the $Q_{\beta\beta}$ energy of $^{136}$Xe. Such background events, producing a topological structure similar to a \NLDBD decay, can only be produced in the active detector volume by gammas with an energy equal to or above the $Q_{\beta\beta}$ value. Some of those gammas end up producing a long electron track of characteristics similar to the \NLDBD signal. Still, a good choice of parameters extracted from the \emph{physical track} provides an excellent instrument to differentiate between this kind of events and electron tracks produced by \NLDBD decays.

Previous studies on \NLDBD pattern recognition~\cite{cebrian2013pattern,Ferrario2016,NEXTRennerNN} have already shown the potential of certain parameters, such as the \emph{number of secondary tracks} or \emph{end-track energies} to differentiate background and signal events. We reach the same conclusions, and we obtain comparable results on their discrimination power. Furthermore, we will investigate the potential of novel parameters never used in previous studies, \emph{viz.} as the end-track \emph{twist parameter} and the \emph{track length}. Below we discuss the respective merits of both, previously mentioned and novel parameters.


\begin{itemize}
    \item {\bf Number of secondary tracks, or track energy ratio.} One of the most powerful discriminators is related to the number of independent tracks identified inside an event taking place in the active detector volume\footnote{The number of tracks is found after using \emph{TRestHitsToTrackProcess}, described in~\ref{sc:evProcesses}, while the conditions in which those are obtained is described in~\ref{sc:dataChain}}. Background events are originated by high energy gammas, and many of those gammas interacting in the detector will produce multi-track events due to Compton scattering combined with a photoelectric interaction absorbing the remaining gamma energy, therefore producing a main long energetic track. Additionally, the higher initial energy of the single electron background compared to the initial energy of each of the two electrons generated in the \NLDBD decay makes so that the probability of producing secondary gamma radiation is higher for single electron background events.

In practice, it is convenient to define the \emph{track energy ratio}, $\theta$, as the ratio between the total energy of secondary tracks and the total energy of the event, $E_{tot}$, expressed as,
\begin{equation}
	\theta = \frac{1}{E_{tot}} \sum_{n_i=1}^{N_i} E_{i,n}
\end{equation}

\noindent where $E_{i,n}$ is the energy of track number $n$ from projection $i$, and $n_i$=0 corresponds to the most energetic track of each projection. If the event does not contain secondary tracks, as it is frequently the case in \NLDBD events, the value of $\theta$ will be zero. Such a definition accommodates the 2D and 3D detector readouts in a single observable, and provides a probability density function (pdf) of the $\theta$ parameter that can be exploited for signal and background separation. It is important to remark that the definition of $\theta$ is done for convenience on the flexibility to tune the signal-to-background ratio by selecting an appropriate $\theta$-value. We have compared ourselves this observable with other approaches used in other studies~\cite{cebrian2013pattern}, obainning very similar results. It must be remarked, that using the $\theta$ parameter does not constrain the number of tracks. However, typically low values of $\theta$ will accept mostly events containing only one or two tracks, and rarely three.



%

\item {\bf End-track energies, or blobs charge.} An electron traveling in a gas medium with energies above $\sim$MeV experiences a constant energy loss, $dE/dx$, along its trajectory in the medium. Once the electron loses most of its energy, at levels well below 1\,MeV, its $dE/dx$ increases suddenly until it loses all of its kinematic energy. This phenomenon produces a high density charge region at the electron track end, or Bragg peak, commonly called \emph{blob}.
Obviously, a \NLDBD track emitting two electrons with a common origin will be similar to a single track with two \emph{blobs} at the track ends, while single electron background tracks will only produce one Bragg peak. The end of single electron tracks where no \emph{blob} is found will be called \emph{tail}.



The energy, or charge, $\mathcal{Q}$, in a certain \emph{blob} is determined by summing up the energy deposited around a sphere of radius $R_o$ centered at each of the high density track end coordinates\footnote{The blob coordinates are obtained by \emph{TRestFindTrackBlobsProcess}, detailed in~\ref{sc:anaPcs}}. The following mathematical expression summarizes our \emph{blob} charge definition,

\begin{equation}
	\mathcal{Q}_{j} = \sum_{d=r_{j,n}<R_o} q_{n}
	 \quad\mbox{and}\quad
	 \mathcal{Q}_{l} = \mbox{min}(\mathcal{Q}_{1} , \mathcal{Q}_{2})
\end{equation}

        where the index $j$=1,2 represents each of the track ends, and the index $n$ corresponds to the hit number inside the \emph{TRestTrackEvent} definition (given in~\ref{sc:REST}). The charge integration at each end-track, $\mathcal{Q}_{1,2}$, is performed for those hits, $q_n$, that satisfy that the hit distance $d$ to the blob coordinates, at each track end, $j$, is below $R_o$. Different combinations of $\mathcal{Q}_1$ and $\mathcal{Q}_2$ may be chosen to construct an observable to be used for pattern recognition. In our posterior analysis we will use the lowest charge blob definition, $\mathcal{Q}_l$, a value which, in principle, is necessarily low for background events in which the electron \emph{tail} is properly identified, while for \NLDBD events cannot be low. In the case of 2-dimensional readouts this observable is independently calculated for each projection, obtaining two values, $\mathcal{Q}^{X}_l$ and $\mathcal{Q}^{Y}_l$, to be combined later.


	\item {\bf End-track twist.} Another feature characterizing the behavior of the electron trajectory near the \emph{blob} is the erratic nature of its trajectory. This erratic behavior, due to higher electron scattering angles, does not manifest at the initial interaction vertex, or \emph{tail}, at single electron background events. On the contrary, energetic electrons, with initial energies of the order of the $Q_{\beta\beta}$ value, will usually produce a clear straight \emph{tail}.

        In order to quantify this effect we introduce the \emph{twist parameter} that measures the angle between consecutive hits belonging to a \emph{top-level track}\footnote{A \emph{top-level track} in our analysis identifies with the \emph{physical track} reconstructed, see also \emph{TRestTrackEvent} definition at~\ref{sc:REST}, } near the track ends. The \emph{twist parameter}, $\xi$, is expressed as follows,

\begin{equation}
	 \xi_j = \frac{1}{N_j} \sum_{i=0}^{N_j} (1-\hat{n_i} \cdot \hat{n}_{i+1})/2,
	 \quad\mbox{and}\quad
	 \xi_{l} = \mbox{min}(\xi_{1} , \xi_{2})
 \label{eq:twist}
\end{equation}

where $\hat{n_i}$ represents the unit vector defined by the $i^{th}$ and $i^{th}+1$ nodes of a reconstructed \emph{top-level track}, and $i=0$ can be identified with the first (or last) node of the \emph{physical track}. The total number of hits, $N_j$, considered in each end-track, $j$=1,2, is relative to the total number of hits, $N_{tot}$, at the \emph{top-level track}. In our analysis we have defined $N_j$ as the 25\% of $N_{tot}$, i.e. a track containing $N_{tot}$=28 hits will average the 6 angles formed by the 7 hits closer to the end-tracks.
The normalization factor 2 in the previous expression is introduced to assure $\xi$ is mathematically contained in the range $(0,1)$, where $\xi$=0 means that the nodes are fully aligned.
As in the previous observable, we define $\xi_l$ as the lowest value between the \emph{twist parameter} obtained at each end-track, and obtain independent values for each projection, $\xi_l^X$ and $\xi_l^Y$, for the 2-dimensional case.


\item {\bf Track length.} We might also expect that two electron \NLDBD tracks will produce a slightly shorter path due to the generation of two blobs, and therefore a higher $dE/dx$ in average. Although this observable is expected to have a weak discriminating power, we still assess its performance.

	The \emph{track length}, $\mathcal{L}$, is simply calculated as the total measured distance between the track ends, following the ordered node sequence at the \emph{top-level track} produced after \emph{TRestTrackReconnectionProcess}. Expressed as,
 \begin{equation}
	 \mathcal{L} = \sum_{i=0}^{N-1} d_{i,i+1}
 \label{eq:length}
 \end{equation}

where $N$ is the total number of hits, or nodes, in the \emph{top-level track}, and $d_{i,i+1}$ is the distance between the hit $i$ and $i+1$, or \emph{connected hits}. Again, for the 2-dimensional case we will obtain 2 independent length measurements for each projection, $\mathcal{L}^X$ and~$\mathcal{L}^Y$.

\end{itemize}

In summary, these observables will allow us to discriminate background by differentiating the main background and signal properties. When identifying a \NLDBD signal event we search for a main energetic track, with almost negligible energy deposition on secondary tracks belonging to the event. Signal events will generally lack the presence of an end-track \emph{tail}, and therefore, the combination of low end-track values for \emph{blob charge}, $\mathcal{Q}_l$ and \emph{twist parameter}, $\xi_l$, will be decisive on the identification of such a feature. These main characteristics can be easily recognized on the background and signal events shown on Figure~\ref{fig:track}, at~\ref{sc:Montecarlo}.

Finally, it is important to remark that these parameters serve also to describe in a quantitative way each event, i.e. they can be exploited to characterize the experimental data and support the construction of a Monte Carlo background model. The nature from different background contributions due to the different contamination sources and components in the detector might be better understood by studying the behavior of these parameters on different populations of events, measured at different data taking conditions. Therefore, even if a parameter does not contribute finally on the signal-to-background significance, it may be valuable in other scenarios to help understanding the nature of the data collected with the detector.







\subsection{Results}\label{sc:results}

We report in this section the results obtained on background reduction considering the topological features of signal and background events. We will take into account the forecasted PandaX-III data taking conditions, i.e. a vessel filled with 10\,bar of xenon+TMA\,1\% gas mixture, and a detailed detector response, carefully described in~\ref{sc:Montecarlo}.

The Monte Carlo \NLDBD signal events, used to define our signal efficiency have been imported in \emph{RestG4} using a \emph{Decay0}\cite{Ponkratenko:2000um} pre-generated event population. The initial 4-momentum definition of each of the two electrons produced by the \NLDBD decay is randomly positioned at the active volume of the detector, or gas volume, for each event, and then tracked using \emph{Geant4} physics processes.

In reference~\cite{chen2017pandax}, the main contamination sources contributing to the background of the experiment were analyzed, allowing to identify the most critical detector components in terms of negative impact to the overall background of the experiment. In this work we go a step further by presenting a more definite background level achievable with the PandaX-III detector. For that purpose, we have chosen representative detector components, as the high pressure copper vessel and the Micromegas readout plane, to evaluate the pattern recognition capability of our event reconstruction algorithms. We have simulated the $^{238}$U and $^{232}$Th decay chains at those detector components. For the copper vessel we generate our events randomly on the bulk of the material, while for the Micromegas contamination we generate surface events at the readout plane location, behind 10\,$\mu$m of copper.
In order to guarantee enough statistics at the end of our topological analysis we have generated near 10$^9$ decays of $^{238}$U and $^{232}$Th isotopes for the vessel contribution, and 5$\cdot$10$^7$ decays for the Micromegas contributions. Our signal efficiency has been calculated over an initial population of 10$^6$ generated \NLDBD events.

\begin{figure}
\centering
\begin{tabular}{cc}
\includegraphics[width=0.49\textwidth]{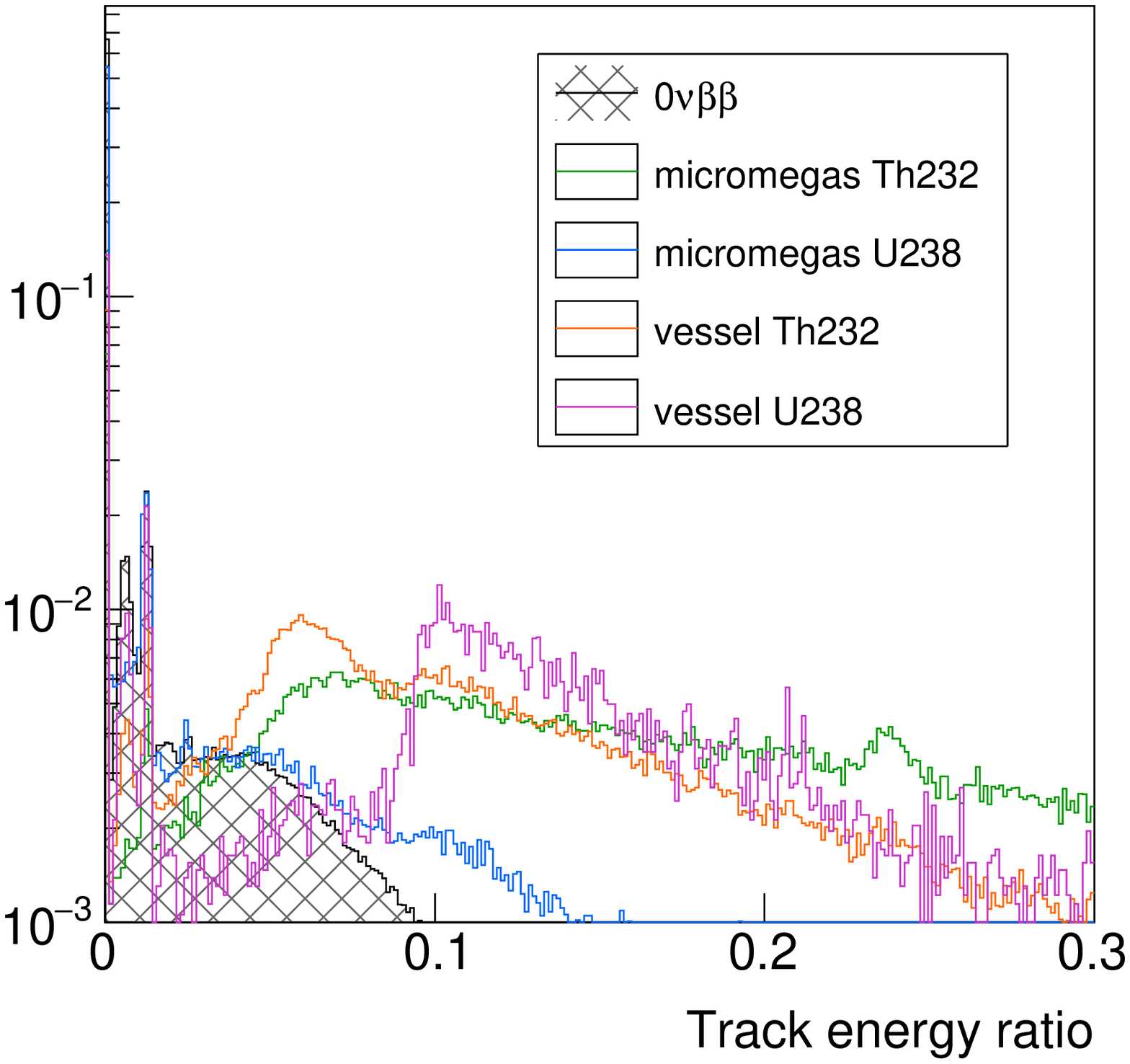} &
\includegraphics[width=0.49\textwidth]{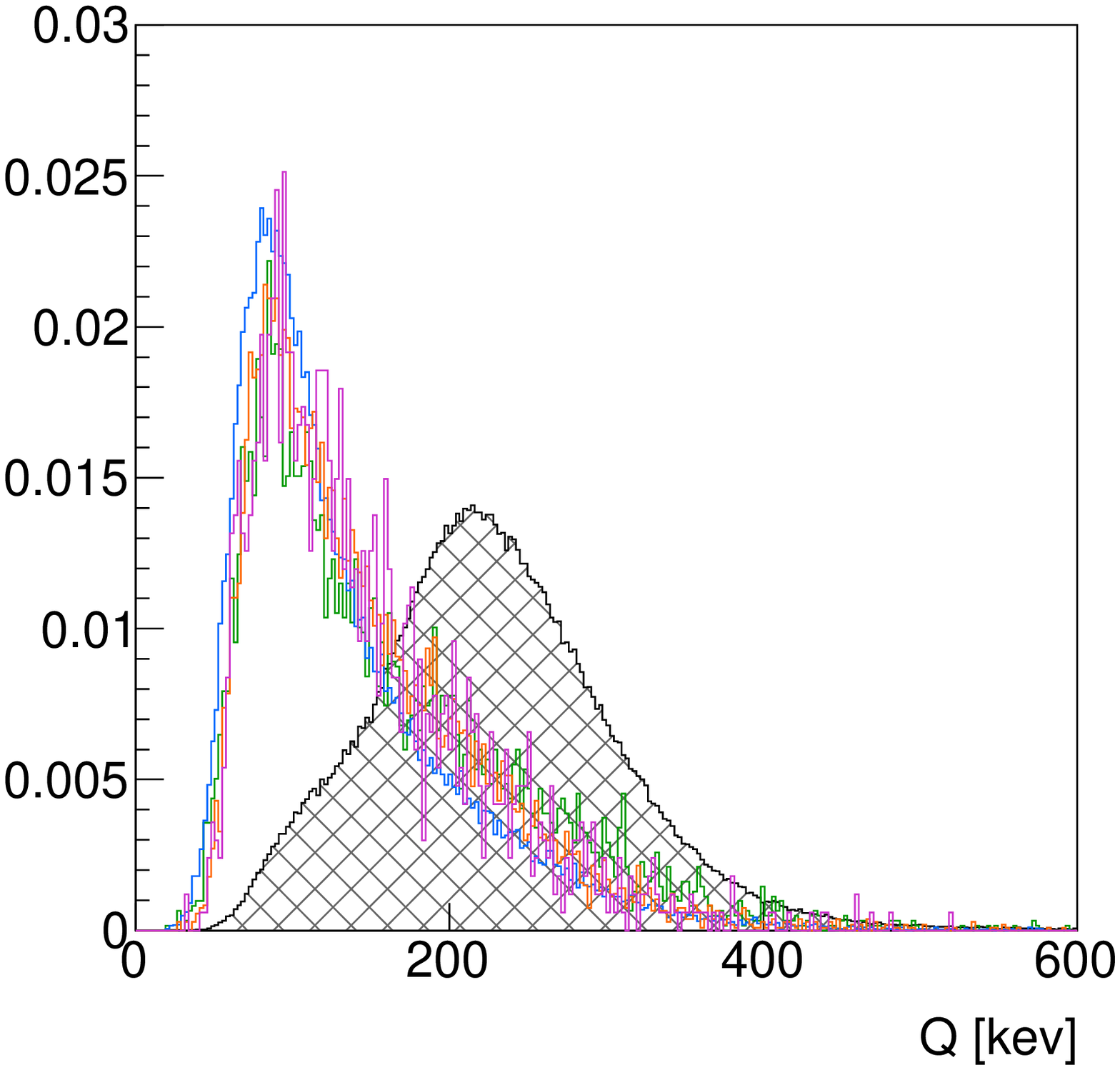} \\
(a) & (b) \\
\includegraphics[width=0.49\textwidth]{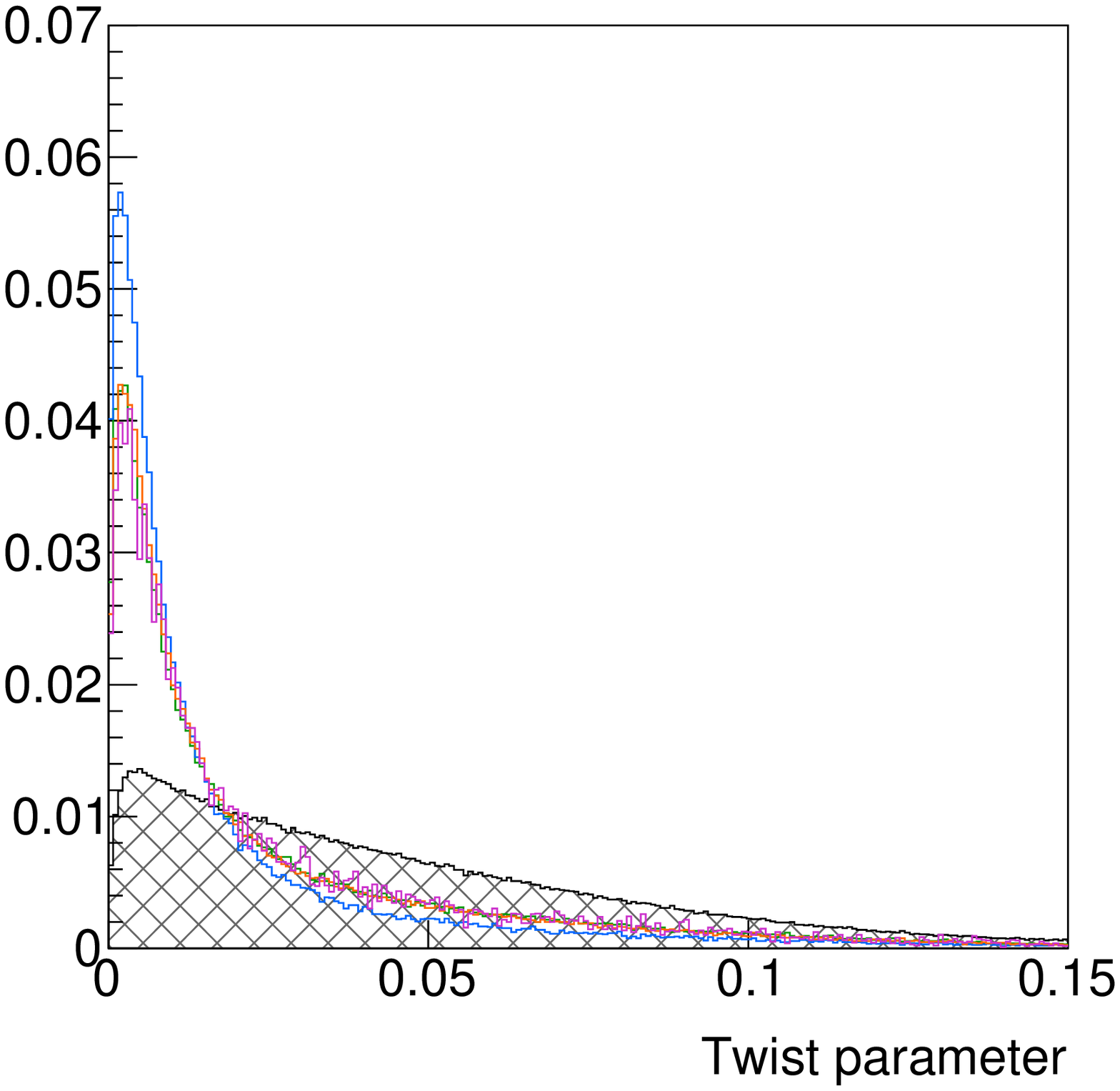} &
\includegraphics[width=0.49\textwidth]{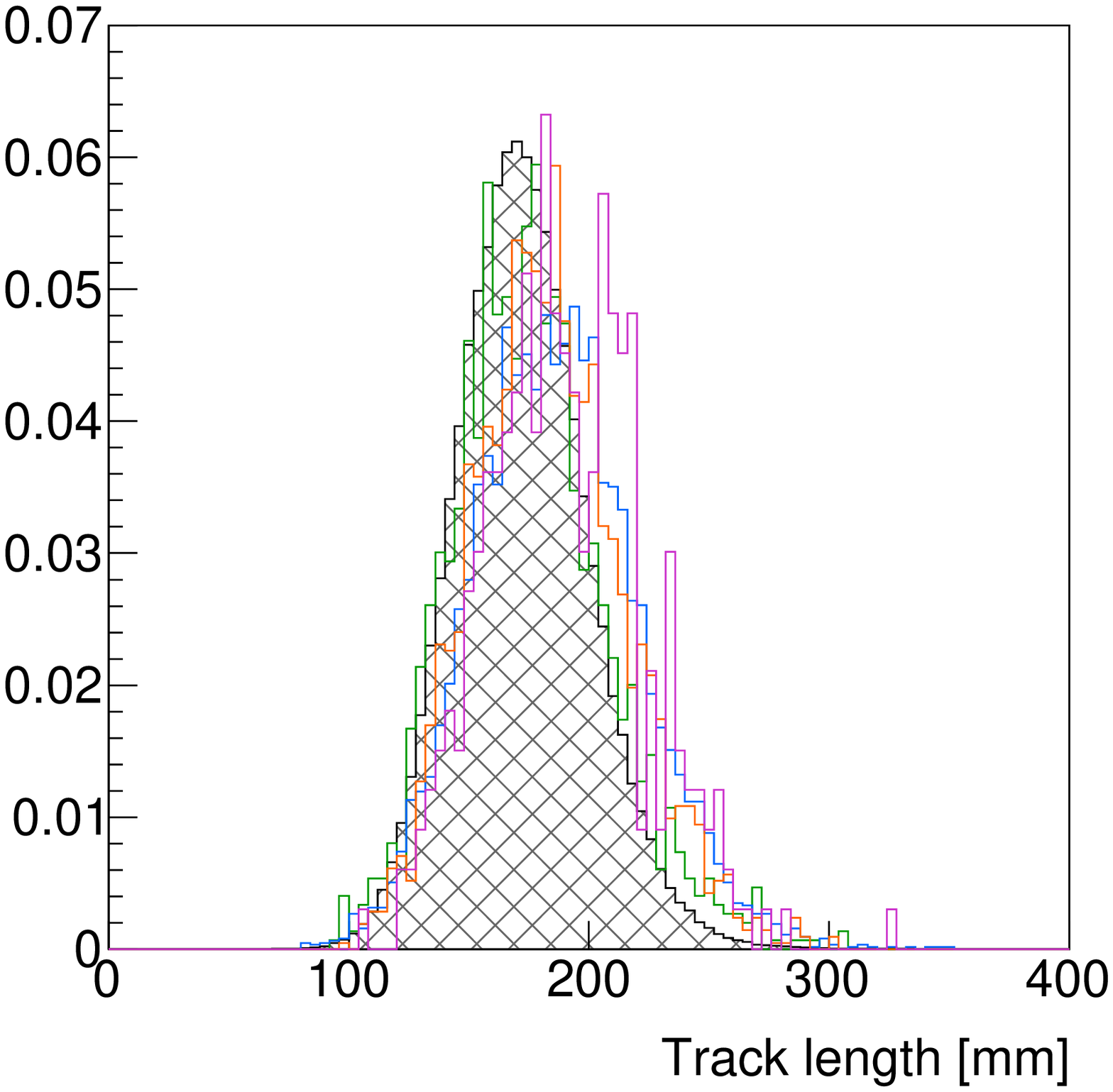} \\
(c) & (d) \\
\end{tabular}
	\caption{The distributions of topological parameters, normalized to unity, obtained from the XZ-projection of 0$\nu\beta\beta$ event tracks (filled curve), compared to the equivalent distributions of the four background datasets considered in our study (colored curves). The distributions are built using only the events contained in the ROI, i.e. in the energy range (2395,2520)\,keV. On top of that, each distribution accumulates the sequential topological criteria we applied in the preceding parameter.
		(a) The track energy ratio, or $\theta$ parameter. Different physics processes contribute to create certain structures on this distribution. Peaks found at low $\theta$-values, $\theta\lesssim0.01$, are related to the escape peaks of xenon, while hills appearing at higher values are related to bremsstrahlung gamma emission produced by high energy electron tracks, or Compton scattering processes that end up on a photoelectric process producing a main electron energetic track.
		(b) The lower track energy blob, $\mathcal{Q}^X_l$. Events satisfying $\theta<$1.33$\cdot$10$^{-3}$.
		(c) The lower \emph{twist parameter}, $\xi_l^X$, satisfying $\theta<$1.33$\cdot$10$^{-3}$ and $\mathcal{Q}^{X}_l<$176\,keV.
		(d) The track length, $\mathcal{L}^X$, satisfying $\theta<$1.33$\cdot$10$^{-3}$, $\mathcal{Q}^{X}_l<$176\,keV and $\xi^{X}_l<$0.0133.
}
    \label{fig:3mmdists}
\end{figure}

Signal and background events have been processed following the scheme described on~\ref{sc:dataChain}. Figure~\ref{fig:3mmdists} shows the parameter distributions obtained during the data processing chain, and described previously on section~\ref{sc:topoParams}. Except for the \emph{track energy ratio}, it must be noted that the other distributions follow similar patterns for different simulated background components.

We apply our criteria to discriminate signal from background sequentially on these parameters, following the order af the list provided in section~\ref{sc:topoParams}. We combine the XZ- and YZ-projections into a single parameter by choosing the minimum value of each projection for the \emph{blob charge}, $\mathcal{Q}^{min}$, and \emph{twist parameter}, $\xi^{min}$, i.e. for each event we only consider the 2-dimensional projection where \emph{tail} identification is more obvious. While we have chosen the maximum \emph{track length} of each projection, $\mathcal{L}^{max}$, supported by the argument that \NLDBD events are expected to be shorter. We have obtained our threshold values, or cuts, such that they maximize the signal-to-background significance by maximizing the figure of merit, $\epsilon_s/\sqrt\epsilon_b$, at each of the four topological criteria used in our analysis, i.e. \emph{track energy ratio}, \emph{blob charge}, \emph{twist parameter} and \emph{track length}. Here, $\epsilon_s$ and $\epsilon_b$ correspond to the signal and background reduction factors at each of those steps. The following topological parameter conditions were obtained after such optimization,

\vspace{-0.5cm}
\begin{eqnarray*}
	\quad\quad\quad \theta < 1.33\cdot10^{-3}\mbox{,}
	\quad\quad & \quad\quad
	\mathcal{Q}^{min}_l > 176\,\mbox{keV} \\[0.5em]
	\quad\quad\quad \xi^{min}_l > 0.0133
	\quad\quad\mbox{and} & \quad\quad 
	\mathcal{L}^{max} < 234.7\,\mbox{mm}.
\end{eqnarray*}

The final selected event population, signal or background events, will necessarily satisfy those conditions. We should mention that these values have been obtained by optimizing the dataset corresponding to the $^{238}$U vessel generated events. However, it must be noted that such a choice is not critical in view of the similar behavior of the different topological parameters produced by different background sources and components, as it can be observed in Figure~\ref{fig:3mmdists}. As a consequence, the threshold values obtained are not strongly dependent on this choice. Anyhow, we must remark that the most sensitive threshold value corresponds to the \emph{track energy ratio} parameter, as it can be induced from the distribution presented in Figure~\ref{fig:3mmdists}(a). Anyhow, in our particular conditions, we observe that the low value of $\theta$, resulting in our particular optimization, constrains our event selection practically to the population of events that contain only one single main track.

Before applying the sequential topological criteria we filter the event population to those events that are found in the ROI, defined as $Q_{\beta\beta}\pm$2$\sigma$, where $\sigma$=31\,keV is given by the assumed 3\%\,FWHM energy resolution, introduced in \emph{TRestHitsSmearingProcess} (see \ref{sc:evProcesses}). 

\begin{figure}[]
\centering
\begin{tabular}{cc}
	\includegraphics[width=0.48\textwidth]{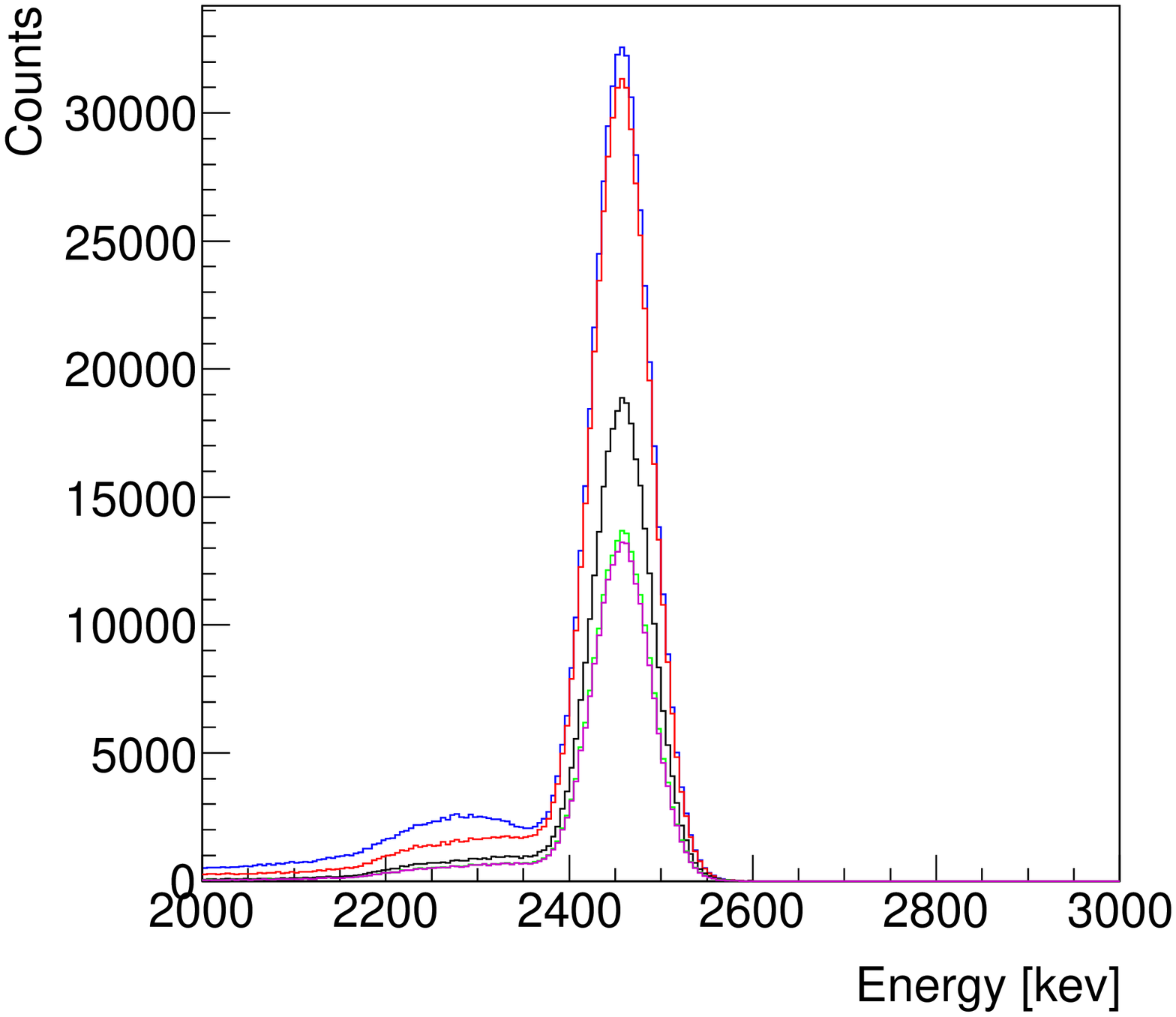} &
	\includegraphics[width=0.48\textwidth]{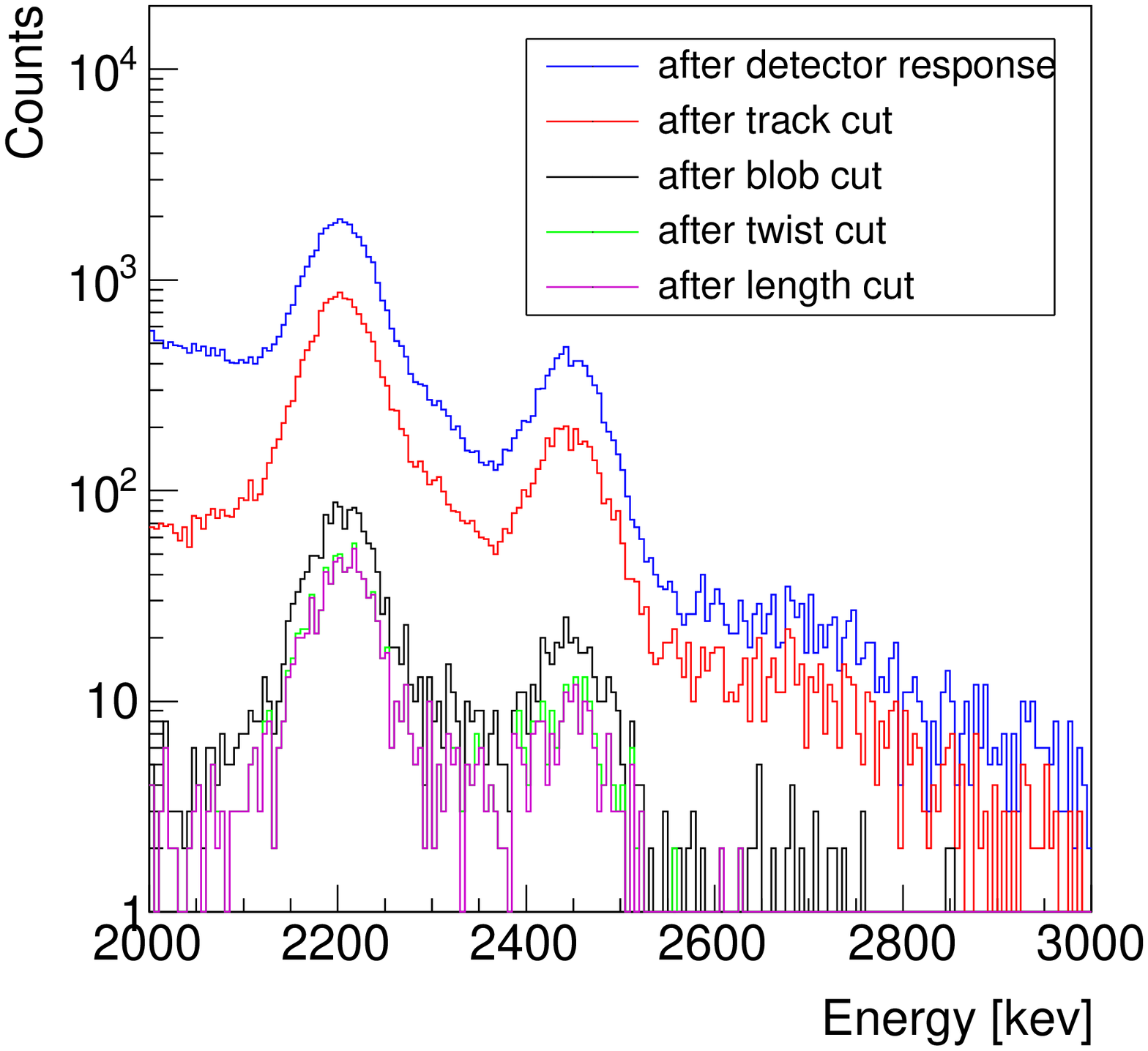} \\
(a) & (b) \\
\end{tabular}
\caption{(a) The simulated energy spectrum response of 0$\nu\beta\beta$ events around the $Q_{\beta\beta}$ value, and the resulting spectra after applying the different topological criteria sequentially, i.e. \emph{track energy ratio}, \emph{blob charge}, \emph{twist parameter} and \emph{track length}. The main peak corresponds to \NLDBD events for which full decay energy was acquired by our detector, while the distribution left tail corresponds to \NLDBD events with partial energy loss outside the detector readout plane boundaries, or partially contained in the detector acquisition window, limited by the electronics.
	(b) The equivalent background spectra resulting from $^{238}$U decays generated from the copper vessel. The peaks observed correspond to two high energy gammas, 2,204\,keV and 2448\,keV, from the spectrum gamma emission of $^{214}$Bi present in the $^{238}$U decay chain.
The 3\%\,FWHM Gaussian smearing introduced by the \emph{TRestHitsSmearingProcess} can be recognized in both figures. }
	\label{fig:xe_spectrum}
\end{figure}

The energy definition considered is the value extracted from \emph{TRestTriggerAnalysisProcess} (see~\ref{sc:anaPcs}) which contains all the effects introduced by the detector readout response. The resulting energy spectrum for \NLDBD, together with one of the background components studied is shown on Figure~\ref{fig:xe_spectrum}, including the accumulated effect of sequential cuts on signal efficiency and background reduction.

Table~\ref{table:pandax_result} shows the percentage of accepted signal events and background reduction factors, i.e. the final number of events surviving after each cut per simulated one. The first row, \emph{Geant4}, is the calorimetric response where we consider all the energy depositions in our active volume, here we only integrate events inside our ROI; the second row, \emph{Response}, corresponds to the population of events that remain in the ROI after applying the detector response, i.e. after applying readout fiducialization and boundaries imposed by the electronics acquisition window. We should remark, that when considering background events with energies above the ROI, they may finally get inside the ROI definition due to the partial event energy registered by our detector. The subsequent rows in the table show the accumulative effect of the topological cuts applied there after. The resulting topological selection contains only contributions from the $^{208}$Tl isotope decay produced inside the $^{232}$Th decay chain, and $^{214}$Bi produced inside the $^{238}$U chain.

\begin{table}
\center
\begin{tabular}{lccccc}
\toprule[0.4mm]
Origin		&	gas volume	&	\multicolumn{2}{c}{lateral vessel}	&	\multicolumn{2}{c}{Micromegas}	\\
Isotope		& 	$^{136}$Xe [\%]	&	$^{238}$U	& $^{232}$Th	&	$^{238}$U	& $^{232}$Th	\\
\bottomrule[0.2mm]
\emph{Geant4} 		& 71.1 &	2.28$\times$10$^{-5}$	&	2.10$\times$10$^{-4}$	&	2.69$\times$10$^{-3}$	&	3.38$\times$10$^{-3}$ \\
\emph{Response} 	& 52.6 &	7.40$\times$10$^{-6}$	&	8.87$\times$10$^{-5}$	&	1.65$\times$10$^{-3}$	&	1.56$\times$10$^{-3}$ \\
\bottomrule[0.1mm]
\emph{Track} 		& 37.7 &	1.71$\times$10$^{-6}$	&	1.13$\times$10$^{-5}$	&	9.60$\times$10$^{-4}$	&	1.14$\times$10$^{-4}$ \\
\emph{Blob} 		& 26.4 &	2.60$\times$10$^{-7}$	&	1.60$\times$10$^{-6}$	&	9.28$\times$10$^{-5}$	&	2.71$\times$10$^{-5}$ \\
\emph{Twist} 		& 22.0 &	1.64$\times$10$^{-7}$	&	1.03$\times$10$^{-6}$	&	4.87$\times$10$^{-5}$	&	1.97$\times$10$^{-5}$ \\
\emph{Length} 		& 17.5 &	7.85$\times$10$^{-8}$	&	6.07$\times$10$^{-7}$	&	3.06$\times$10$^{-5}$	&	1.43$\times$10$^{-5}$ \\
\bottomrule[0.4mm]
\end{tabular}
\caption{Percentage of accepted $^{136}$Xe signal events, or signal efficiency, together with the absolute background reduction factors at the ROI for the PandaX-III CDR baseline readout, after applying the detector response and sequential topological cuts described in the text. Each row includes the accumulative reduction of all previous cuts. For the sake of simplicity we avoid to express statistical errors in this table, although we will propagate them in Table~\ref{table:background_rate}, where final background rates are given. Systematic errors related to the Geant4 generator were studied previously in reference~\cite{chen2017pandax} using two independent Monte Carlo generators and geometry implementations. Any other source of uncertainty will be discussed in our conclusions. }
\label{table:pandax_result}
\end{table}

\begin{figure}
\centering
\begin{tabular}{cc}
	\includegraphics[width=0.48\textwidth]{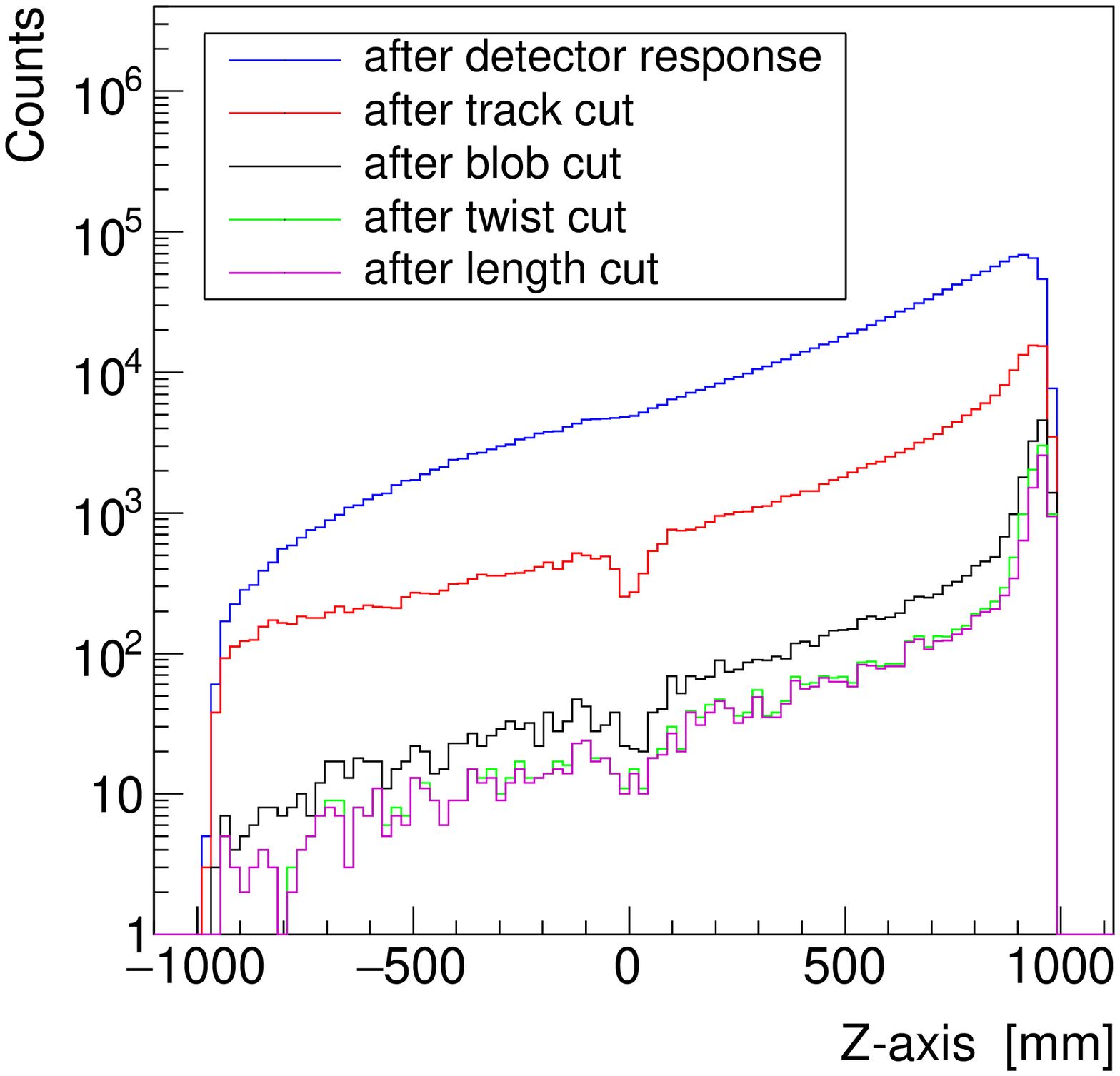} &
	\includegraphics[width=0.48\textwidth]{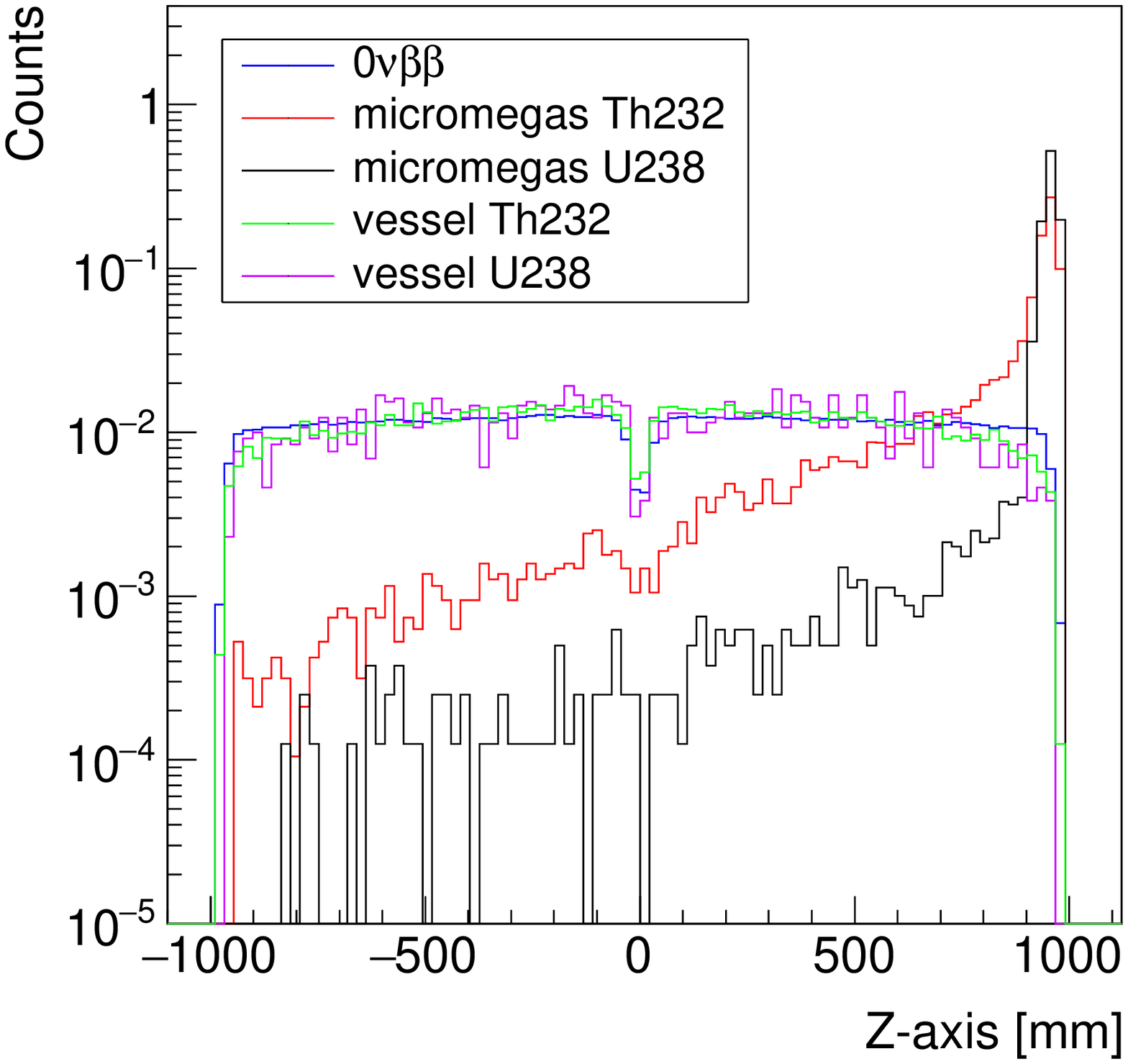} \\
	(a) & (b) \\
\end{tabular}

\caption{(a) The distribution of the average $z$-position for tracks from $^{232}$Th generated events originated from the Micromegas readout plane, placed at $z$=1000\,mm, after different steps of the topological cuts.
(b) The distribution of the average $z$-position for signal and the four background contributions studied, after all the topological cuts have been applied.}
    \label{fig:zmean_spectrum}
\end{figure}

We discover looking into Table~\ref{table:pandax_result} that the topological background reduction is about a factor 100, and it is comparable for all the background components except for the $^{238}$U Micromegas dataset, which is worse by about a factor 2. This result is explained mainly due to the failure of our topological cuts to discriminate surface events, as revealed by Figure~\ref{fig:zmean_spectrum}(a). The surface background contamination originated from the Micromegas readout is related to $^{214}$Bi decays producing $\beta$-tracks emanating from the readout plane. The $^{214}$Bi decay is the only isotope, inside the $^{238}$U decay chain, producing a $\beta$-decay with $Q_\beta$ end-point energy above our ROI. The surface contamination coming from the $^{232}$Th Micromegas decays is also present, but as it is observed in Figure~\ref{fig:zmean_spectrum}(b), there is also an important volume component which is due to a 2,614\,keV gamma which is produced at practically each $^{208}$Tl decay in the $^{232}$Th chain. The volume component due to $^{214}$Bi in the Micromegas is negligible, compared to its surface contribution, related to the fact that only about 1.5\% of the decays produce a 2,447\,keV gamma.

Table~\ref{table:background_rate} shows the final background contribution of the detector components studied, after renormalization with the corresponding material activities of $^{238}$U and $^{232}$Th for Micromegas and vessel components. We use the same values as reported in reference~\cite{chen2017pandax}. The surface contamination upper limit of Micromegas being \textless45\,nBq/cm$^2$ and \textless14\,nBq/cm$^2$, and the radiopure copper vessel bulk contamination being 0.75\,$\mu$Bq/kg and 0.2\,$\mu$Bq/kg\,\cite{Abgrall:2016cct}, for $^{238}$U and $^{232}$Th, respectively.

\begin{table}
\center
\begin{tabular}{ccccl}
\toprule[0.4mm]
		\multicolumn{2}{c}{Lateral vessel}	&	\multicolumn{2}{c}{Micromegas}	& \multirow{2}{*}{Units}						\\
	$^{238}$U	& $^{232}$Th	&	$^{238}$U	& $^{232}$Th	              &						\\
\bottomrule[0.2mm]
	6\,$\pm$\,1	&	13$\pm$1	&	\textless 1534	&	\textless 223  &	10$^{-3}$\,cpy	\\
0.365\,$\pm$\,0.042	& 0.752\,$\pm$\,0.032	& \textless 310	&	\textless 45.0                                                             &	10$^{-6}$\,keV$^{-1}$kg$^{-1}$y$^{-1}$	\\

\bottomrule[0.4mm]
\end{tabular}
\caption{Final background contribution after applying all the topological cuts, i.e. renormalizing the results of the last row presented on Table~\ref{table:pandax_result} using the expected $^{232}$Th and $^{238}$U isotope contamination for the different detector components. The results are presented in the usual units, counts per year (top row) and keV$^{-1}$kg$^{-1}$y$^{-1}$ (bottom row). Measurements of Microbulk Micromegas surface contamination provide results within the sensitivity limit of the measuring device, and therefore, our result for this component is given as an upper limit of the final background contribution to the experiment. }
\label{table:background_rate}
\end{table}

We conclude that the vessel contamination contribution is negligible compared to the contribution of the Micromegas readout plane, being its estimated overall contribution at the level of 1\,cpy. Of course, the final value will ultimately depend on the real surface contamination present at the Micromegas readout. Considering that the contamination levels used in our Monte Carlo are limits obtained experimentally we may still remain optimistic about this contribution. In section~\ref{sc:fiducial} we will show how we can achieve an additional reduction on the surface contamination near the readout planes by exploiting the effect of electron diffusion on the time signals registered by our TPC~\cite{Lin:2018mpd}. Finally, we must also consider our result as conservative from the point of view that there will be probably room for improvement by enhancing the existing parameter correlations, e.g. exploiting more sophisticated multivariate methods.

We reserve any discussion on the discrimination potential of the topological parameters used in our analysis for section~\ref{sc:granularity}, where we will also argue on the benefit of introducing the \emph{twist parameter} for our particular readout granularity, and we will assess the appropriateness of the 2-dimensional 3mm-pitch detector choice as a baseline for the PandaX-III experiment.

%% file: granularity.tex
\section{Readout topology and granularity studies}\label{sc:granularity}

\begin{table}[]
\center
\begin{tabular}{lccccccccc}
\toprule[0.6mm]
Origin		&	\multicolumn{3}{c}{1mm}	&	\multicolumn{3}{c}{2mm}	&	\multicolumn{3}{c}{3mm}	\\
Isotope		& 	$^{136}$Xe	&	$^{238}$U	& $^{232}$Th	&   $^{136}$Xe  &	$^{238}$U	& $^{232}$Th	&   $^{136}$Xe  &	$^{238}$U	& $^{232}$Th		\\
\bottomrule[0.2mm]
\emph{Track} 		& 69.6 &    9.92	&	 2.68	&	 63.9 & 8.91	&	 2.19&	 69.2 & 11.2	&	 2.79\\
\emph{Blob} 		& 58.5 &	9.13    &	 9.40	&	 59.4 & 10.2    &	 10.7&	 65.4 & 12.9    &	 14.0\\
\emph{Twist} 		& 85.1 &	60.6    &	 65.0	&	 93.9 &	78.7    &	 80.0&	 94.2 &	80.6    &	 81.1\\
\bottomrule[0.1mm]
\emph{Total} 		& 34.6 &	0.55    &	 0.16	&	 35.6 &	0.71    &	 0.19&	 42.6 &	1.16    &	 0.32 \\
\bottomrule[0.4mm]
\end{tabular}
\begin{tabular}{lccccccccc}
            &                           &                           \\
\toprule[0.4mm]
Origin		&	\multicolumn{3}{c}{4mm}	&	\multicolumn{3}{c}{6mm}	\\
Isotope		& 	$^{136}$Xe	&	$^{238}$U	& $^{232}$Th	&   $^{136}$Xe  &	$^{238}$U	& $^{232}$Th		\\
\bottomrule[0.2mm]
\emph{Track} 		& 73.1 &    13.2	&	 3.33	&	 79.2 & 16.7	&	 4.37\\
\emph{Blob} 		& 64.2 &	14.1    &	 14.5	&	 54.2 & 12.3    &	 12.2\\
\emph{Twist} 		& 94.0 &	83.2    &	 81.6	&	 90.2 &	82.5    &	 74.4\\
\bottomrule[0.1mm]
\emph{Total} 		& 44.1 &	1.55    &	 0.39	&	 38.7 &	1.69    &	 0.40\\
\bottomrule[0.6mm]
\end{tabular}
\caption{Relative signal efficiency and background reduction factors, in percentage, after each topological criterion, at different pitch values (1\,mm, 2\,mm, 3\,mm, 4\,mm and 6\,mm) for the 2-dimensional stripped readout. The initial population of events considered at each step are the events in the ROI surviving from the previous criteria.}\label{table:2D}
\end{table}

This section extends our study on background discrimination to different readout topologies and granularities. We will evaluate our new event reconstruction algorithms -detailed in~\ref{sc:REST}- using two different readout layouts, a 2-dimensional stripped layout and a 3-dimensional pixel layout, to assess the potential gain on signal efficiency and/or background reduction. The results presented in this section will provide insight on the optimum readout scheme to be used in the PandaX-III data taking conditions, i.e. a vessel filled with 10\,bar of xenon+TMA~1\% gas mixture. We will discuss then the appropriateness of the PandaX-III baseline readout choice -detailed in~\ref{sc:readout}- considering the advantages and disadvantages of 2-dimensional versus 3-dimensional event reconstruction, for different sizes of the readout channels, or detector granularity.

We have re-processed the same datasets analyzed in section~\ref{sc:panda3mm}. However, this time we focus only on the $^{238}$U and $^{232}$Th background contributions originated in the vessel. The same processing chain -as described in~\ref{sc:dataChain}- has been applied, being the readout used for event reconstruction the only modified element. We have systematically produced different readout structures in 2D and 3D at different pitch values. For the \emph{2-dimensional case}, the 3\,mm pitch PandaX-III baseline readout module design has been generalized to allow the definition of readout channels of any pitch value. When defining a different pitch value we keep the size of the module unchanged, e.g. when we define a 1\,mm pitch readout we increase the number of readout channels from 64 to 192. For the \emph{3-dimensional case} we have simply designed a readout module using squared pixels, and for practical reasons we adapt the number of pixels to keep constant the size of the readout module as a function of the pixel size. We must remark anyhow that minor adjustments are required at the processing chain when dealing with different pitch sizes, e.g. the \emph{cluster distance} used to identify tracks (see \emph{TRestHitsToTrackProcess} description at~\ref{sc:evProcesses}) was set to 2.5$\times$ the pitch size, and the radius of the \emph{blob charge} definition was fine tuned, being as low as $R=$1\,cm for 1\,mm pitch, and as high as $R=$2\,cm for 1\,cm pitch.

\begin{table}[]
\center
\begin{tabular}{lccccccccc}
\toprule[0.6mm]
Origin		&	\multicolumn{3}{c}{2mm}	&	\multicolumn{3}{c}{4mm}	&	\multicolumn{3}{c}{6mm}	\\
Isotope		& 	$^{136}$Xe	&	$^{238}$U	& $^{232}$Th	&   $^{136}$Xe  &	$^{238}$U	& $^{232}$Th	&   $^{136}$Xe  &	$^{238}$U	& $^{232}$Th		\\
\bottomrule[0.2mm]
\emph{Track} 		& 69.6 &    5.84	&	 1.60	&	 74.0 & 8.20	&	 2.20	&	 76.8 & 10.3	&	 2.73\\
\emph{Blob} 		& 79.1 &	13.4    &	 13.0	&	 74.8 & 13.2    &	 12.9	&	 65.2 & 10.4    &	 10.6\\
\emph{Twist} 		& 83.4 &	48.7    &	 46.5	&	 77.3 &	53.8    &	 49.0	&	 74.4 &	58.6    &	 52.6\\
\bottomrule[0.1mm]
\emph{Total} 		& 45.9 &	0.38    &	 0.097	&	 42.8 &	0.58    &	 0.14   &	 37.3 &	0.63    &	 0.15\\
\bottomrule[0.4mm]
\end{tabular}

\begin{tabular}{lccccccccc}
            &                           &                           \\
\toprule[0.4mm]
Origin		&	\multicolumn{3}{c}{8mm}	&	\multicolumn{3}{c}{10mm}	\\
Isotope		& 	 $^{136}$Xe  &	$^{238}$U	& $^{232}$Th	&   $^{136}$Xe  &	$^{238}$U	& $^{232}$Th		\\
\bottomrule[0.2mm]
\emph{Track} 			& 78.8 & 	12.0	&	 3.3	&	 80.3 & 13.6	&	 3.75\\
\emph{Blob} 			& 68.0 & 	10.6    &	 10.7	&	 64.1 & 10.8    &	 9.72\\
\emph{Twist} 			& -    &	-	    &	 -	 	&	 - 	  &		-   &	 -	 \\
\bottomrule[0.1mm]
\emph{Total} 			&  53.6   &	1.20    &	 0.35	&	51.5  &	1.47    &	 0.36\\
\bottomrule[0.6mm]
\end{tabular}
\caption{Relative signal efficiency and background reduction factors, in percentage, after each topological criterion, at different pitch values (2\,mm, 4\,mm, 6\,mm, 8\,mm and 10\,mm) for the 3-dimensional pixel readout. The initial population of events considered at each step are the events in the ROI surviving from the previous criteria.}\label{table:3D}
\end{table}

In this particular study we seek to compare the rejection potential using different readout topologies. We must remark that the fiducialization response is not relevant for such goal, and we are interested to determine the topological rejection potential without those constraints. Therefore, an important difference with respect to the analysis of the previous section~\ref{sc:panda3mm} is that we consider the full event interacting in the gas volume. In addition, our readout plane will fully cover the gas medium, or active area\,\footnote{Instead of using the 41-modules readout plane scheme described in~\ref{sc:readout}, we extend the readout definition limits by adding readout modules as needed to cover the full active area.}, so that the event reconstruction limits are just bound by the field cage wall defined in the \emph{Geant4} geometry.

\afterpage{
\begin{landscape}
\begin{figure}[p]
\centering
\begin{tabular}{cccc}
    \includegraphics[totalheight=5.25cm]{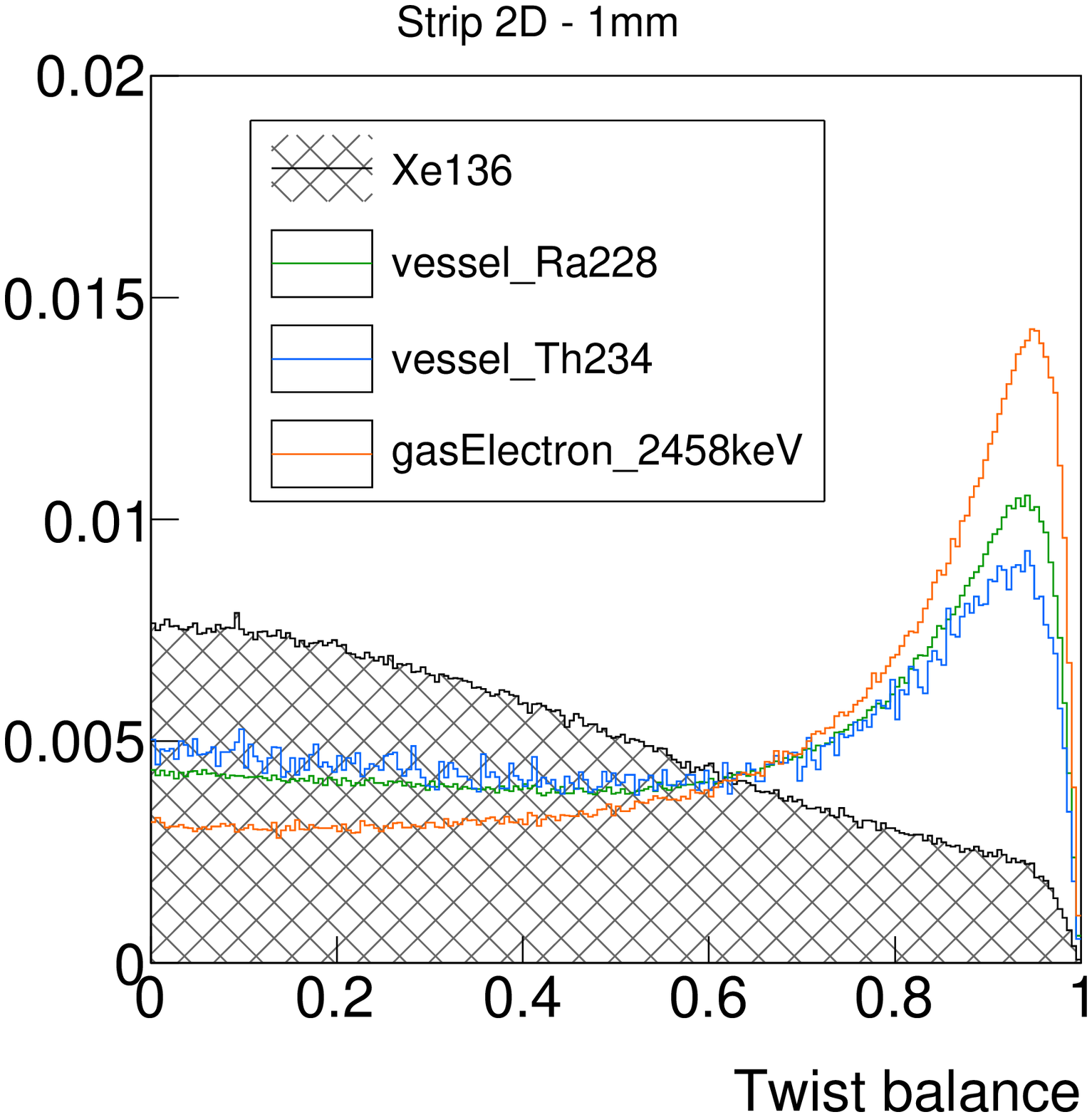} &
    \includegraphics[totalheight=5.25cm]{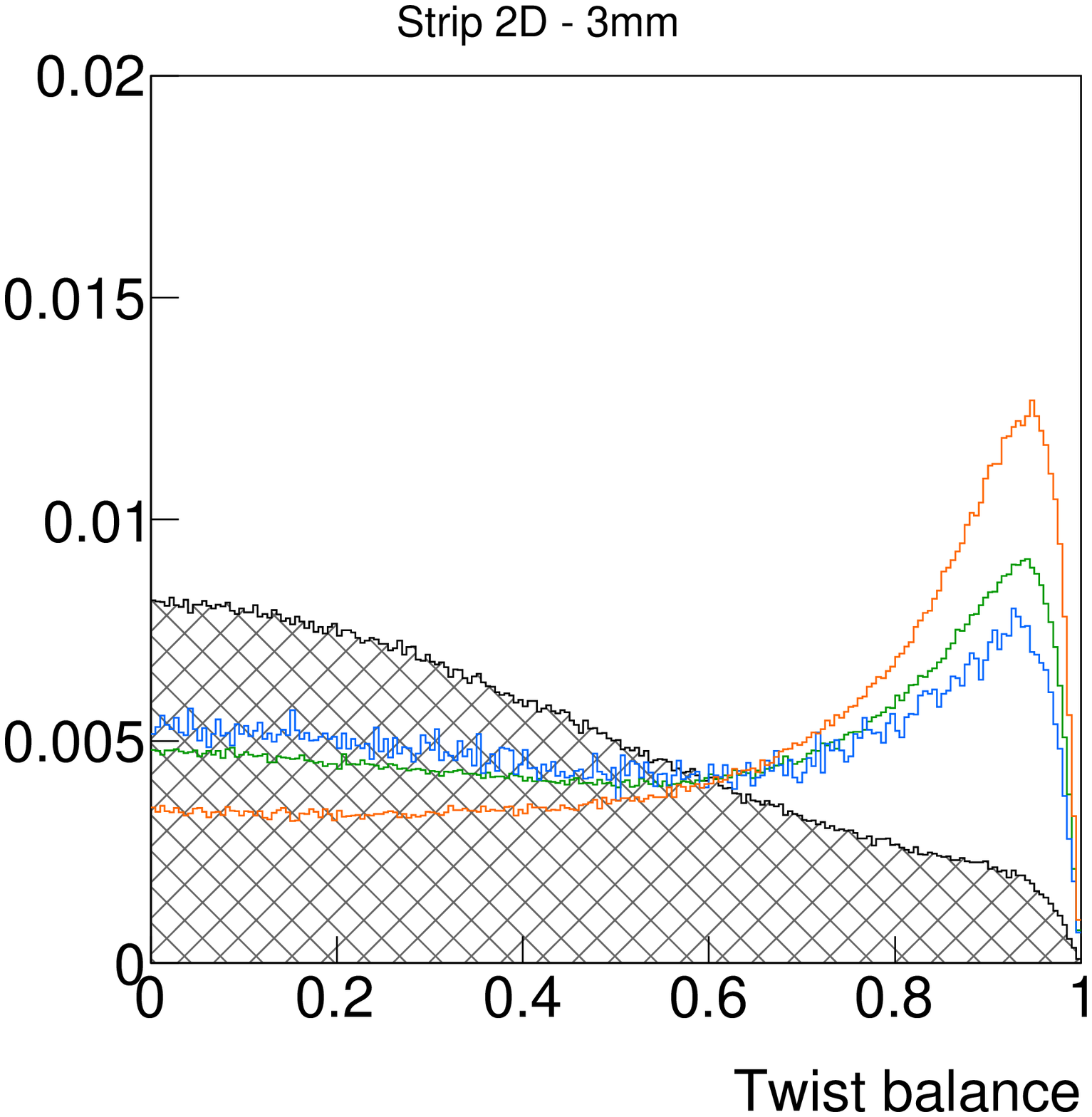} &
    \includegraphics[totalheight=5.25cm]{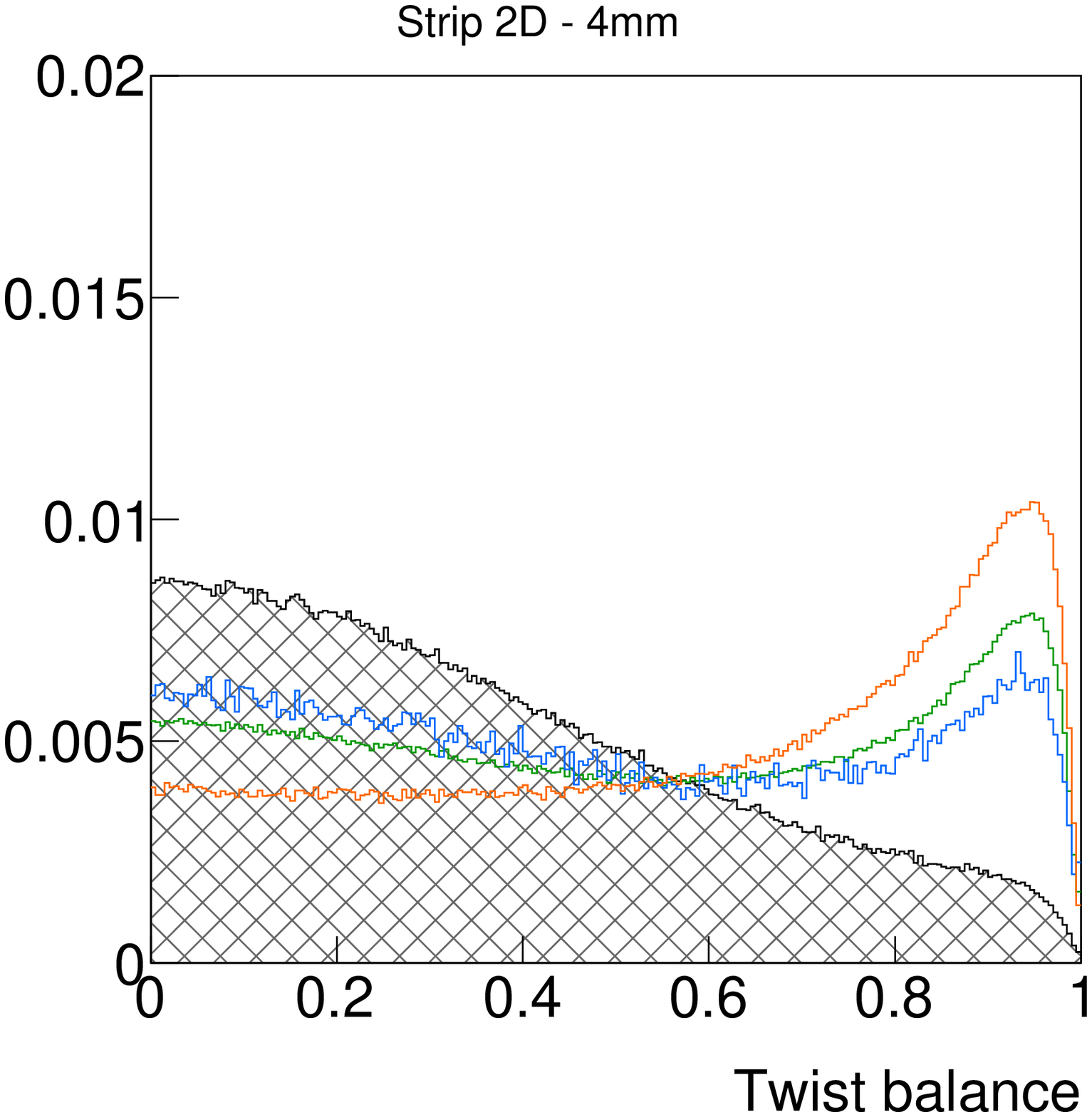} &
    \includegraphics[totalheight=5.25cm]{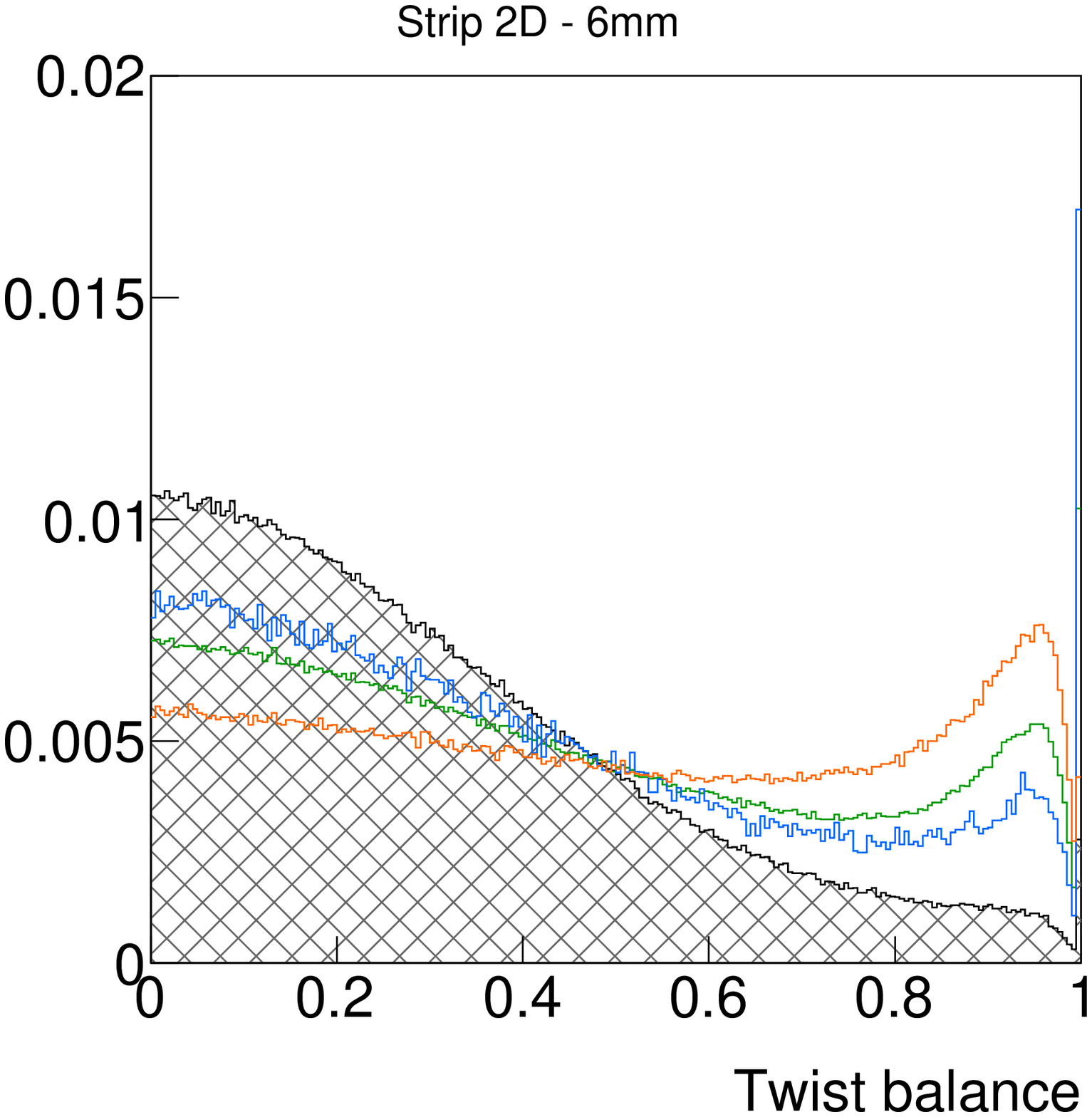} \\
	(a) & (b) & (c) & (d) \\
	 &  &  &  \\

    \includegraphics[totalheight=5.25cm]{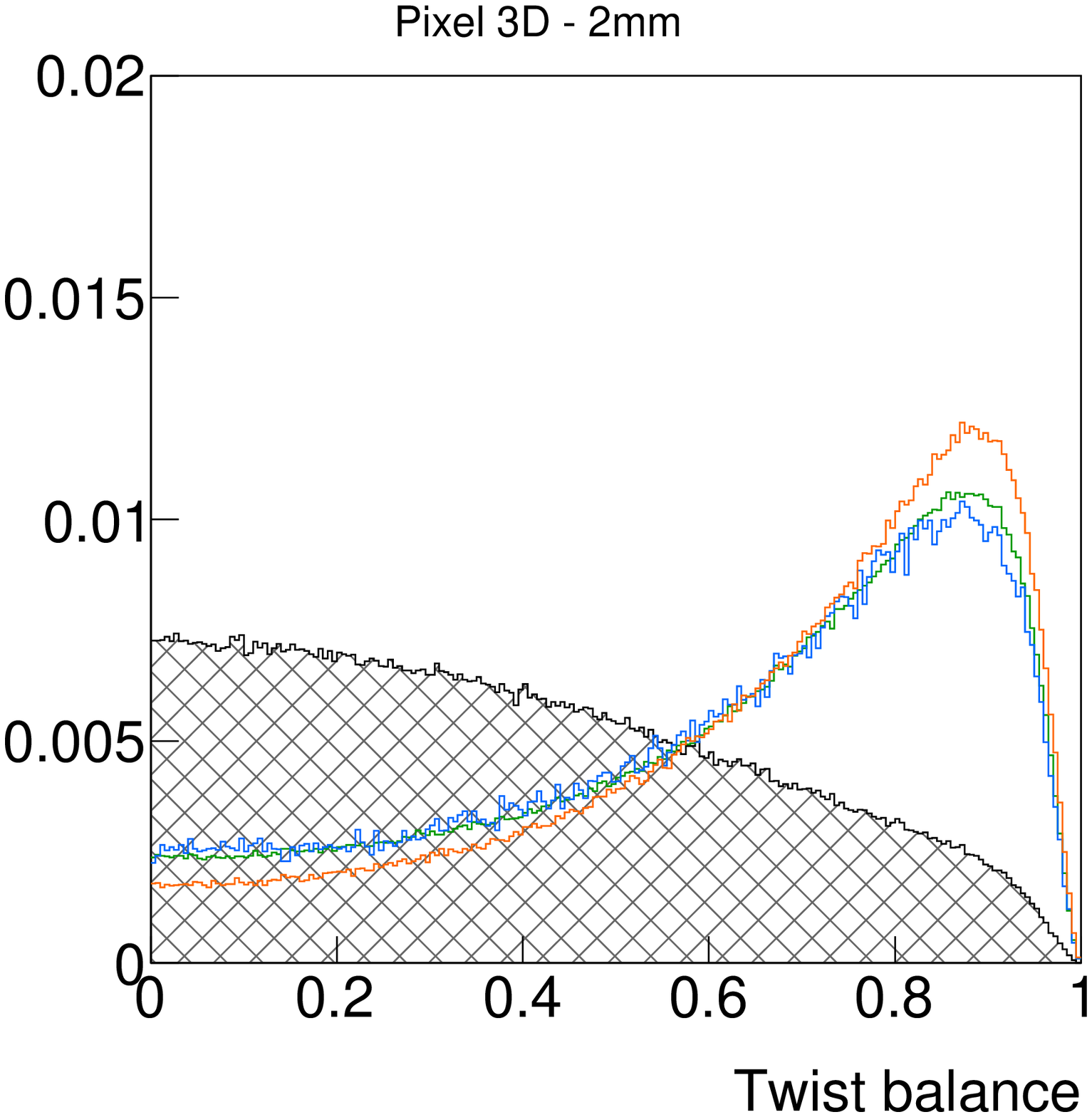} &
    \includegraphics[totalheight=5.25cm]{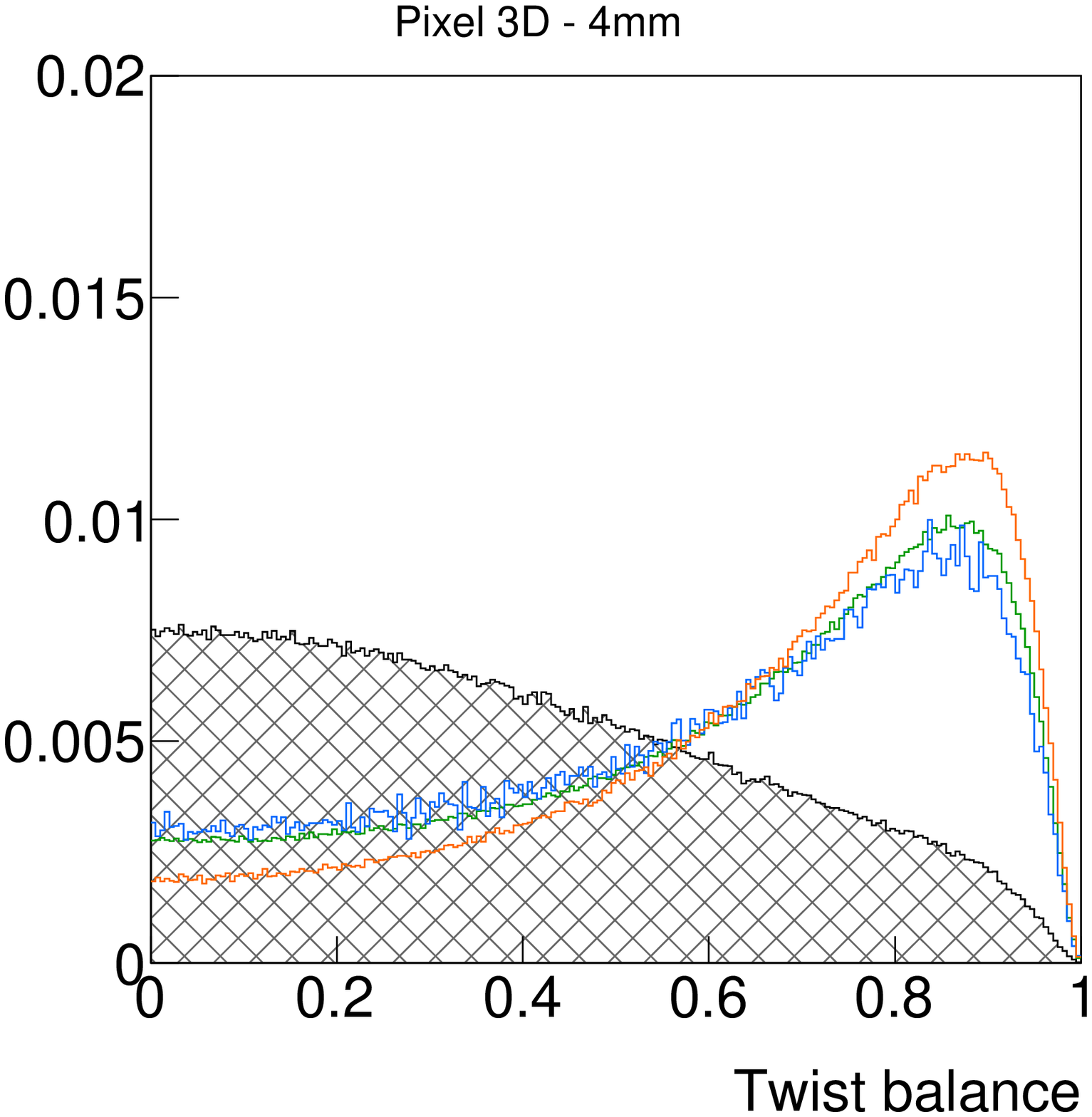} &
    \includegraphics[totalheight=5.25cm]{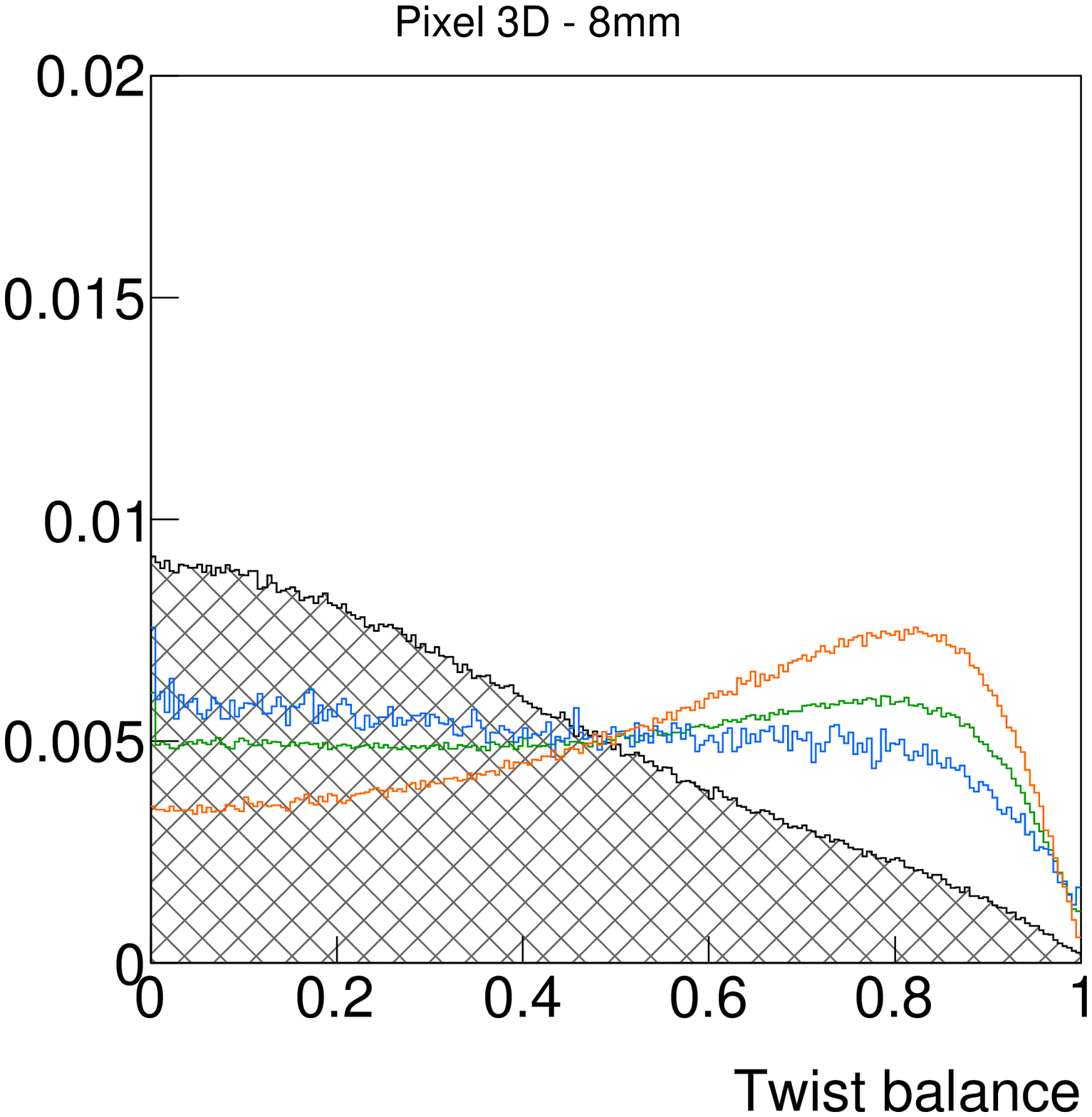} &
    \includegraphics[totalheight=5.25cm]{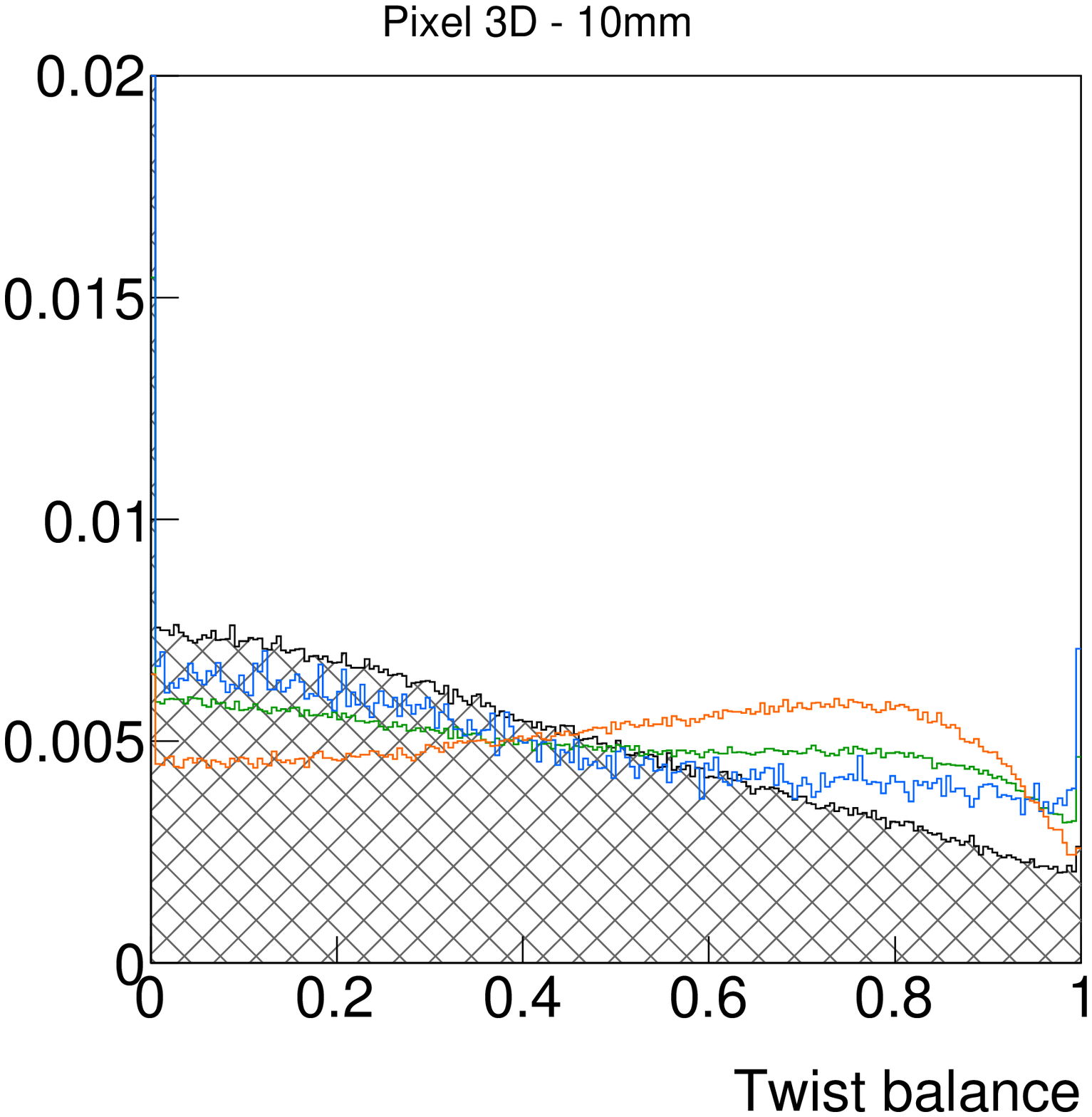} \\
	(e) & (f) & (g) & (h) \\
\end{tabular}
	\caption{ The distribution of the twist balance, $\xi_b$, parameter, a variant of the \emph{twist parameter} calculated as $\xi_b = (\xi_h-\xi_l)/(\xi_h + \xi_l)$, being $\xi_h$ the higher end-track \emph{twist parameter} value, and $\xi_l$ the lower end-track \emph{twist parameter} value. The different plots show the evolution of this parameter for different detector granularities, and different readout topologies, 2D strips (top figures), and 3D pixels (bottom figures). The distribution from \NLDBD signal events (filled curve) is compared to the background distribution of different contaminations as described in the text. For the 2D strips distributions only one of the projections is represented. }
    \label{fig:twistBalance}
\end{figure}
\end{landscape}
}

Tables~\ref{table:2D} and \ref{table:3D} show the effects of topological criteria, similar to those applied on section~\ref{sc:panda3mm}, for the simulated signal and background populations with different readout layouts and granularities. For each criterion (\emph{track energy ratio}, \emph{blob charge} and \emph{twist parameter}) we provide the relative reduction, in percentage, with respect to the surviving event population from the previous criterion, or cut. We maximize our signal significance at each step by maximizing the quantity, $\epsilon_s/\sqrt\epsilon_b$. We observe that the reduction potential of the \emph{track energy ratio} criterion is slightly better compared to the version where we included the fiducial detector response in section~\ref{sc:panda3mm}, i.e. by comparing the 3mm-pitch stripped readout results shown in Table~\ref{table:2D} with the values that can be deduced from Table~\ref{table:pandax_result}. This difference is mostly due to the fact that multi-track events have left the ROI after applying the detector response, and a subset of the background event population loses its multi-track features when we apply the detector response.

One interesting finding is that the newly introduced \emph{twist parameter} starts to be efficient below 6\,mm pitch, for both, the 2-dimensional and the 3-dimensional readout versions. As observed in Table~\ref{table:2D}, the 8\,mm and 10\,mm pitch pixel readouts studied did not improve the significance of the signal when using the \emph{twist parameter} criterion. Figure~\ref{fig:twistBalance} shows the evolution of the \emph{twist parameter} and it confirms the advantage to differentiate signal and background distributions as the pitch value is reduced. Furthermore, we have added to this figure a third dataset where we generate single electrons with energy equal to the $Q_{\beta\beta}$ of $^{136}$Xe, demonstrating a small dependency on the quality of the topological observables related to the nature of the background source.

The \emph{twist parameter} provides the less powerful topological criterion (as it is deduced from Tables~\ref{table:2D} and~\ref{table:3D}). However, we should take into account that this criterion operates after the event selection of \emph{track energy ratio} and \emph{blob charge} criteria. It is remarkable then that the \emph{twist parameter} criterion is still capable to improve the signal significance, meaning that the \emph{twist parameter} definition is an additional track feature that can be explored, exploited, and optimized in a future multivariate analysis that combines the track observables more efficiently.

\begin{figure}[h!]
\centering
\includegraphics[width=0.75\textwidth]{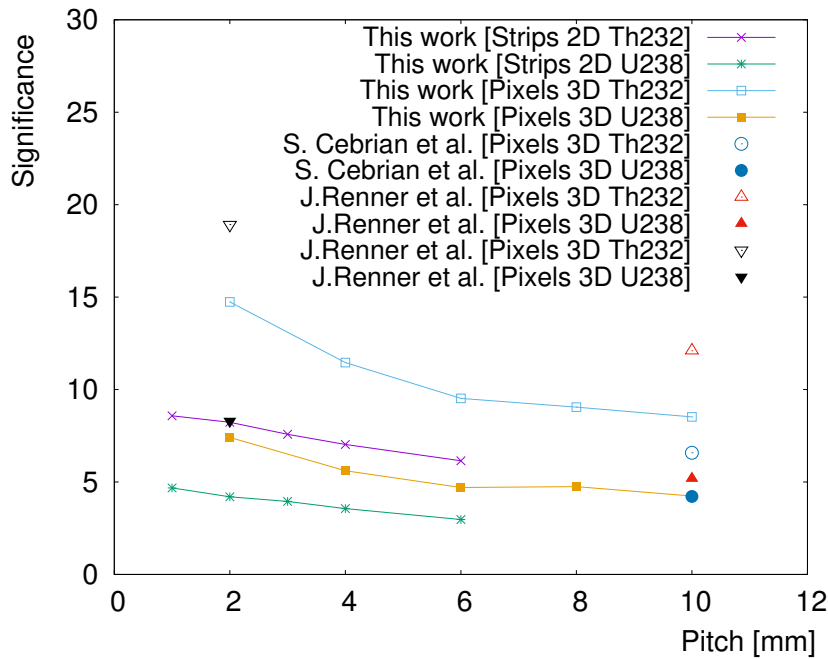}
\caption{ Total significance of topological criteria for the different detector granularities and topologies studied. The lines interconnecting data points show our results for the two different contaminations studied ($^{232}$Th and $^{238}$U), and the two different readout topologies (Strips 2D and Pixels 3D). We add to this figure the significances obtained in previous studies, calculated from the background and signal efficiencies reported at references~\cite{cebrian2013pattern,NEXTRennerNN}, and discussed in the text. }
    \label{fig:sig2d3d}
\end{figure}

We summarize our results in Figure~\ref{fig:sig2d3d}, where we present the total signal significance, measured as $\epsilon_s/\sqrt\epsilon_b$, obtained with our topological criteria for different readouts. This figure allows us to assess now the gain on experimental sensitivity when reducing the detector granularity, and quantify the impact of using a 2-dimensional readout compared to a 3-dimensional one. We conclude from these results that the negative impact of using a stripped 2-dimensional readout is counterbalanced by using a lower detector granularity, i.e. a 3\,mm pitch 2-dimensional readout is equivalent to a 1\,cm pitch 3-dimensional readout, with the advantage that the number of channels is reduced considerably in the 2-dimensional version.

Moreover, we use Figure~\ref{fig:sig2d3d} to compare our results to previously published studies - where they use similar conventional discrimination techniques. As it is observed, we find remarkably good agreement with reference~\cite{cebrian2013pattern} on the final significance at 1\,cm 3D pixel readout, for both vessel contaminations. While, for reference~\cite{NEXTRennerNN}, although we find reasonably compatible results for the $^{238}$U contamination chain (or $^{214}$Bi), we find their result on $^{232}$Th (or $^{208}$Tl) surprisingly high when compared to our data, being the origin of the main discrepancy at the level of the \emph{track energy ratio} criterion. The difference between our results and those reported in reference~\cite{NEXTRennerNN} are even larger when we consider that their 1\,cm results correspond to pure xenon, and it was claimed in reference~\cite{cebrian2013pattern} that background reduction due to blob identification is worse by a factor 3 due to the higher electron diffusion in pure xenon, and that finally implies almost a factor 2 worse on significance. The source of this discrepancy might be related to the non-independent treatment of electron diffusion and detector granularity in~\cite{NEXTRennerNN}, and the higher gas pressure, 15\,bar, of their setup. On the other hand, we should be cautious, since direct comparison of different setups is not always obvious. The significance of the criteria may depend not only on the origin, or the particular geometry of the setup, but also on the nature of the background and the target density, or detector pressure. We have seen, for example in Figure~\ref{fig:twistBalance}, how a single electron track artificially launched from the gas volume, as it is done e.g. in reference~\cite{Ai_2018}, slightly generates better results in terms of the topological parameters quality.






%% file: fiducial.tex
\setcounter{footnote}{0}


\section{Fiducial background rejection in charge based TPC readouts.}\label{sc:fiducial}

The results obtained in section~\ref{sc:panda3mm} unveiled the weakness of our topological cuts to discriminate surface contamination events originated from the Micromegas readout planes, as revealed by Figure~\ref{fig:zmean_spectrum}(a). Our setup lacks an absolute reference time, $t_o$, from the interactions taking place in the detector volume. Without such a reference we cannot directly determine the event absolute z-position, making it hard to discriminate surface events originated on the readout planes, or the cathode.

In this section we develop a basic technique to demonstrate the remaining potential for further background reduction by attenuating the negative impact of surface contamination, even in the absence of $t_o$. For that we take advantage of the readout signal dependence on the diffusion of ionized electrons drifting towards the readout planes. The time signals induced on the Micromegas readout, due to a particular energy deposition on the TPC volume, will be broader when the electron cloud has drifted a longer distance towards the readout plane. However, the resulting primary electrons from $\beta$-tracks generated near the readout plane will drift a very short distance, and therefore the original charge distribution will be almost unaffected by the electron drift.

\begin{figure}[t]
\centering
\begin{tabular}{cc}
\includegraphics[width=0.49\textwidth]{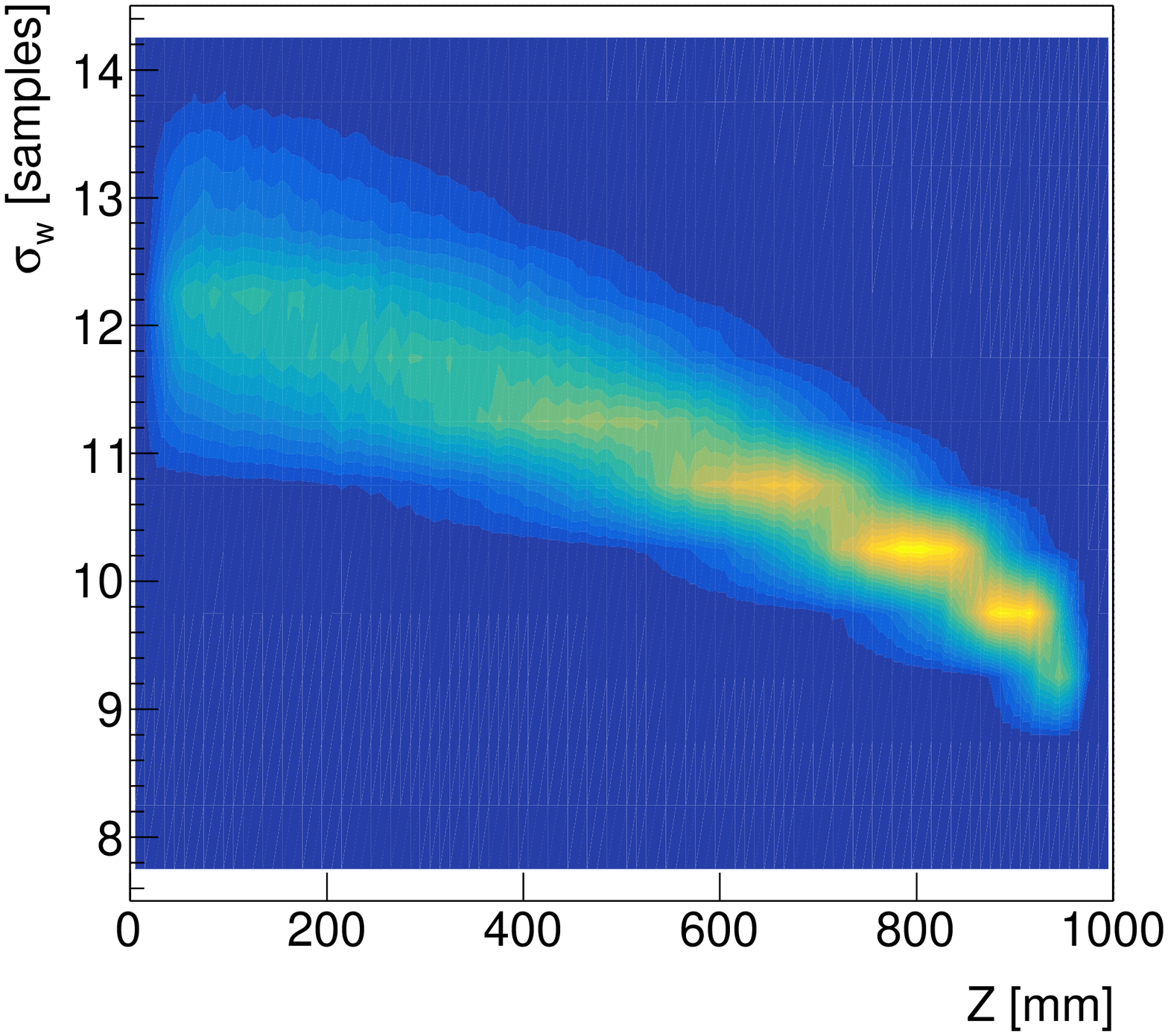} &
\includegraphics[width=0.49\textwidth]{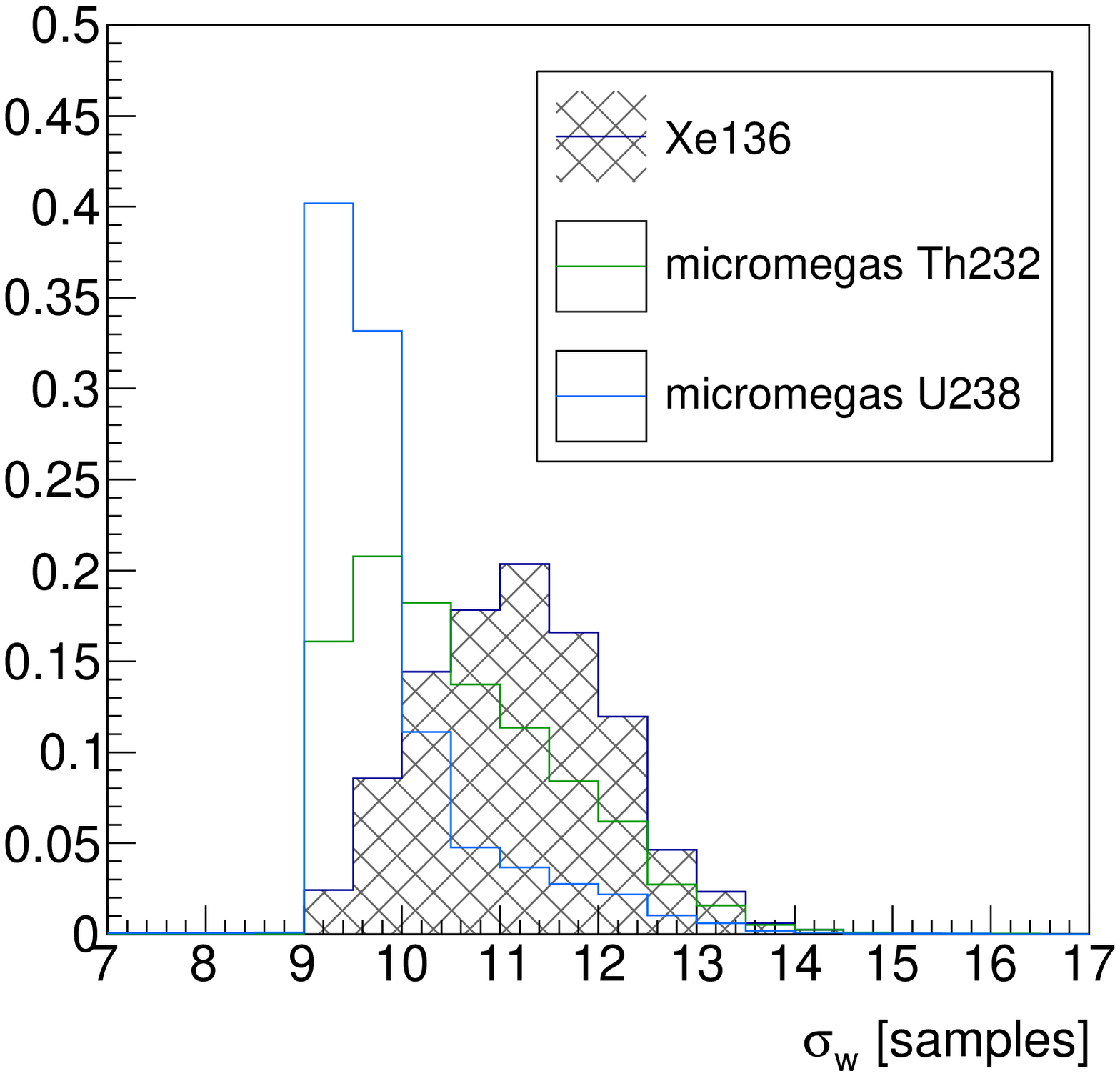} \\
(a) & (b) \\
\end{tabular}
\caption{(a) A contour plot produced using the \NLDBD signal event population, and showing the diffusion parameter, $\sigma_w$, as a function of the event position. Only one half of the symmetric TPC is shown here. Lower $\sigma_w$ values correspond to events taking place near the readout plane, at z=1000\,mm, and larger values correspond to events taking place near the cathode, at z=0\,mm. (b) The distribution of the $\sigma_w$ parameter for \NLDBD signal events (filled curve), and the two background Micromegas components studied.}
    \label{fig:widthDistribution}
\end{figure}

We have re-processed the \NLDBD signal and Micromegas background datasets analyzed in previous sections using the same data chain parameters as defined in section~\ref{sc:panda3mm}. However, now we need to include few additional effects in order to extract a parameter, $\sigma_w$, that correlates to the diffusion of electrons in the gas. The obtention of this parameter is performed under realistic conditions including the signal shaping introduced by electronics and a typical electronic noise level (further details on the extraction of this parameter are given in~\ref{sc:RawProcessing}). In Figure~\ref{fig:widthDistribution} we observe the desired dependence of $\sigma_w$ with the z-position of the simulated \NLDBD event, together with the $\sigma_w$ distribution produced by the two Micromegas background components compared to the \NLDBD signal one. We observe clearly how background surface events peak at low values of $\sigma_w$.

Therefore, we exploit the properties of $\sigma_w$ to introduce an additional selection criterion in our data by accepting only the population of events that are above a certain diffusion threshold, as quantified by $\sigma_w$. Table~\ref{table:diffusionCuts} presents the results on event acceptance after applying cuts for different threshold values, where we observe that we manage to reduce by an additional factor 0.15 the contribution from the $^{238}$U chain, due to $^{214}$Bi decays, while keeping our signal efficiency near 75\%. We must remember that the Micromegas contribution of $^{238}$U was a dominant component in Table~\ref{table:background_rate}. The background reduction on the $^{238}$U chain is more significant than in the $^{232}$Th chain, due to the fact that the 2,614\,keV gamma from $^{208}$Tl is produced in almost any decay, generating finally a higher ratio of volume to surface events compared to the $^{214}$Bi contribution, which produces a $\beta$ decay with high energy gamma emmission only at 1.5\% of the decays. We should remark that the values presented on Table~\ref{table:diffusionCuts} have been obtained without applying topological cuts. Still, these results are independent on previous topological cuts, since we do not observe any dependence on the z-distribution after cuts, as it would have been revealed by the \NLDBD population in Figure~\ref{fig:zmean_spectrum}(b), after the topological analysis performed in section~\ref{sc:panda3mm}.

\begin{table}[]
\center
\begin{tabular}{lccc}
\toprule[0.4mm]
& 	$\sigma_w>$9	& 	$\sigma_w>$10 	& 	$\sigma_w>$10.5 \\
\bottomrule[0.2mm]
$^{232}$Th 		& 93.3 &	53.7    &	 44.4	\\
$^{238}$U		& 80.9 &	20.1    &	 15.1	\\
$^{136}$Xe 		& 99.4 &    83.0	&	 74.5	\\ 
\bottomrule[0.4mm]
\end{tabular}
\caption{Overall background reduction, in percentage, for events generated from the Micromegas readout plane, $^{232}$Th and $^{238}$U contaminations, and the corresponding signal efficiency, after applying different cuts on the diffusion parameter, $\sigma_w$. }\label{table:diffusionCuts} 
\end{table}

\begin{figure}
\centering
\begin{tabular}{cc}
\includegraphics[width=0.49\textwidth]{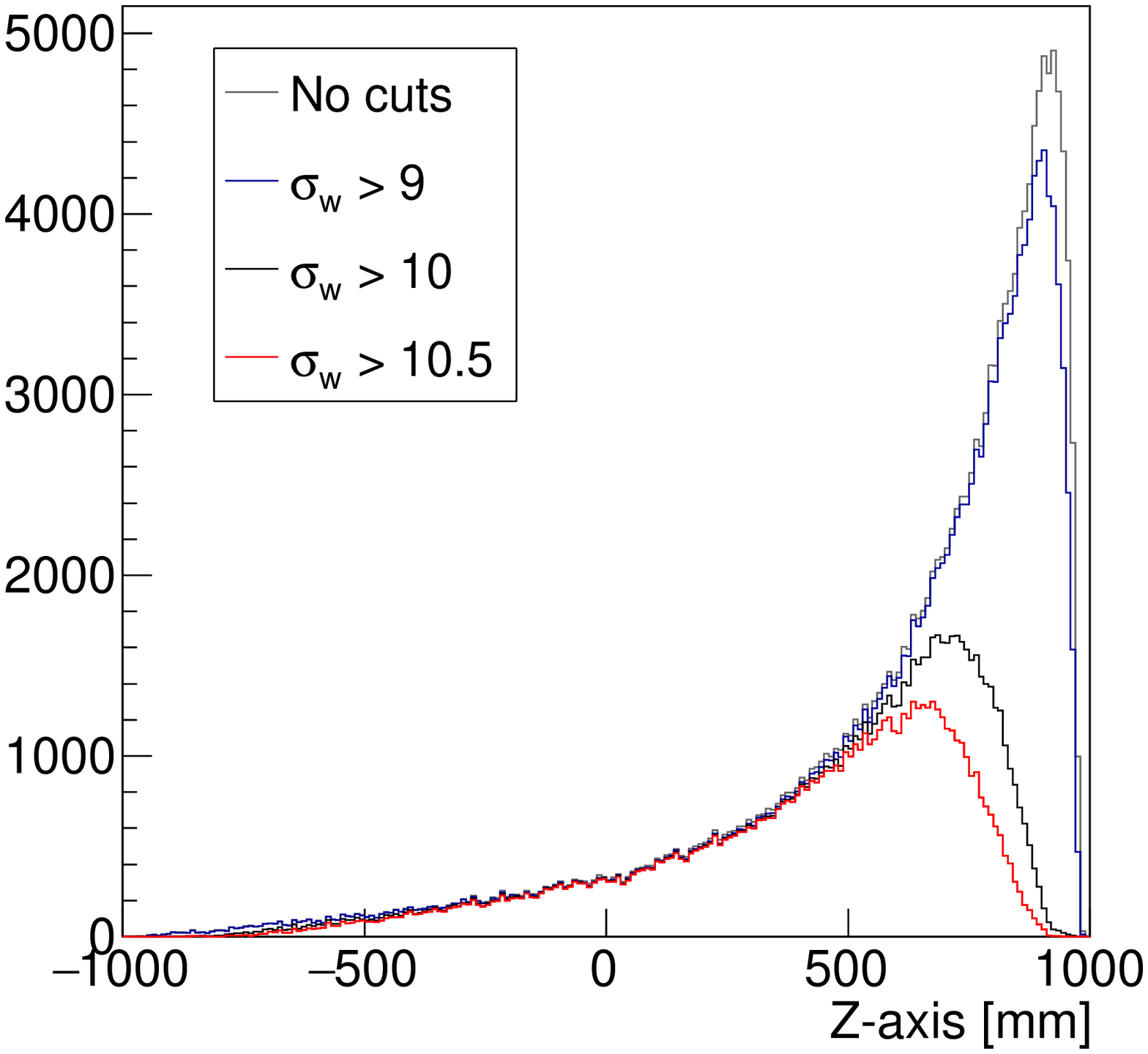} &
\includegraphics[width=0.49\textwidth]{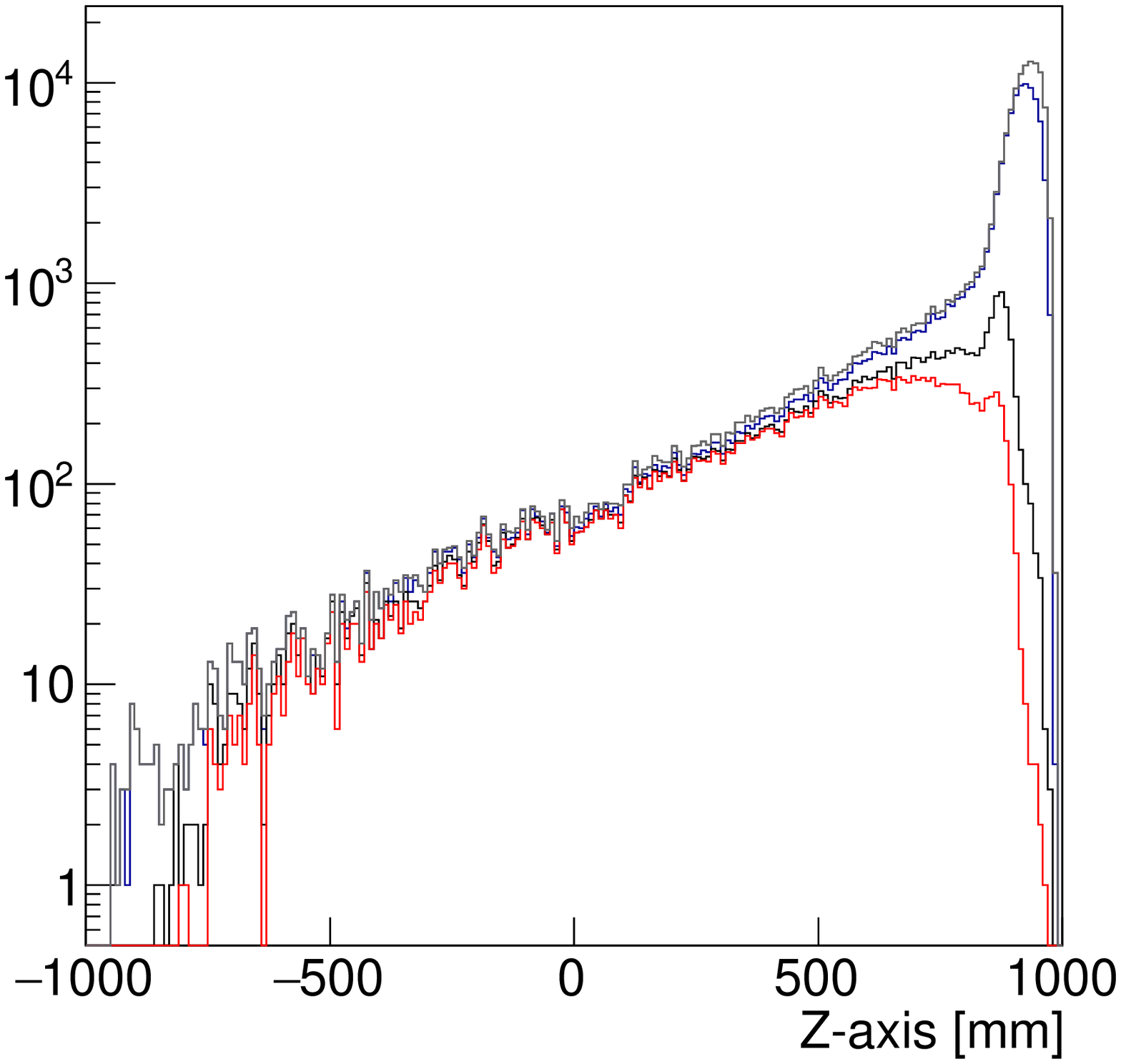} \\
(a) & (b) \\
\end{tabular}
\caption{ Distribution of the average z-position for events produced by $^{232}$Th (a), in linear scale, and $^{238}$U (b), in logarithmic scale, generated from the Micromegas readout plane, placed at z=1000\,mm. Different curves represent the background reduction after applying different threshold values on the diffusion parameter, $\sigma_w$. The initial population of events, tagged as "No cuts", are the events found in our ROI definition. }
    \label{fig:zMeanDiffCuts}
\end{figure}

Finally, in Figure~\ref{fig:zMeanDiffCuts} we show the background distribution as a function of the event position, for each of the two Micromegas isotope contaminations, after applying different $\sigma_w$ threshold values. In these figures we observe how all the events from surface nature have been completely removed already for $\sigma_w>$10, and how the remaining background event population on Table~\ref{table:diffusionCuts}, i.e. 20.1\% for $^{238}$U and 53.7\% for $^{232}$Th, are only due to volume events that can only be removed by exploiting topological event features.

%% file: conclusions.tex

\section{Conclusions}\label{sc:conclusions}

In this work, we have presented results on topological background discrimination considering a realistic detector response for the PandaX-III TPC baseline design. The realistic response includes an accurate description of the existing Micromegas detector layout used for the reconstruction of events in a similar way as it is done with experimental data. However, the most important element on the detector response relies on an appropriate definition of the energy of each event by considering the electronic acquisition window and the readout plane boundaries of the PandaX-III design. 

We observe a negative effect on our topological background discrimination when we apply the detector response to the Monte Carlo event data, hence the importance to include those effects in our study. The results obtained show that a promising final background level is achievable. Using topological arguments, we find that for a signal reduction of $\sim$1/3 we can achieve a reduction factor of about 100 on the background contamination. Even after including the detailed detector response, this result is still in good agreement with the preliminary estimate used to calculate the sensitivity of the first PandaX-III module at reference~\cite{chen2017pandax}.

It is remarkable that the overall background contribution of the detector vessel will be negligible for the lifetime of the experiment. While an acceptable value for the contribution of one of the most uncertain sources of background in the experiment - the detector readout plane - results in a conservative contribution of 1-2 counts per year. 

Furthermore, we demonstrate the capability of our setup to perform a fiducialization in the z-axis even in the absence of $t_o$, and remove a considerable amount of surface background events that cannot be eliminated by other means. By exploiting this technique we are able to reduce the Micromegas background contribution to levels well below 1 count per year, preserving 80\% of the \NLDBD signal.

We have also addressed the question relative to the impact of a 2-dimensional readout on the pattern recognition of \NLDBD. It is obvious that the lack of a full 3-dimensional event reconstruction leads to a reduced discrimination power. In the 2-dimensional case, our event identification and classification must be based on 2 projections of the event, and necessarily some of those event projections hide information that can only be revealed by a 3-dimensional reconstruction. Occasionally, a secondary track will be hidden by the projection and counted as energy of the main track, affecting the \emph{track energy ratio} criterion, while in other events, the integration of the charge of an electron moving orthogonally to the projection may mimic a \emph{blob}, reducing the rejection power of the \emph{blob charge} and the \emph{twist parameter} criteria. We have systematically studied the significance of our topological criteria for different detector granularities at 3-dimensional pixel and 2-dimensional stripped readout designs, concluding that the negative impact on 2-dimensional readout pattern recognition is counterbalanced by a reduced detector granularity, i.e. a 1\,cm pixel readout reaches an event topological rejection power comparable to a 3\,mm stripped readout. Consequently, the 3\,mm pitch stripped baseline readout choice for PandaX-III detector is much more convenient, not only in terms of detector design complexity due to the reduced number of channels required, but also minimizing the impact of typically non-radiopure electronics equipment near the detector readout planes.

We have introduced novel track parameters, as the \emph{end-track twist} and the \emph{track length}, never exploited before on the pattern recognition of \NLDBD using conventional techniques. We have assessed their discrimination potential, and observed how the \emph{end-track twist} feature is still contributing to the signal-to-background significance, even after filtering the event population previously with other topological criteria, as the \emph{track energy ratio} and the \emph{blob charge}. Furthermore, we have found that this parameter is relevant for pattern recognition when the detector granularity is below 6\,mm, concluding that such granularity or better is necessary to reveal the aforementioned track features.

We believe our results are conservative, and that there exists room for improvement and optimization through multivariate methods that better exploit the different parameter correlations. Additionally, exploring the use of novel track parameters, or improving the existing ones, can push the conventional background discrimination further. Additional parameters, as the $dE/dx$ along the track, previously studied at reference~\cite{Zeng:2016rcz}, and its peculiarities at the end-tracks could allow to improve these results by properly combining them with the parameters studied in this work.

Recently, we have seen how machine learning methods provide enhanced capabilities for pattern recognition, as reported in references~\cite{NEXTRennerNN,Qiao2018,Ai_2018}. These techniques, which provide excellent signal-to-background discrimination ratios, could never replace a conventional analysis. In a conventional analysis the different topological parameters serve as a mechanism to control the goodness of the Monte Carlo event reconstruction. In other words, we will be able to validate the different event parameters obtained by comparing with the experimental data, allowing us to be confident on our event reconstruction and the effectiveness of our topological criteria. The use of machine learning techniques, without further or complementary analysis, will not be sufficient by itself to prove the background of the detector.

The present work served to demonstrate the existing software infrastructure implemented in REST for event reconstruction, and the flexibility to add a detailed detector response at different levels, as necessary. REST has been conceived to provide a high degree of modularity allowing to connect or disconnect different \emph{event processes} at any place of the \emph{event data} processing chain. We have designed dedicated \emph{event processes} and \emph{metadata} structures to construct the PandaX-III Monte Carlo processing chain. Still, additional details can be included in future studies, that may implement even more realistic event reconstruction. A future Monte Carlo event processing may consider further details ignored in this work, as e.g. the gap inter-distance between neighbour \emph{readout modules}, or the necessary unconvolution of the electronic signal shaping. 

Finally, when comparing to real experimental conditions we will need to consider different systematic effects due to different parameters that may fluctuate, or change, during the data taking periods of the experiment, such as the electronic noise, the drift field intensity or irregularities, the concentration of TMA and its effect on the gas properties, the gain inhomogenities on each detector module, the effect of readout module gaps on the event reconstruction, possible dead channels in the detectors, etc. All those studies must be addressed in the near future in order to have full control on the detector performance. The present work represents a necessary step towards those studies.

%% file: REST.tex
\section{REST as a data analysis framework.}\label{sc:REST}

RESTSoft (Rare Event Searches with TPCs Software) is a collaborative software effort providing a common framework for acquisition, simulation, and data analysis for gaseous Time Projection Chambers (TPCs)\,\cite{tomas2013development}. REST is composed of a set of libraries written in C++ and is fully integrated in ROOT~\cite{ROOT}, i.e. all REST classes inherit from TObject and can be read/accessed/written using the ROOT I/O interface. The only structural dependence is related to ROOT libraries, while other packages, as \emph{Geant4}~\cite{Agostinelli:2002hh} or \emph{Garfield++}~\cite{Garfield}, can be optionally integrated and used within the REST framework when generating or processing Monte Carlo data.

During the last years, major upgrades took place on the REST core libraries\,\cite{Galan_8thTPC}. An important feature introduced in REST is that \emph{metadata} and \emph{event data} are stored together in a unique file. We understand by \emph{metadata} any information required to give meaning to the data registered in the \emph{event data}, as it can be the initial run data taking conditions, the geometry of the detector, the gas properties, or the detector readout topology. Additionally, any input or output parameters, required during the processing or transformation of \emph{event data}, using \emph{event processes}, will be stored as \emph{metadata}. Any previous existing \emph{metadata} structures inside the REST input file will be transferred to any future output, assuring full traceability, as well as reproducibility of results obtained with a particular dataset.

An ambitious feature of REST is its capability to analyze Monte Carlo and experimental data using common \emph{event processes}. This is possible by using existing REST \emph{event processes} to condition the input data generated, for example, by a \emph{Geant4} Monte Carlo simulation. After an appropriate \emph{event data} conditioning, our Monte Carlo generated event will reproduce the \emph{rawdata} of the detector (as it is shown in~\ref{sc:RawProcessing}). Once we are at that stage, we can benefit from using the same \emph{event processes} to analyze Monte Carlo and real experimental data. A realistic Monte Carlo \emph{rawdata} reconstruction will allow us to assess, validate and optimize the processes that will be involved in the real event reconstruction and analysis even before the start of the physics run of the experiment.

In the following subsections, we recall the definitions of the different components of REST, \emph{viz.} \emph{event types}, \emph{event processes}, and \emph{analysis tree}. These definitions will serve as a reference for the article. Note that we do this having in mind the case where the physics variables of interest are local energy deposits, called hits, and the signal is digitized by a sampling ADC. The REST software is versatile enough, though, to handle many other cases. We include here only those components of REST that are relevant to our study.

\subsection{Event types}

REST defines basic \emph{event types}, or data structures, commonly used to store \emph{event data} generated during the data acquisition, production and/or \emph{event data} processing. The definition of a reduced set of basic \emph{event types} allows for better inter-process connectivity. The following data structures defined in REST are used in our study.

\begin{itemize}
\item \emph{TRestHitsEvent:} This structure contains an arbitrary number of elements or hits used to store a physical variable, $\phi_i$, defined in a 3-dimensional coordinate system $(x_i, y_i, z_i)$. In our study, we use this \emph{event type} to describe the energy depositions on the volume of the detector.
\item \emph{TRestG4Event:} It is a natural extension of \emph{TRestHitsEvent} containing additional information gathered during the \emph{Geant4} simulation, such as the physical interaction associated to the hit, and the volume of the simulation geometry where the interaction took place.
\item \emph{TRestTrackEvent:} A more sophisticated structure where hits are grouped into tracks, following e.g., a proximity criterion. Tracks can, in turn, be grouped into higher level structures. An \emph{identification number} and a \emph{parent number} are assigned to each object and allow to keep the record of its genealogy. A track that is not associated to a parent track will be denominated as an \emph{origin track}, while a track with no daughter tracks will be denominated as a \emph{top-level track}. Therefore, in this scheme we may distinguish different track levels related to the number of track generations.

\item \emph{TRestRawSignalEvent:} Stores the digitized signal samples, in the shape of fixed size arrays. Each array corresponds to one front-end electronics channel, and is assigned a logical \emph{signalId}. The set of \emph{signalId}'s is mapped to the geometry of the detection with help of a \emph{readout metadata} table, typically known as \emph{decoding}. The arrays generally describe the time evolution of the signal.
\item \emph{TRestTimeSignalEvent:} It contains an arbitrary number of non-fixed size arrays that define a physical quantity, $\phi_i$, at arbitrary times, $t_i$, expressed in physical time units. Each array dataset may contain a different number of data points. This kind of \emph{event type} can be exploited, e.g. for experimental data reduction or storage of timed signals produced in Monte Carlo generated data. As in the case of a \emph{TRestRawSignalEvent}, each array is also identified using a \emph{signalId} associated to the electronic channels described inside the \emph{readout metadata} description.

\end{itemize}

We stress again that the \emph{event types} defined in REST aspire to be as generic and abstract as possible. It is the role of \emph{processes}, and other \emph{metadata} information used during the data processing, to provide a physical meaning, or final interpretation, to the information that is contained in our \emph{event type}. In the scenario where certain specialization is required, as it is the case of a \emph{TRestG4Event}, an existing REST \emph{event process} will be capable to transform this specialized, or dedicated, \emph{event type} into an existing and more basic one, as it is a \emph{TRestHitsEvent}.

The \emph{event types} described are not only containers to store \emph{event data} in REST, but they implement methods associated to the nature of each data structure, as e.g. spatial rotation or translation of coordinates in a \emph{TRestHitsEvent} or time signal processing methods in a \emph{TRestRawSignalEvent}, as e.g. differentiation, smoothing or Fast Fourier Transform (FFT) methods, together with prototyped methods, i.e. common to all the \emph{event types}, for visualizing and printing the contents of a particular stored event.

\subsection{Event processes}\label{sc:evProcesses}

REST offers a large variety of \emph{event processes} operating on the basic \emph{event types} in order to manipulate the \emph{event data}, transform it and extract relevant information in the intermediate steps taking place during a particular \emph{event data} processing chain. An \emph{event process} in REST is modular, and it can be integrated in REST by fixing only its input and output \emph{event types}. In other words, any \emph{event process} that participates in a data processing chain must satisfy that its input \emph{event type} corresponds with the output \emph{event type} of the previous \emph{event process}, although there are exceptions as will be described in~\ref{sc:anTree}.

Therefore, \emph{event processes} can be classified as processes that transform an \emph{event type} into another, i.e. \emph{conversion processes}, and those which operate in a particular \emph{event type}, i.e. \emph{hit}, \emph{raw signal} or \emph{track processes}, in which the output \emph{event type} remains unchanged (although its content - the \emph{event data} - will usually be transformed). In the following sub-sections we provide a description of the different processes involved in our data processing. A full, detailed and up-to-date list of documented processes will be available at~\cite{RESTsultan} for further reference.

\subsubsection{ Conversion processes }\label{sc:convPcs}

\begin{itemize}
	\item \emph{TRestG4ToHitsProcess:} A simple process to transform a \emph{TRestG4Event} into a \emph{TRestHitsEvent}, visualized in Figure~\ref{fig:EvReconstruction}(a). Any specific hits information related to \emph{Geant4} will be lost after this step. Then, the only way in REST to have this information available at the end of the data processing chain will be through the \emph{analysis tree}, described in~\ref{sc:anTree}.

\item \emph{TRestHitsToSignalProcess:} This process produces a time projection of the primary electron positions in the TPC volume into the readout plane of the detector, producing the result shown in Figure~\ref{fig:EvReconstruction}(c). The coordinates of hits $(x_i, y_i, z_i)$ stored in a \emph{TRestHitsEvent} are transformed into a \emph{TRestTimeSignalEvent}. This process uses the REST \emph{readout metadata} structure (described in detail on~\ref{sc:readout}) to associate each hit coordinates $(x_i, y_i)$ with a detector \emph{readout channel}\footnote{As an argument to support this description we assume here that the field drifting the electrons is defined along the $z$-axis, although the readout plane can be placed in any arbitrary orientation.}, and perform a spatial to time conversion of the $z_i$ coordinate using the readout plane position and the drift velocity of electrons in the gas. The gas properties can be introduced directly as an input metadata parameter, or extracted from a specific metadata class, named \emph{TRestGas}, that interfaces with \emph{Garfield++} to obtain the properties of any gas mixture as calculated by \emph{Magboltz}. A sampling time, $\delta t$, can be provided optionally to discretize the resulting time values at the output \emph{TRestTimeSignalEvent}.

\item \emph{TRestSignalToHitsProcess:} The inverse process of a \emph{TRestHitsToSignalProcess}, producing the result shown in Figure~\ref{fig:EvReconstruction}(d). We recover, or reconstruct, the hit coordinates using the time information inside a \emph{TRestTimeSignalEvent} and the gas properties. The reconstructed coordinates are obtained using the \emph{readout metadata} description and the corresponding \emph{readout channel} number, associated to each \emph{signalId}.

\begin{figure}[h]
\centering
\begin{tabular}{cc}
    \includegraphics[totalheight=6.5cm]{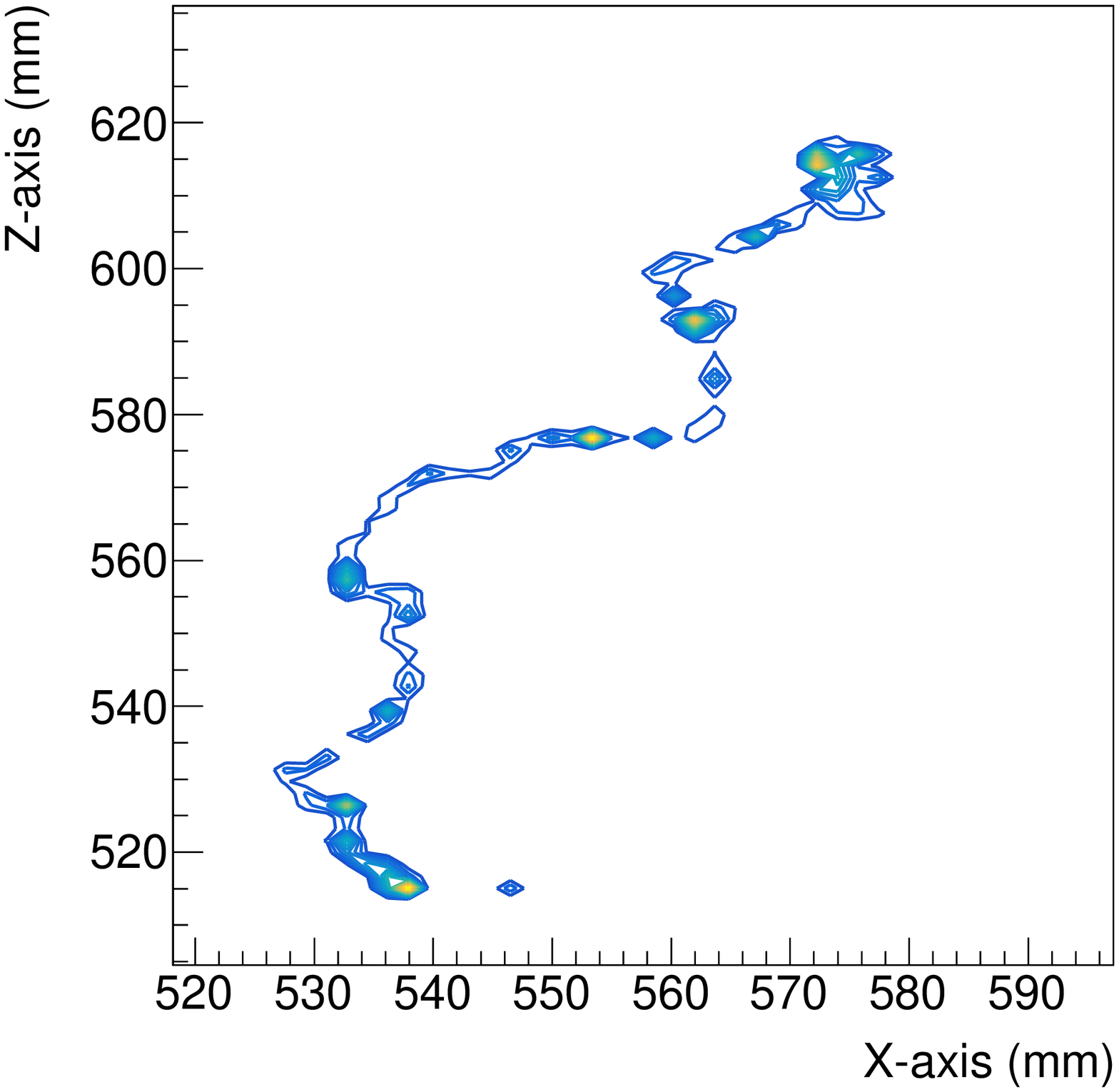} &
    \includegraphics[totalheight=6.5cm]{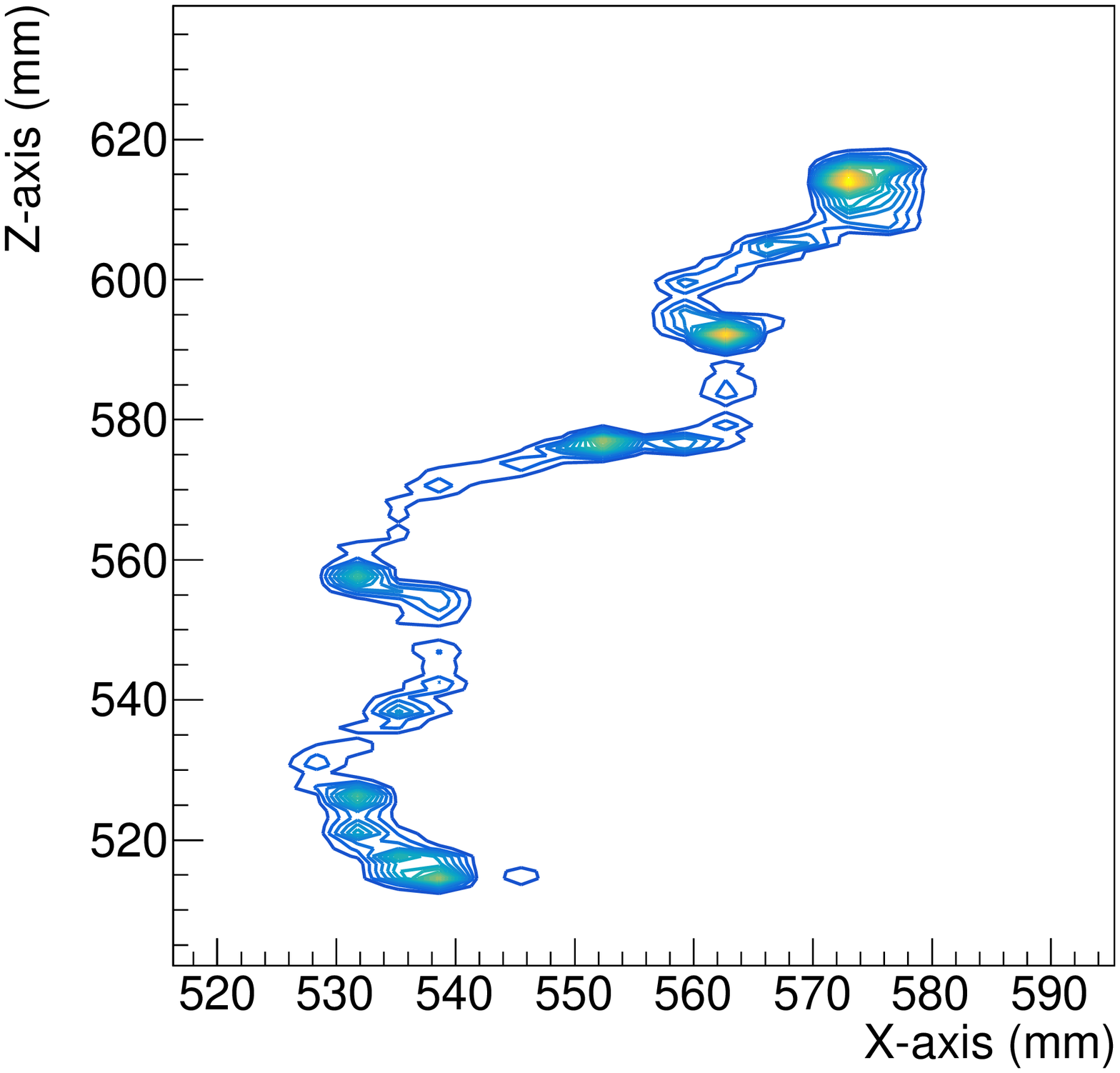} \\
	(a) & (b) \\
	& \\
    \includegraphics[totalheight=6.5cm]{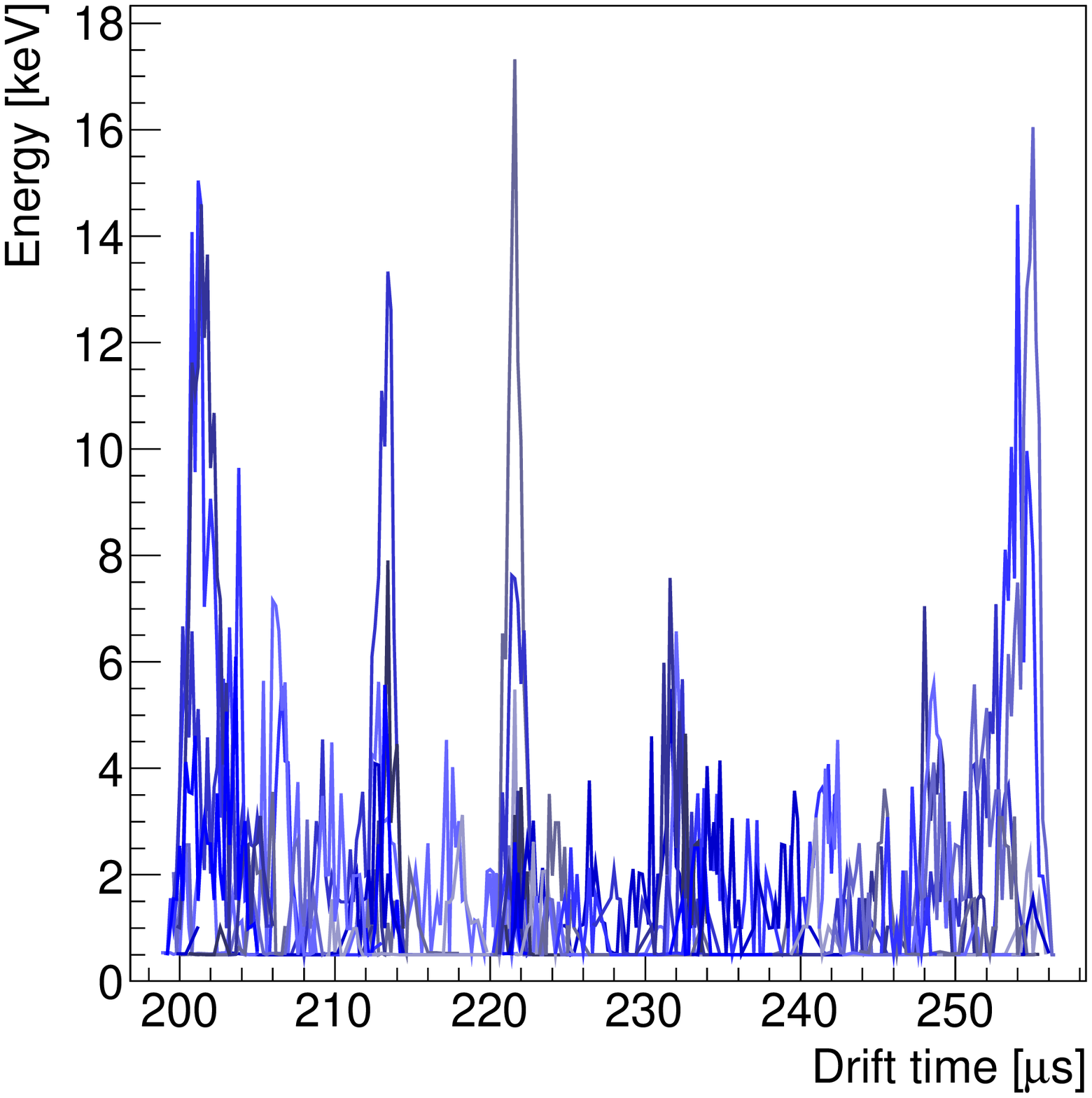} &
    \includegraphics[totalheight=6.5cm]{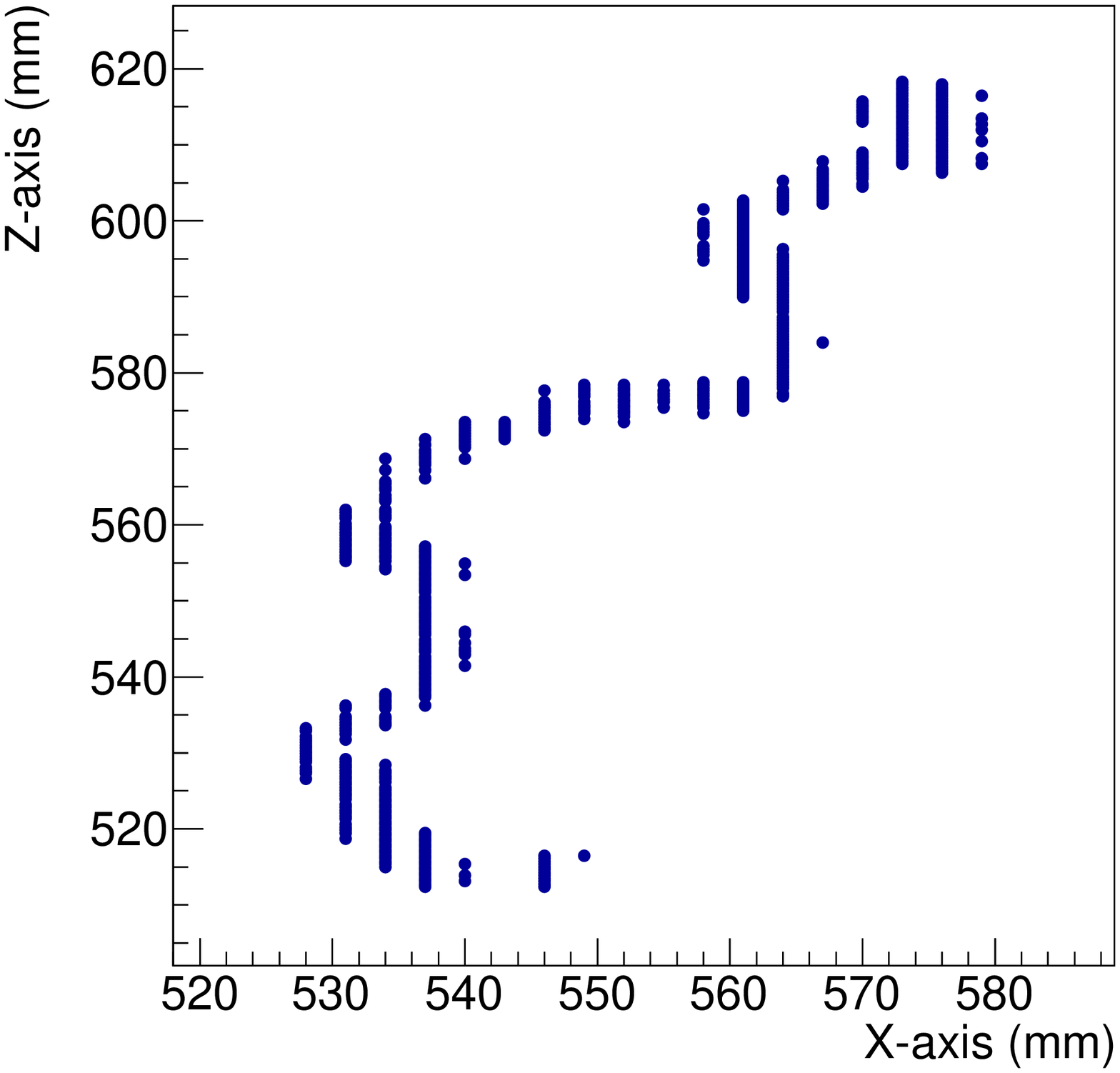} \\
	(c) & (d) \\
	& \\
\end{tabular}
\caption{Representation of different event type outputs after different \emph{event processes} on the simulation of \XeNLDBDEvt. (a) Top-left panel visualizes a \emph{TRestG4Event} type, and represents the XZ-plane projection of the charge density of the 3-dimensional event. (b) Top-right panel shows a \emph{TRestHitsEvent} event type after applying the \emph{TRestElectronDiffusionProcess} where we can appreciate the effect of electron diffusion in a xenon + 1\%TMA gas mixture at 10\,bar, the readout plane is placed at $z=990$\,mm. (c) Bottom-left panel shows a \emph{TRestTimeSignalEvent} type after the conversion by a \emph{TRestHitsToSignalProcess}. (d) Bottom-right panel shows a \emph{TRestHitsEvent} type after the reconstruction using \emph{TRestSignalToHitsProcess}. A scatter plot is used in this case to emphasize the effect introduced by the 3mm-pitch detector readout that can be observed along the x-axis, and the 200\,ns sampling rate of the electronics on the z-axis. At this stage the \emph{TRestHitsEvent} is not anymore a 3-dimensional event, since we used here the PandaX-III stripped readout described on~\ref{sc:readout}.}
    \label{fig:EvReconstruction}
\end{figure}

\item \emph{TRestSignalToRawSignalProcess:} This process takes as input a \emph{TRestTimeSignalEvent} type and samples its contents into a fixed data array compatible with the \emph{TRestRawSignalEvent} output type, i.e. fixing the number of data points and sampling time, $\delta t$, which are provided as an input parameter to this process. A reference time, or \emph{trigger definition}, is used to define the physical time corresponding to the first sample of the resulting data array. Figure~\ref{fig:triggerDefinition} describes one of the trigger methods available on this process which we use later in this study.

\begin{figure}[]
\centering
\includegraphics[totalheight=6.5cm]{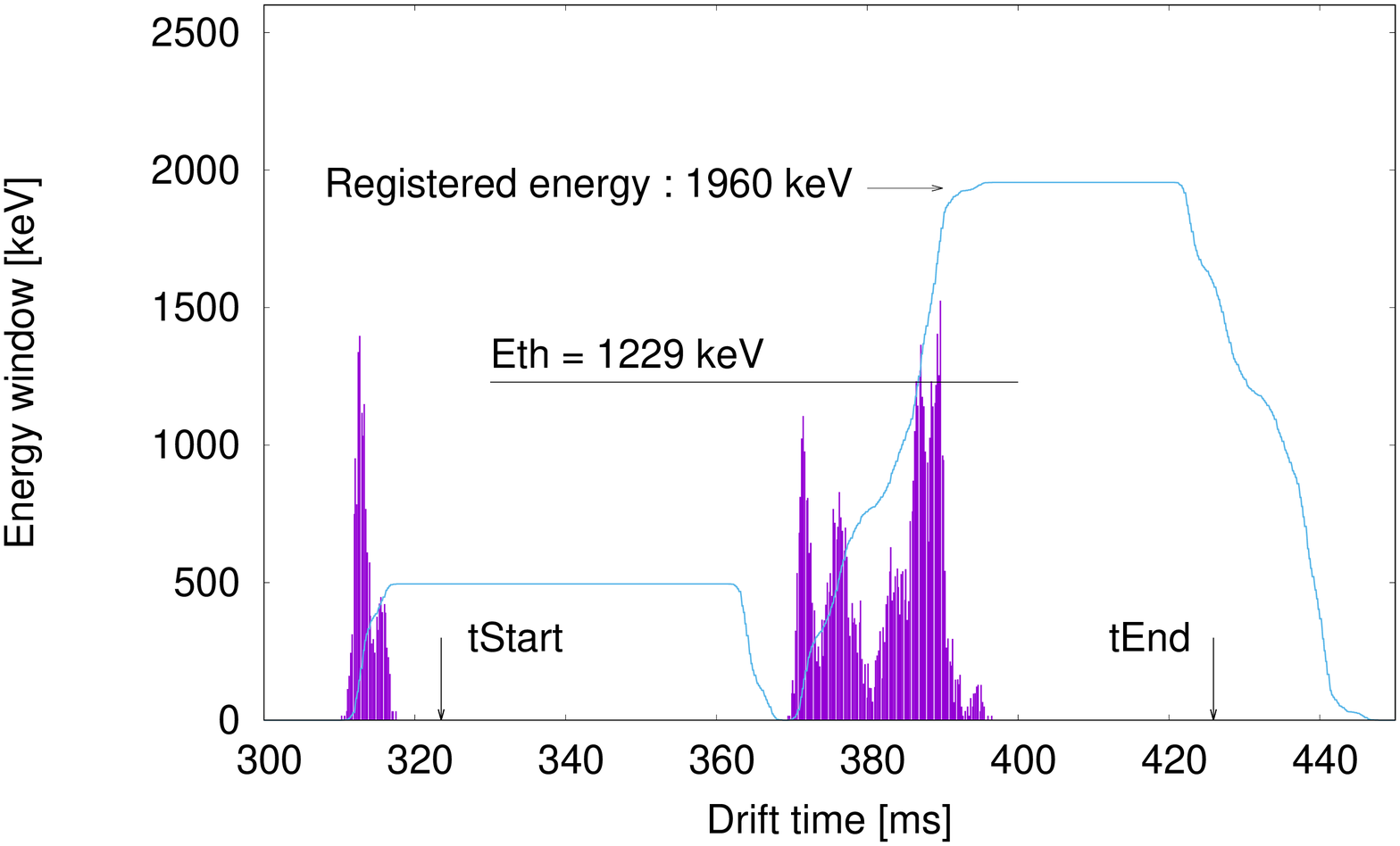}
\caption{A figure illustrating the trigger definition using the threshold method implemented in \emph{TRestSignalToRawSignalProcess} for \XeNLDBDEvt. The filled curve, in red, represents the charge distribution (in arbitrary units not represented on this plot) as a function of the electron drift time. The blue curve is constructed with the integral of the charges (with magnitude represented on the y-axis of this plot) in a fixed time width, here of about $\sim$50\,$\mu$s, and corresponding to the half size of the acquisition window of the electronics, of about $\sim$100\,$\mu$s. When the integration exceeds a certain threshold $E_{th}$, in this case equal to $Q_{\beta\beta}/2$ of $^{136}$Xe, the center of the acquisition window is fixed. The resulting acquisition window is represented by $t_{Start}$ and $t_{End}$. An additional offset is introduced on the definition of $t_{Start}$ to assure few time bins will be available for \emph{baseline} definition during the \emph{raw signal} event processing. } 
\label{fig:triggerDefinition}
\end{figure}


\item \emph{TRestHitsToTrackProcess:} This process takes as input a \emph{TRestHitsEvent} type and transforms it into a \emph{TRestTrackEvent}. The hit inter-distances inside the input event are evaluated, and those which fall within a certain distance will be assigned to the same identifiable \emph{track}. This distance is introduced as a metadata input parameter to the process, named \emph{cluster distance}. The algorithm starts creating a first track by adding an arbitrary and unassociated hit\footnote{A hit that does not belong yet to any \emph{track}}. The \emph{track} definition will only end when no more unassociated hits are found to satisfy the inter-distance relation with - and this is important - \emph{any} of the hits already added to the \emph{track}, i.e. the inter-distances are evaluated recursively. If after ending the definition of the first track there are still unassociated hits, the process continues to define a second track, and successively. The process ends when no more unassociated hits are found inside the \emph{TRestHitsEvent}.

\end{itemize}

\subsubsection{ Hit processes } 
\begin{itemize}

	\item \emph{TRestElectronDiffusionProcess:} This process uses the longitudinal and transverse coefficients of a particular gas mixture to emulate the relative deviation of electrons from their original positions in \emph{TRestHitsEvent}, presumably produced by primary ionization. The energy of each hit found inside the input \emph{TRestHitsEvent} is converted to the corresponding number of primary electrons that would be produced in the ionization process. Each electron will be a new hit in the output \emph{TRestHitsEvent} structure, and their coordinates will be randomly deviated following a Gaussian distribution related to the gas parameters, longitudinal and transverse diffusion coefficients, and the distance to the readout plane, which is given as metadata input. The result of applying this process is shown in Figure~\ref{fig:EvReconstruction}(b). This process may optionally add the possibility to include the effect of electron attachment.

\item \emph{TRestHitsSmearingProcess:} This process smears the energy of the input \emph{TRestHitsEvent} according to a Gaussian distribution described by parameters given as metadata input.
\end{itemize}

\subsubsection{ Track processes for $0\nu\beta\beta$ topology }\label{sc:trackProcesses}

\begin{figure}
\centering
\begin{tabular}{ccc}
    \includegraphics[totalheight=4.9cm]{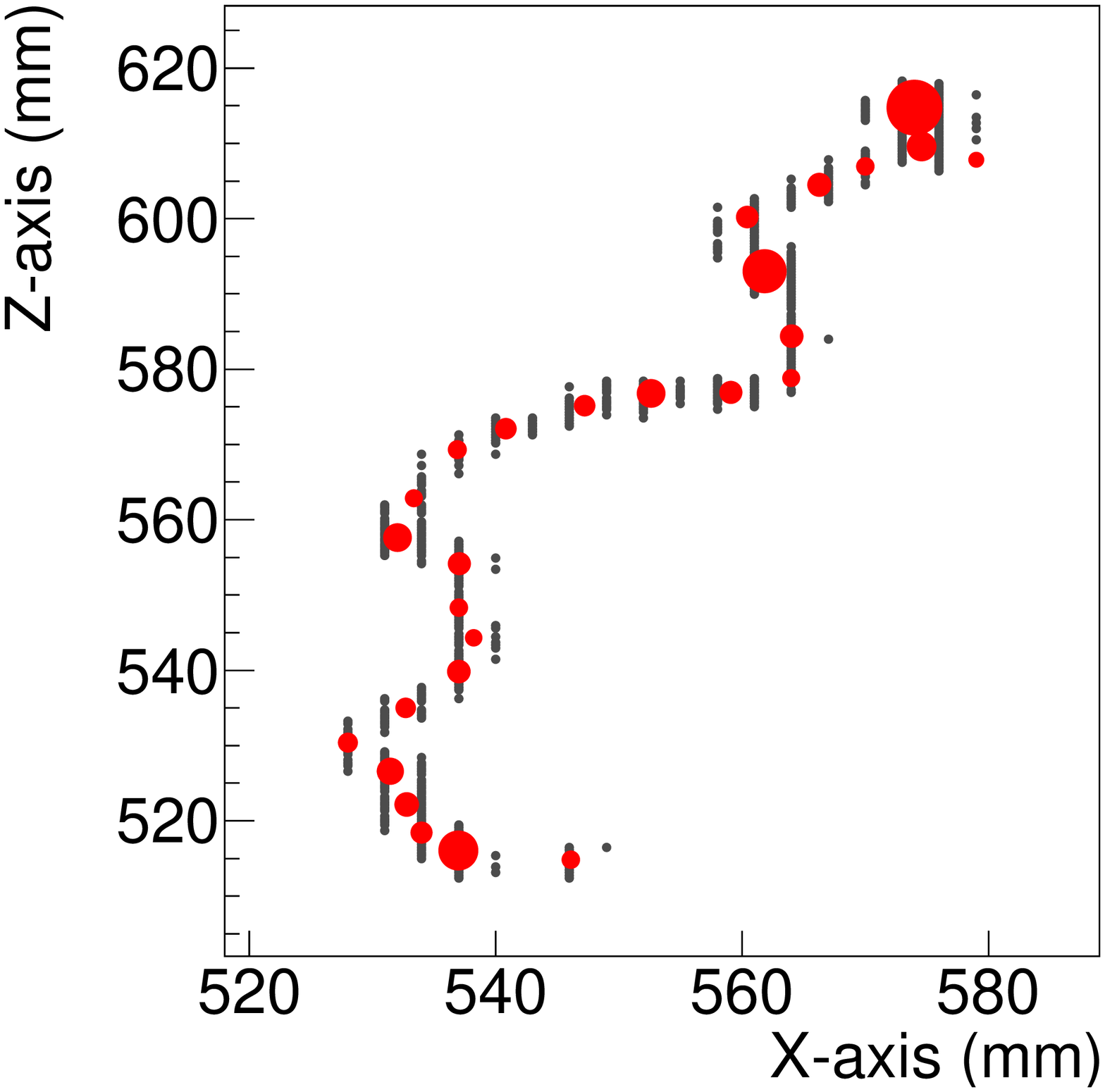} &
    \includegraphics[totalheight=4.9cm]{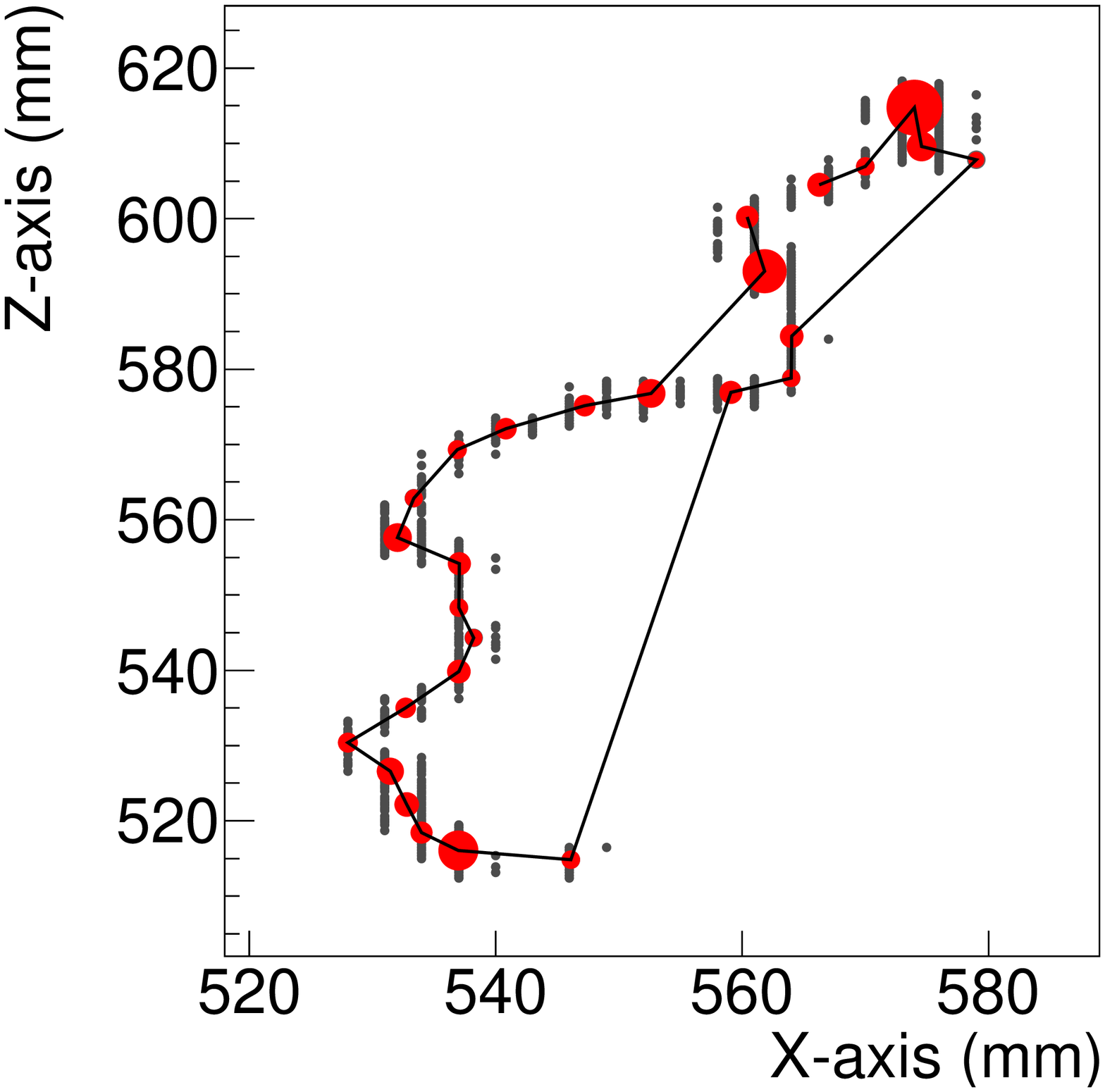} &
    \includegraphics[totalheight=4.9cm]{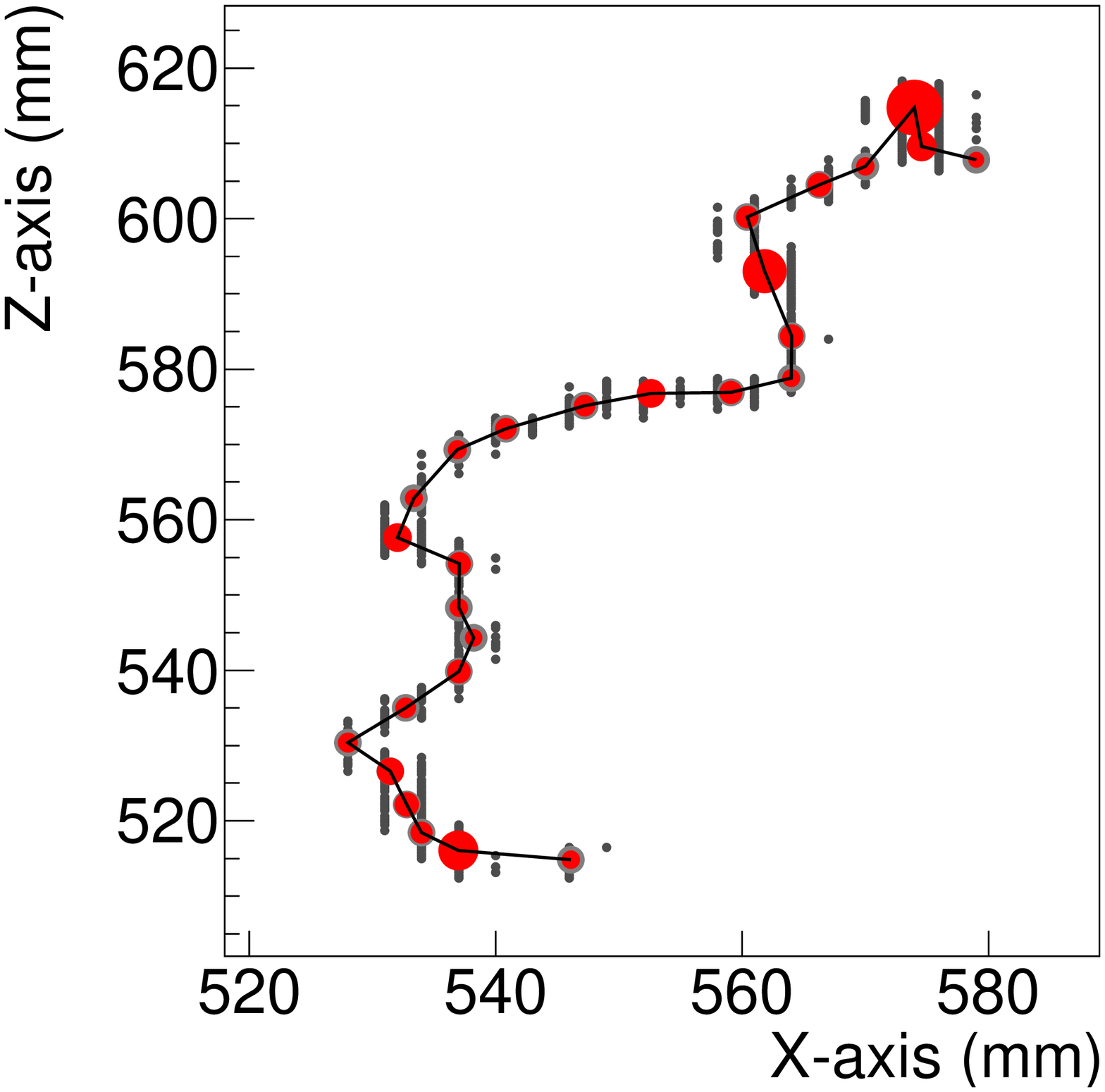} \\
	(a) & (b) & (c) \\
\end{tabular}
\caption{A \emph{TRestTrackEvent} representation of \XeNLDBDEvt~after the different track processes used for physical track identification. This event corresponds to the same event used in Figure~\ref{fig:EvReconstruction}. (a) An image of the hit reduction produced by \emph{TRestTrackReductionProcess}. The red circles represent the final position of reduced hits, which size corresponds with their energy value. The small grey circles on the background represent the hits of the \emph{parent track} used as input. (b) A polyline is added to this representation to visualize the hits inter-connectivity after the \emph{TRestTrackPathMinimizationProcess}. If path minimization works on the whole, it produces at times obviously unphysical connections, as our example illustrates. (c)~The unphysical connections are corrected using \emph{TRestTrackReconnectionProcess}.}
    \label{fig:TopologyProcess}
\end{figure}

\begin{itemize}

	\item \emph{TRestTrackReductionProcess:} This process reduces the number of hits in each of the tracks stored in a \emph{TRestTrackEvent}, as seen in Figure~\ref{fig:TopologyProcess}(a). An input metadata parameter, $N_{max}$, specifies the maximum number of hits on each of the reduced tracks at the output \emph{TRestTrackEvent}. The closest hits are merged iteratively till the $N_{max}$ condition is satisfied. When \emph{two} hits are merged, the resulting hit position is calculated weighting the energy of each merged hit. The input \emph{parent tracks} are also stored in the output \emph{TRestTrackEvent}, together with the reduced tracks which acquire a relation of inheritance.

	\item \emph{TRestTrackPathMinimizationProcess:} This process operates only on each of the \emph{top-level tracks} found in \emph{TRestTrackEvent} and - using graph theory - finds the shortest path that connects all the hits within each track, as seen in Figure~\ref{fig:TopologyProcess}(b). We have integrated in this process a \emph{heldkarp} algorithm~\cite{held1962dynamic} optimized for problems that contain between 25 and 35 nodes. This algorithm was extracted from the \emph{concorde} Travel Sales Problem (TSP) libraries~\cite{Applegate:2007:TSP:1374811,concorde}. It is important to notice that the minimization is performed in a closed loop, i.e. the extremes of the physical track are not properly identified at the end of this process. From now on, we will denominate two consecutive hits as \emph{connected hits}.

	\item \emph{TRestTrackReconnectionProcess:} This process will be commonly used in combination with \emph{TRestTrackPathMinimizationProcess}. This process will detect unphysical path connections between the hits, or nodes. We understand by unphysical connection a distance between nodes that is larger than the average by a certain number of sigmas, or where no hits from the \emph{origin track} are found in between, as seen in Figure~\ref{fig:TopologyProcess}(c). The track is split at those unphysical connections and reconnected with the closest hit. Using this technique, the extremes of the physical track are found naturally. Although it was not used in our analysis, this process allows to identify secondary track branches of complex event tracks when reconnection is not possible.

\end{itemize}

\subsubsection{ Raw signal processes }

\begin{itemize}
\item \emph{TRestRawSignalShapingProcess:} This process realizes the simulation of the acquisition electronics signal shaper in a \emph{TRestRawSignalEvent}. It implements the convolution of an analytical response function with the input \emph{TRestRawSignalEvent}, presumably containing the original charge distribution produced in the detector. The analytical response function requires a single parameter, which is the \emph{shaping time}. However, this process offers the possibility to introduce any arbitrary wavefunction as response function, adding realism to the generated output.

\item \emph{TRestRawSignalAddNoiseProcess:} It is used to include the effect of electronic noise and emulate the fluctuations on the acquired time signal. A value is introduced as input metadata to define the amplitude of the noise. We use a basic noise method implemented in this process that assigns an independent random value, Gaussian distributed, to each of the bins inside a \emph{TRestRawSignalEvent}.
\end{itemize}

\subsection{The analysis tree}\label{sc:anTree}

The different \emph{event processes} can be combined in sequence to build a more complex processing chain. In REST, only the \emph{event data} produced in the last \emph{event process} will be saved to the output file\footnote{However, other \emph{metadata} information used in the processing chain, e.g. process parameters, simulation conditions, readout definition, or gas properties, will be stored without exception, including previous historical \emph{metadata} information}. There are two main reasons why it is not desirable to stock intermediate \emph{event data} into a single data file\footnote{Although no restrictions apply to produce intermediate files with a snapshot of the \emph{event data} at a particular step of the processing chain.}. First, we risk to pile-up, or replicate unnecessarily, data that might require to allocate a non negligible amount of disk space. Second, the traceability of changes introduced in the data might be lost by storing data at different \emph{levels} of processing inside a unique file. This possibility would add an undesirable degree of complexity to the future tracking of the data processing, if for example, we decide to continue processing data using a previously processed file with \emph{event data} at different processing levels but we start our \emph{event data} processing at an intermediate state.


However, the inconvenience of this scheme resides on the fact that, during the processing, the \emph{event data} is transformed in a way that information might be lost, as when we transform a \emph{TRestGeant4Event} into a \emph{TRestHitsEvent}. Therefore, it will be not available at the end of the \emph{event data} processing. The \emph{analysis tree} is a REST component that emerges to solve this problem. It is based on a ROOT TTree structure, and it is accessible at any stage of the \emph{event data} processing by any \emph{event process}. Any process in REST is allowed to create a new entry, or variable, inside the \emph{analysis tree} which will be always available in any REST file. Future processing may add new variables to the \emph{analysis tree} preserving the existing ones from previous processing levels. Therefore, the \emph{analysis tree} is used to gather relevant information along the processing chain that might not be anymore available at the \emph{event data} output of the last process.


In this context, and just for classification purposes, we can distinguish two new types of \emph{event processes}, the \emph{analysis processes} and the \emph{pure analysis processes}. Both type of processes have in common that they do not modify or transform the input \emph{event data}, and are just dedicated to add new entries to the \emph{analysis tree}. Furthermore, the \emph{pure analysis processes} will not even require access to the \emph{event data}, being these processes the only ones that can be connected at any place of the processing chain without input/output \emph{event type} restrictions. For example, a \emph{pure analysis process} might be a process that reads a variable (or branch) in the \emph{analysis tree} to perform a multi-peak fit, and write the results of the fit in a \emph{metadata} structure.

%% file: MCDataChain.tex

\section{PandaX-III Monte Carlo event simulation in REST}\label{sc:Montecarlo}

The first step of Monte Carlo simulation is to consider PandaX-III as a simple calorimeter and simulate the energy deposition of particles interacting in the gaseous medium, or active volume of the TPC. The calorimetric response of the conceptual PandaX-III design was studied extensively in reference~\cite{chen2017pandax} using the \emph{Geant4}-based simulation packages \emph{BambooMC} and \emph{RestG4}. In \Geant, detailed tracks of the ionizing particles in the TPC active volume are generated. As a particle travels in the gas medium, it deposits energy along the trajectory via multiple scattering and other physical processes. \Geant~tracks the trajectory in pre-defined steps.  In each step, information such as timestamp, particle type, momentum, energy deposition, position, and physical process involved in the interaction are registered.

In our study, we use the \emph{RestG4} package as an interface to define the simulation conditions through a REST metadata structure named \emph{TRestG4Metadata}, and to store all event information in a \emph{TRestG4Event} type that can be further processed inside REST, i.e., \emph{RestG4} serves to generate a first event dataset. Event tracks can be reconstructed using this original dataset, and we call those \emph{MC-true tracks}. The \emph{MC-true tracks} have been generated in \Geant~with a spatial precision of 0.2\,mm, implying that every step of the simulation of the movement of a particle is no more than 0.2\,mm, with particle information being recorded after each step. The \emph{G4EmLivermorePhysics} physics list - describing the electromagnetic processes in \Geant~- was used in the event generation.

The main features of the PandaX-III TPC are faithfully represented in our detector geometry. The cylindrical TPC is 1.5\,m in diameter and 2\,m in length. The cathode in the middle divides the TPC into two symmetrical drift volumes, in which electrons are drifted away from the cathode and collected by the charge readout planes located at both ends. When running at 10\,bar pressure the TPC can hold 200\,kg xenon gas. The active volume is contained by a 3\,cm thick copper vessel, and placed in a water shielding tank in the laboratory. The geometry description, written in GDML~\cite{Chytracek:2006be}, is the same as the one used at reference~\cite{chen2017pandax}. Therefore, we refer to that publication for further details on the different detector components implemented, materials used, and drawings.

\subsection{Data processing chain for PandaX-III.}\label{sc:dataChain} 

The generated \emph{MC-true tracks} are introduced in the REST processing scheme to achieve a realistic detector response, including electron drift diffusion, energy resolution, charge readout segmentation, and signal sampling.

\begin{figure}
\centering
\includegraphics[width=\textwidth]{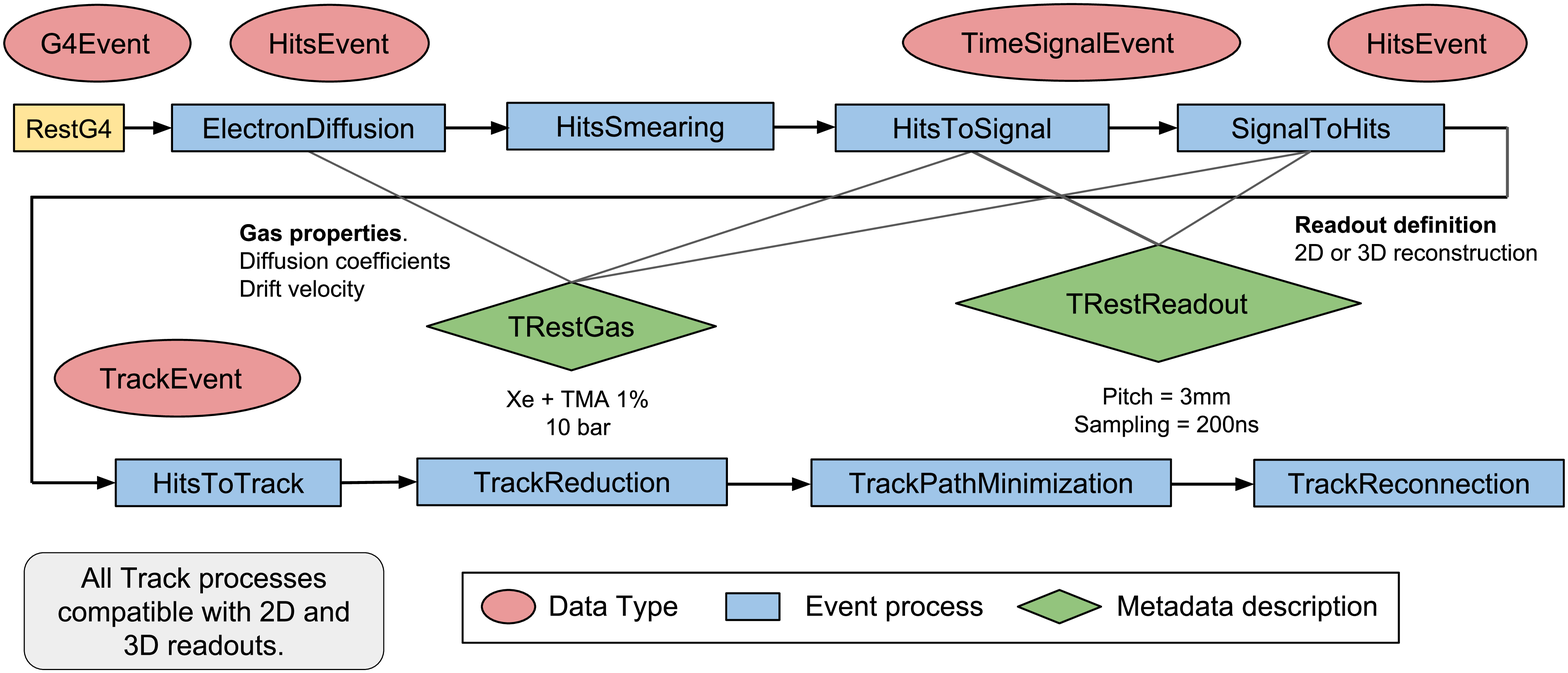}
	\caption{Data processing chain used to manipulate the Monte Carlo events as described in the text. This figure gives an overview of the data flow through the successive \emph{event types} (red ovals). Few \emph{event processes} (blue rectangles) have been omitted, for simplicity's sake. The chain starts with \Geant~generated data (yellow rectangle at top-left). The flow is from left to right. We can distinguish two rows of \evProcesses. The top row (\emph{TRestElectronDiffusionProcess}, \emph{TRestHitsToSignalProcess}, etc) represent processes used to include the detector response (or \emph{first phase}) and event reconstruction (or \emph{second phase}), while the bottom row represents the track processes used for \emph{physical track} reconstruction (or \emph{third phase}). Just in this figure, the \emph{TRest} prefix and \emph{Process} termination has been removed from the REST process and event names.}
    \label{fig:basicDataChain}
\end{figure}

Figure~\ref{fig:basicDataChain} shows the complete data chain, or \emph{event data} flow, used to process \emph{Geant4} Monte Carlo generated data, and includes different effects related to the detector response. Obvious \emph{event type} conversion processes detailed on~\ref{sc:convPcs}, and analysis processes described later on in this chapter, have been omitted in this drawing in order to facilitate focusing on the key aspects of the chain. In our data chain we can differentiate up to three different phases, or stages. In the \emph{first phase} the \emph{event data} is conditioned to take into account the physical response of the electrons drifting in the gas medium, in the \emph{second phase} the readout topology and electronics sampling is considered, and finally in the \emph{third phase} a physical track\footnote{We will use the term \emph{physical track} to refer to the final state of the event reconstruction, where the reconstructed track has physical meaning.} reconstruction takes place to condition the data and make it suitable for a topological track study. The description of \emph{event processes} required for a full understanding of the processing chain is detailed in~\ref{sc:evProcesses}.



The \emph{first phase} of the processing takes place exclusively at the \emph{TRestHitsEvent} level. The input generated by \emph{RestG4}, encapsulated inside a \emph{TRestG4Event} type, is transformed into a \emph{TRestHitsEvent} type using \emph{TRestG4ToHitsProcess}. Two processes are responsible to include the physical detector response, we use \emph{TRestElectronDiffusionProcess} to emulate the electron diffusion in the gaseous medium, and \emph{TRestHitsSmearingProcess} to include the stochastic effect of the detector energy resolution on each independent event. The gas properties used in \emph{TRestElectronDiffusionProcess} were obtained from \emph{Magboltz}\cite{Magboltz} through the \emph{Garfield++}\cite{Garfield} interface integrated in \emph{TRestGas}. In our study we used a gas mixture of xenon+1\%TMA at 10\,bar pressure, setting the TPC drift field to 1\,kV/cm leads to an electron drift velocity of 1.86\,mm/$\mu$s, and 1.46$\times$10$^{-2}$\,cm$^{1/2}$ and 1.01$\times$10$^{-2}$\,cm$^{1/2}$, for the longitudinal and transverse electron diffusion coefficients, respectively. The number of primary electrons yielded at each deposition is correlated with the energy deposited according to the work-function, or $W$-value, i.e. the energy required to extract an electron from an atom. We use a $W$-value of 21.9\,eV~\cite{borges1996experimental} for our gas mixture. The detector energy resolution introduced in \emph{TRestHitsSmearingProcess} was defined as 3\%-FWHM at $Q_{\beta\beta}=2457.83$\,keV.


During the \emph{second phase} the conversion processes \emph{TRestHitsToSignalProcess} and \emph{TRestSignalToHitsProcess} are responsible to introduce the detector granularity and electronics sampling of the time signal induced on the detector readout. These processes are strongly dependent on the \emph{TRestReadout} definition, detailed in~\ref{sc:readout}. For now, it is enough to mention that \emph{TRestHitsToSignalProcess} discretizes our \emph{event data} into time signals in steps of 200\,ns, using a 2-dimensional readout of 3\,mm pitch. At this stage, other details related to time signal conditioning could be added, despite the fact that it is not the aim of this work to include those effects, and any future improvement will be later discussed in our final conclusions. Then, \emph{TRestSignalToHitsProcess} reconstructs the event recovering the \emph{TRestHitsEvent} structure including the effects introduced by the previous process, as it is shown in Figure~\ref{fig:EvReconstruction}(d).







In the final \emph{third phase}, the reconstructed \emph{TRestHitsEvent} is classified into tracks through the \emph{TRestHitsToTrackProcess} using a \emph{cluster distance} value of 7.5\,mm, i.e. two tracks are considered to be independent if the closest hits distance from one track to another is above 7.5\,mm. In this phase the processes described in~\ref{sc:trackProcesses} are used to determine the \emph{physical track}, evaluating the hit distances and connectivity. Figure~\ref{fig:TopologyProcess} shows the result of the three \emph{track processes} implemented in this phase of the processing chain; \emph{TRestTrackReductionProcess} (where we use $N_{max}$=35), \emph{TRestTrackPathMinimizationProcess} and \emph{TRestTrackReconnectionProcess}. At the end of this processing chain we end up with a reconstructed \emph{physical track} where track ends can be recognized naturally. It is the responsibility of the \emph{analysis processes} to extract information, or observables, along the data processing chain, to be used in a final pattern recognition analysis.

It is important to remark that the \emph{event data} processing chain discussed here is fully compatible with 2 and 3-dimensional event representation, i.e., a \emph{TRestHitsEvent} structure is able to identify if it contains 2D or 3D hits, and the \emph{hit} and \emph{track processes} will act on the \emph{event data} accordingly, e.g. \emph{TRestTrackPathMinimizationProcess} will independently minimize the paths of different projections and tracks produced by a 2-dimensional readout. Ultimately, it is the sole responsibility of the \emph{TRestReadout} metadata structure to define the topological nature of the \emph{event data}.



\subsection{PandaX-III readout metadata description}\label{sc:readout}

\begin{figure}
\centering
\begin{tabular}{cc}
	\includegraphics[width=0.46\textwidth]{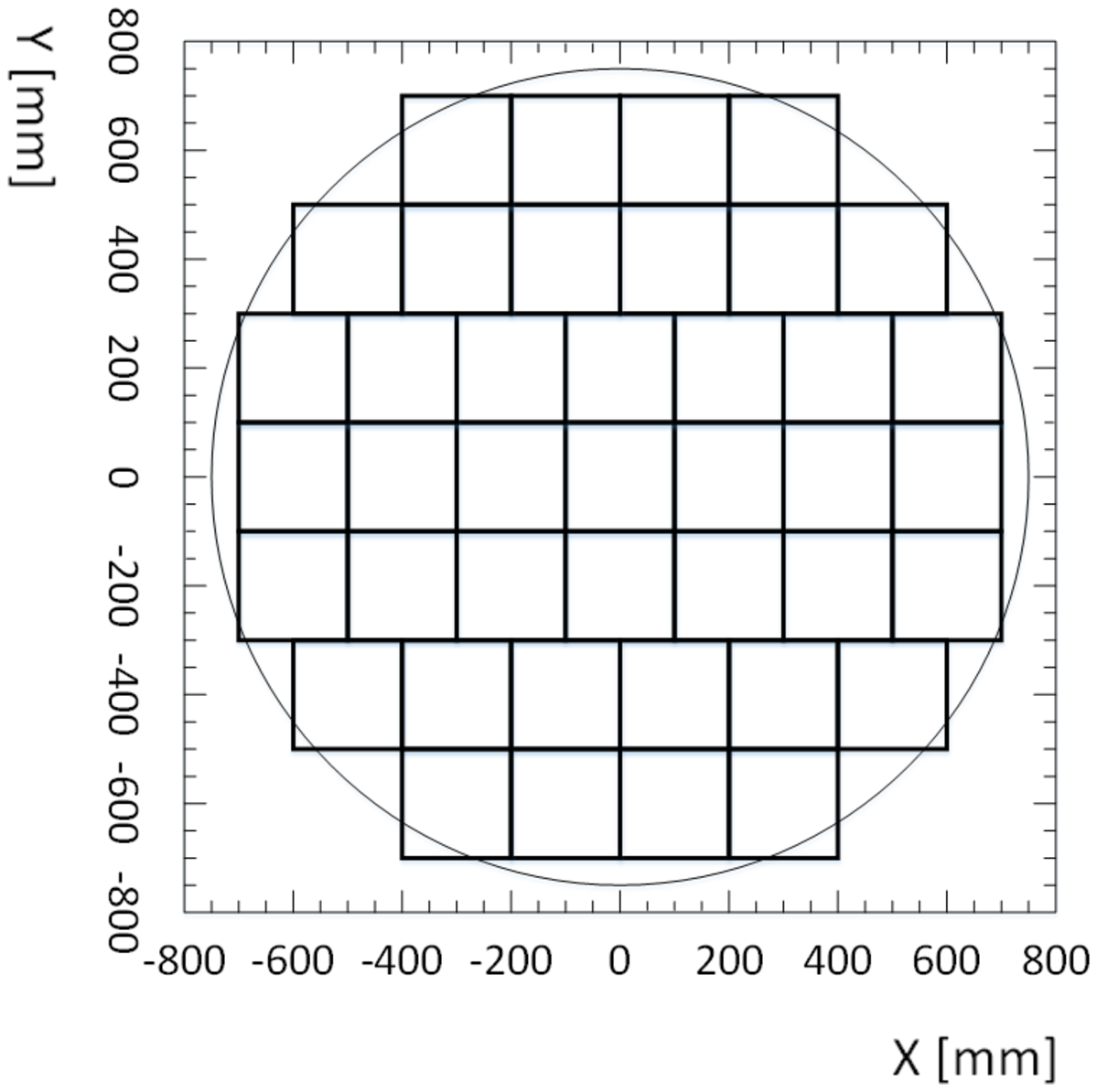} &
	\includegraphics[width=0.44\textwidth]{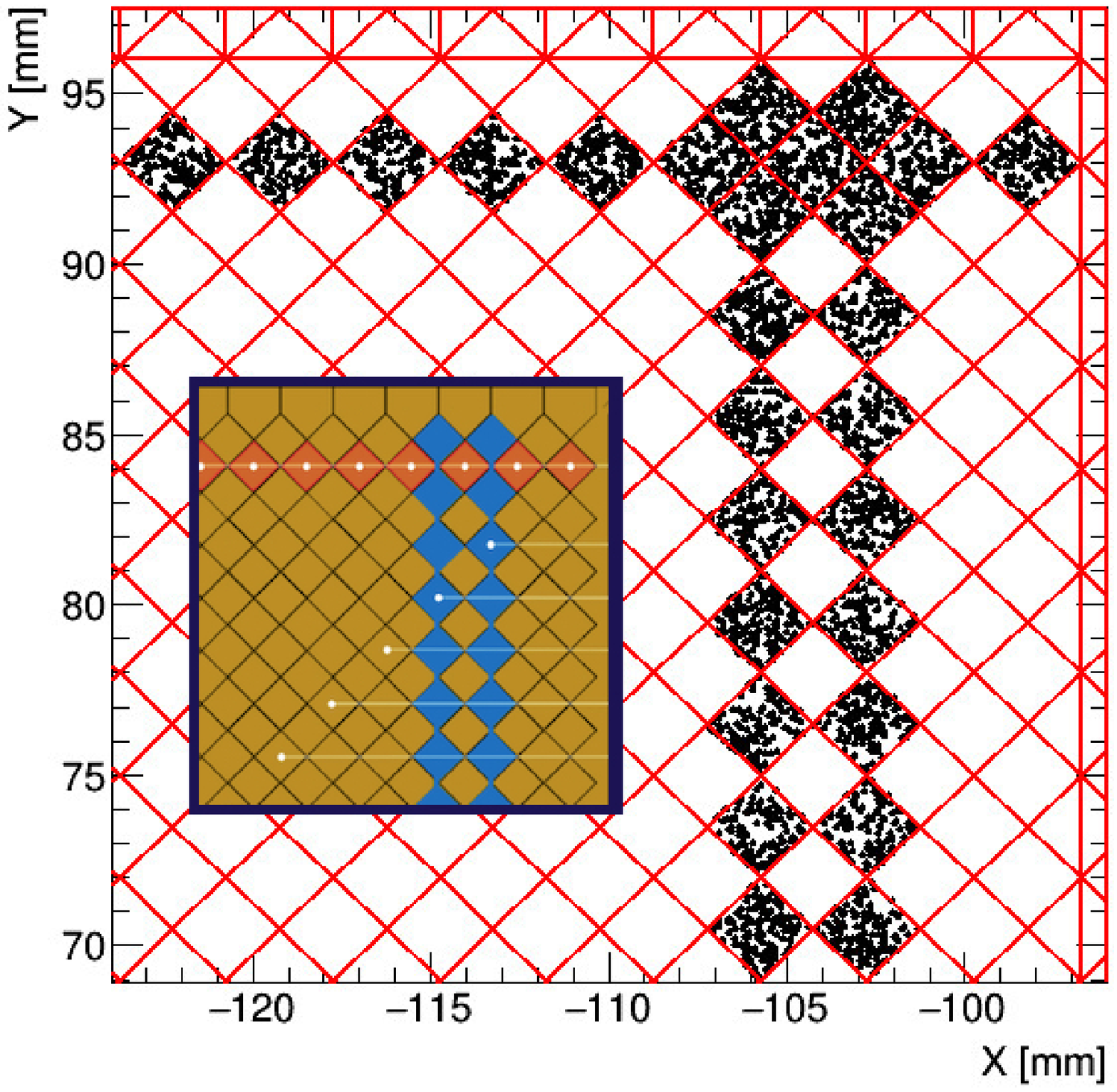} \\
	(a) & (b) \\
\end{tabular}

\caption{(a) The PandaX-III readout plane description in REST, built using 41 Micromegas modules of size 192\,mm$\times$192\,mm. The circle of size R=75\,cm defines the inner radius where the $^{136}$Xe gas is contained. The active detector area is formed by the region of the modules found inside that circle. Small corners inside this circular region are not covered by any readout module, and therefore, they will translate into a signal efficiency loss in our final detector response. (b) The detail of a Micromegas readout module region as implemented in REST, red lines define the limits of \emph{readout pixels}. Two \emph{readout channels} on the X-axis and one \emph{readout channel} on the Y-axis are drawn (black dots) using REST readout validation routines, which allow to select, or activate, a group of \emph{readout channels} and - after generating a number of random coordinates - paint only those generated coordinates that are found inside the active channels. The small embedded image is the corresponding area extracted from the original \emph{Gerber}~\cite{GerberWeb} design used for fabrication. The \emph{Gerber} representation has been modified to identify the two X-axis (blue pixels) and one Y-axis (orange pixels) \emph{readout channels} active for the readout validation.}
	\label{fig:readout}
\end{figure}

For charge readout, PandaX-III baseline design relies on Microbulk Micromegas technology due to its good intrinsic radiopurity levels, good energy resolution and capability to operate at high pressure~\cite{Gonzalez-Diaz:2015oba,Lin:2018mpd}. A modular design has been conceived to build each detector \emph{readout plane} in the PandaX-III TPC design due to the Microbulk size limitation imposed by its fabrication process\,\footnote{Mainly due to the existing equipment used for production.}. In order to cover the full active area of the detector, each \emph{readout plane} consists of 41 independent Micromegas modules, as shown in Figure~\ref{fig:readout}(a). The total number of acquisition channels is limited to a manageable level\,\footnote{Readout channel reduction plays an important role to simplify the design, reduce costs and minimize the impact of typically non-clean electronics from the radiopurity point of view, on a detector component that must be placed as close as possible to the readout plane to minimize electronic noise induction on the readout signal.} by physically interconnecting pixels along the horizontal or vertical axis orientations and they are read out as a single \emph{readout channel}, or strip, as shown in Figure~\ref{fig:readout}(b). Each Micromegas module contains 64 $X$-strips and 64 $Y$-strips of 3\,mm pitch.

The integration of a realistic complex readout scheme in REST is done through the \emph{TRestReadout} metadata structure. \emph{TRestReadout} is a sophisticated structure that defines an arbitrary number of \emph{readout planes}, containing itself any number of \emph{readout modules} composed of \emph{readout channels} that are identified with the data acquisition (DAQ) channels registered by the electronics. A \emph{readout channel} itself is built of one or more \emph{readout pixels}. A \emph{readout pixel} is the most elementary component of the readout description in REST, and it defines its position relative to the module coordinates, its size, orientation, and shape. The combination of \emph{readout pixels} with different sizes, orientations, and shapes at different positions allows to construct any desired readout topology in REST.


Different methods are available at \emph{TRestReadout} and related classes to access the readout description, and determine for a given hit coordinates which is the corresponding \emph{readout channel} in an efficient way. These methods are accessed by related \emph{event processes} and they are exploited to translate a given \emph{TRestHitsEvent} coordinates into \emph{TRestTimeSignalEvent} channels, and vice versa. The \emph{TRestReadout} definition is a crucial element in the construction of the data processing chain and its generic implementation provides versatility to study a common dataset with different detector readout topologies and granularities, as it is done in section~\ref{sc:granularity}.

\begin{figure}
\centering
\begin{tabular}{c}
	\includegraphics[width=0.98\textwidth]{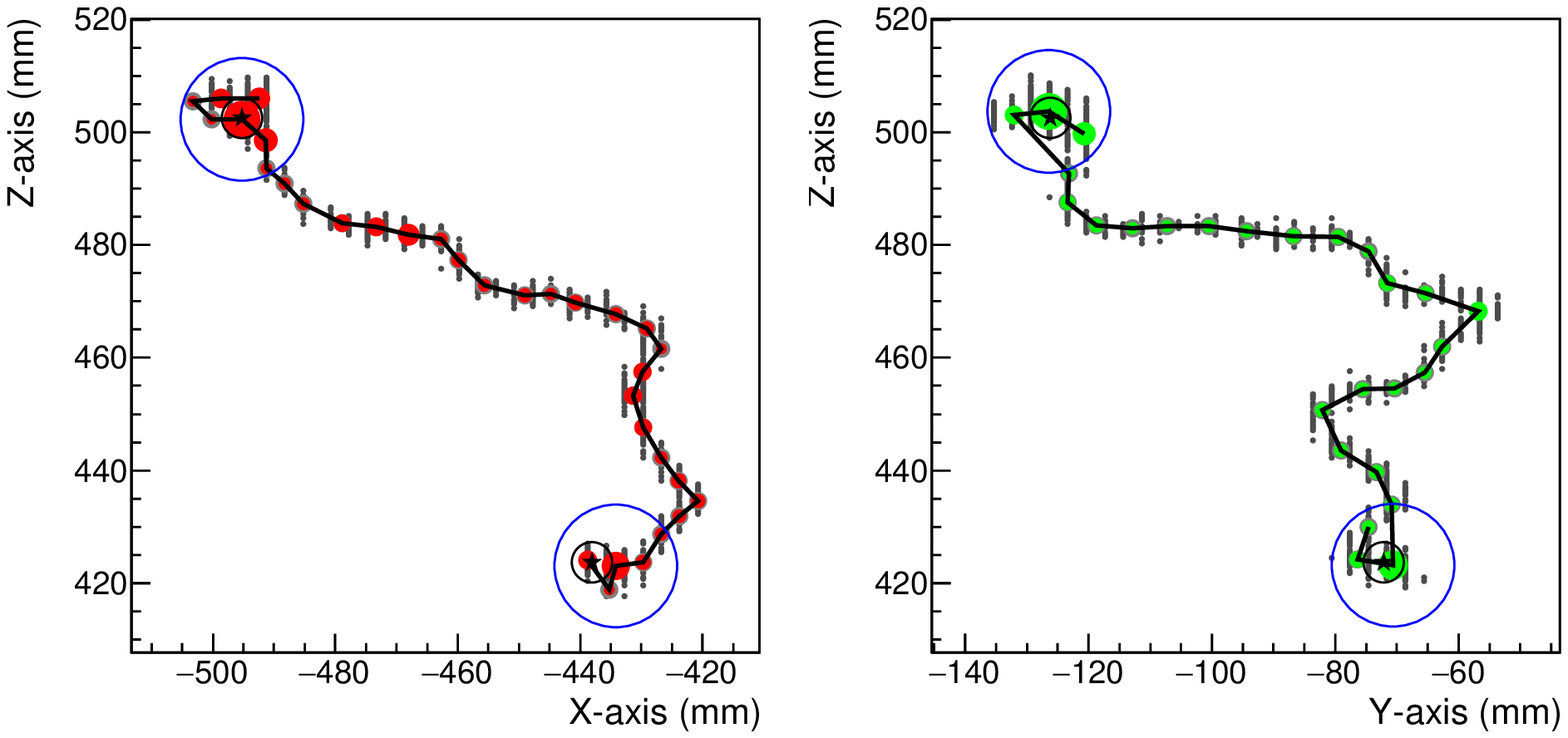} \\
	(a) \\
	\includegraphics[width=0.98\textwidth]{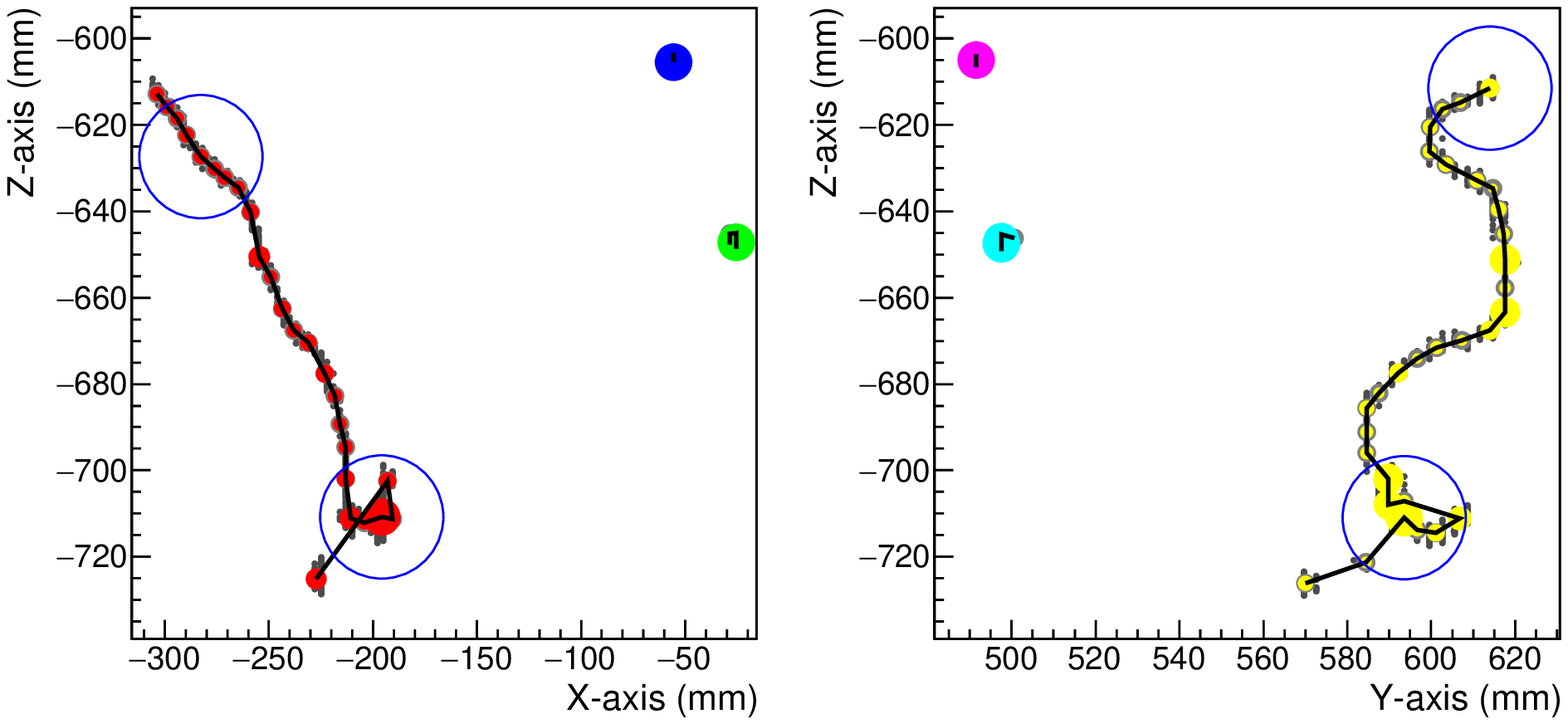} \\
	(b) \\
\end{tabular}
\caption{A \emph{TRestTrackEvent} representation of the 2-dimensional projected electron tracks for a 0$\nu\beta\beta$ (a) and a background event produced in the $^{238}$U decay chain (b) at the end of the data processing chain. The colored filled circles correspond to the \emph{top-level track} of a \emph{TRestTrackEvent}, while the grey scattered points in the background - where the 3\,mm pitch granularity can be discerned - represent the hits in the \emph{origin track}. The size of the circle in the \emph{top-level tracks} is proportional to the hit energy, and different colors serve to identify independent tracks classified after the \emph{TRestHitsToTrackProcess}. A black polyline shows the \emph{top-level track} hits interconnectivity resulting after the \emph{TRestTrackReconnectionProcess}, and helps to visualize the \emph{physical track}. Large \emph{blue circles} correspond with our \emph{blob charge} definition (R=12\,mm) obtained from \emph{TRestFindTrackBlobsProcess}, described in~\ref{sc:anaPcs}. The \emph{starred black markers} found in the \NLDBD event correspond with the positions obtained from the \emph{TRestFindG4BlobProcess} representing the last electron energy deposition for each of the primary simulated electron tracks. }
	
\label{fig:track}
\end{figure}

In particular, the 2-dimensional stripped baseline readout for PandaX-III cannot perform a univocal 3-dimensional reconstruction of the original event topology (see discussion at reference~\cite{Qiao2018}). In our design, a drifted electron that reaches a Micromegas module will induce a signal on either a X-strip or a Y-strip. Therefore, the \emph{TRestHitsEvent} reconstruction after \emph{TRestSignalToHitsProcess} is necessarily a combination of 2-projections, a projection formed by XZ-hits (i.e. hits with valid $X$ and $Z$ coordinates and undefined $Y$ coordinate) derived from signals corresponding to the X-strips, and a projection formed by YZ-hits resulting from the signals found at Y-strips. Figure~\ref{fig:track} shows the resulting projections for a signal and a background event. End-track identification using the \emph{analysis processes} described in the following subsection have been also represented in this figure. The main characteristics of signal and background events is discussed in section~\ref{sc:topoParams}.





\subsection{Analysis event processes}\label{sc:anaPcs}

Additional \emph{event processes} are used during the data processing chain in order to extract required event information in our posterior analysis. The \emph{analysis processes} used, the place of the processing chain where those processes are used, the particular information extracted and/or other parameters required are detailed in the following list.

\begin{itemize}
	\item \emph{TRestFindG4BlobProcess:} It is used at the beginning of the \emph{first phase} of the data processing chain. It operates in a \emph{TRestG4Event} and it is used to extract the \emph{MC-true track} coordinates of each electron track end in a \NLDBD event. These values are used to assess the goodness of our algorithms to identify the track ends after the physical event reconstruction.
	\item \emph{TRestG4AnalysisProcess:} It is also used at the beginning of the data chain, after the previous process. This process is used to add different observables related with the \emph{TRestG4Event} to the \emph{analysis tree}, as e.g. the interaction types involved, or the number and type of particles involved in the event. In our particular case we use this process to extract the total energy deposited in the active volume of the detector, without any fiducialization.
	\item \emph{TRestTriggerAnalysisProcess:} It is used at the \emph{second phase} of the processing chain, at the \emph{TRestTimeSignalEvent} level. This process uses the same trigger definition implemented in \emph{TRestSignalToRawSignalProcess} shown previously in Figure~\ref{fig:triggerDefinition}. We apply a sampling rate of 5\,MHz with a total number of 512 sampled points, as fixed by the PandaX-III electronics system, and a trigger energy threshold, $E_{th}$, of 1229\,keV corresponding to $Q_{\beta\beta}/2$. This process generates a new observable in the \emph{analysis tree}, defining the energy contained in the virtual acquisition window defined. It must be noted that this process, as any other \emph{analysis process}, does not modify the \emph{event data}.
	\item \emph{TRestTrackAnalysisProcess:} It is used at the end of the \emph{third phase} of the processing chain. It extracts parameters used for pattern recognition of \NLDBD events, such as the number of tracks, the track energy ratio, or the twist parameter described in section~\ref{sc:topoParams}.
	\item \emph{TRestFindTrackBlobsProcess:} It is also used at the end of the processing chain. The tracks ends have been naturally identified after the \emph{TRestTrackReconnectionProcess}, and this process registers the coordinates or positions of the end-tracks in the \emph{analysis tree}. This process searches for the highest density region within a 20\% of the track length at the track end, in order to allow an additional degree of freedom identifying the final blob position. This process is used to generate the observables related to the \emph{blobs charge} parameter, described in section~\ref{sc:topoParams}.
\end{itemize}


%% file: RawDataChain.tex

\section{Full Monte Carlo detector response}\label{sc:RawProcessing}

\begin{figure}
\centering
    \includegraphics[width=\textwidth]{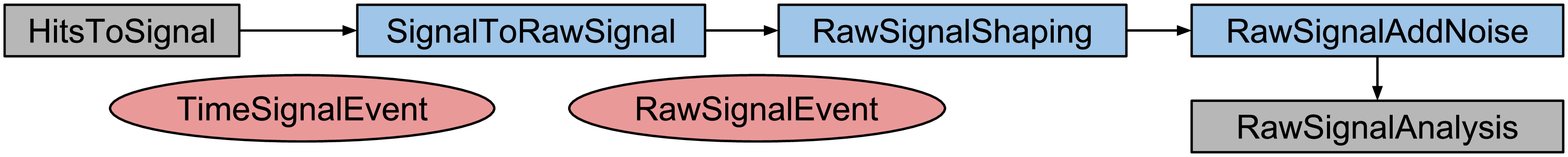}
    \caption{Extension of the event data chain to include appropriate detector readout signal conditioning, \emph{trigger definition}, \emph{electronics shaping} and \emph{electronic noise}. The first process \emph{TRestHitsToSignalProcess} - inside the first grey box - corresponds to the process shown previously in Figure~\ref{fig:basicDataChain}. The event processing ends with the \emph{TRestRawSignalAnalysisProcess} after we extract the parameters of interest for the fiducial analysis carried out in section~\ref{sc:fiducial}. The \emph{TRest} prefix and \emph{Process} termination has been removed from the REST process and event names, keeping the same nomenclature as in Figure~\ref{fig:basicDataChain}. }
    \label{fig:signalDataChain}
\end{figure}

In order to test the detector\'s capability to distinguish surface events we need to extract a new parameter connected with the diffusion of electrons, from the \emph{event data} processing chain presented on~\ref{sc:dataChain}. We could extract such a parameter at different stages in that chain, however, a straightforward parameter characterizing the spread of the charge is the width of the pulses as recorded by the readout electronics. We require then additional event processing to accommodate our Monte Carlo generated events to a time signal similar to the recorded with the detector acquisition. Figure~\ref{fig:signalDataChain} shows the additional \emph{event processes}, detailed also in~\ref{sc:evProcesses}, that we have introduced after \emph{TRestHitsToSignalProcess}, at the \emph{second phase} of the data chain described in~\ref{sc:dataChain}.

\begin{figure}
\centering
\begin{tabular}{cc}
\includegraphics[width=0.49\textwidth]{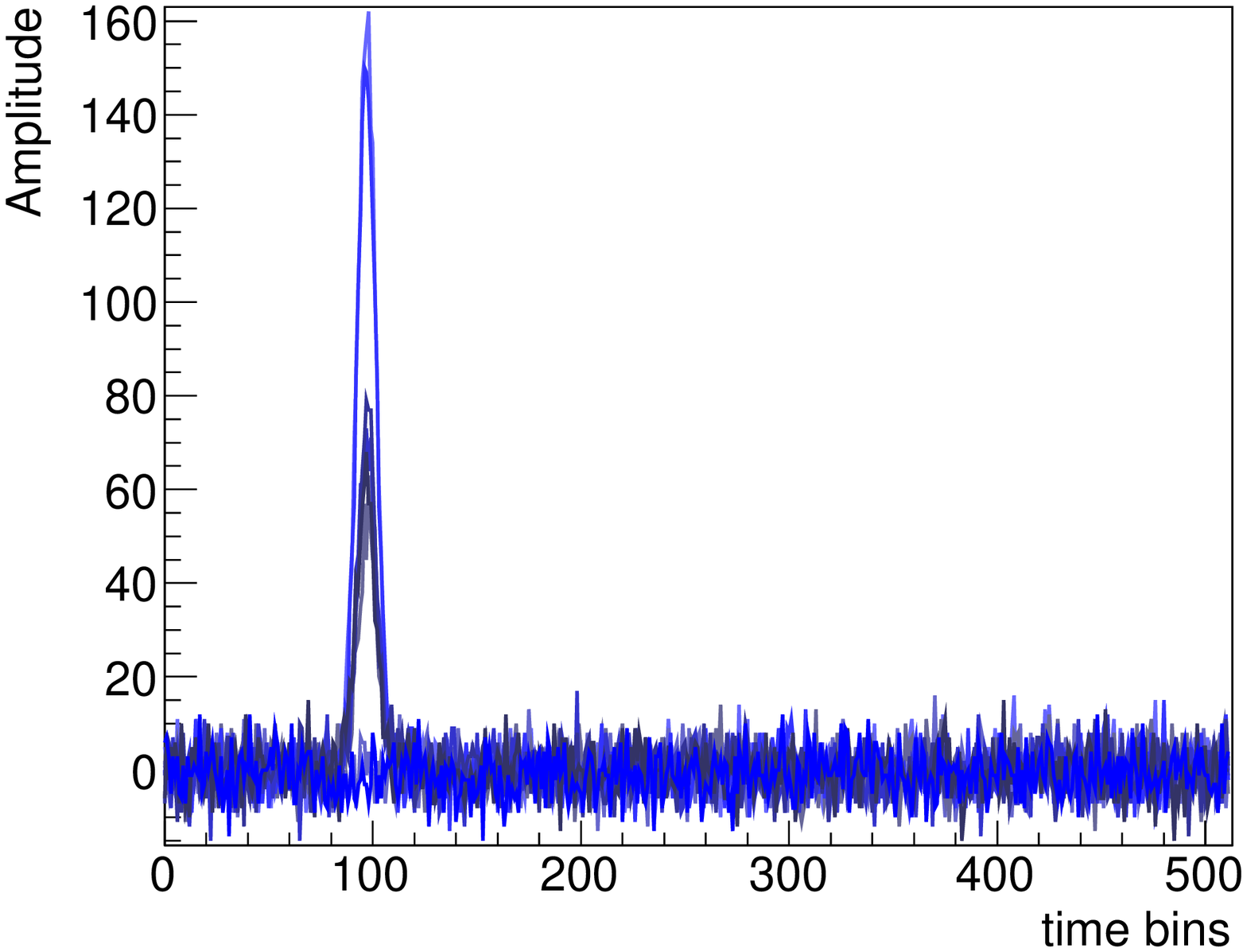} &
\includegraphics[width=0.49\textwidth]{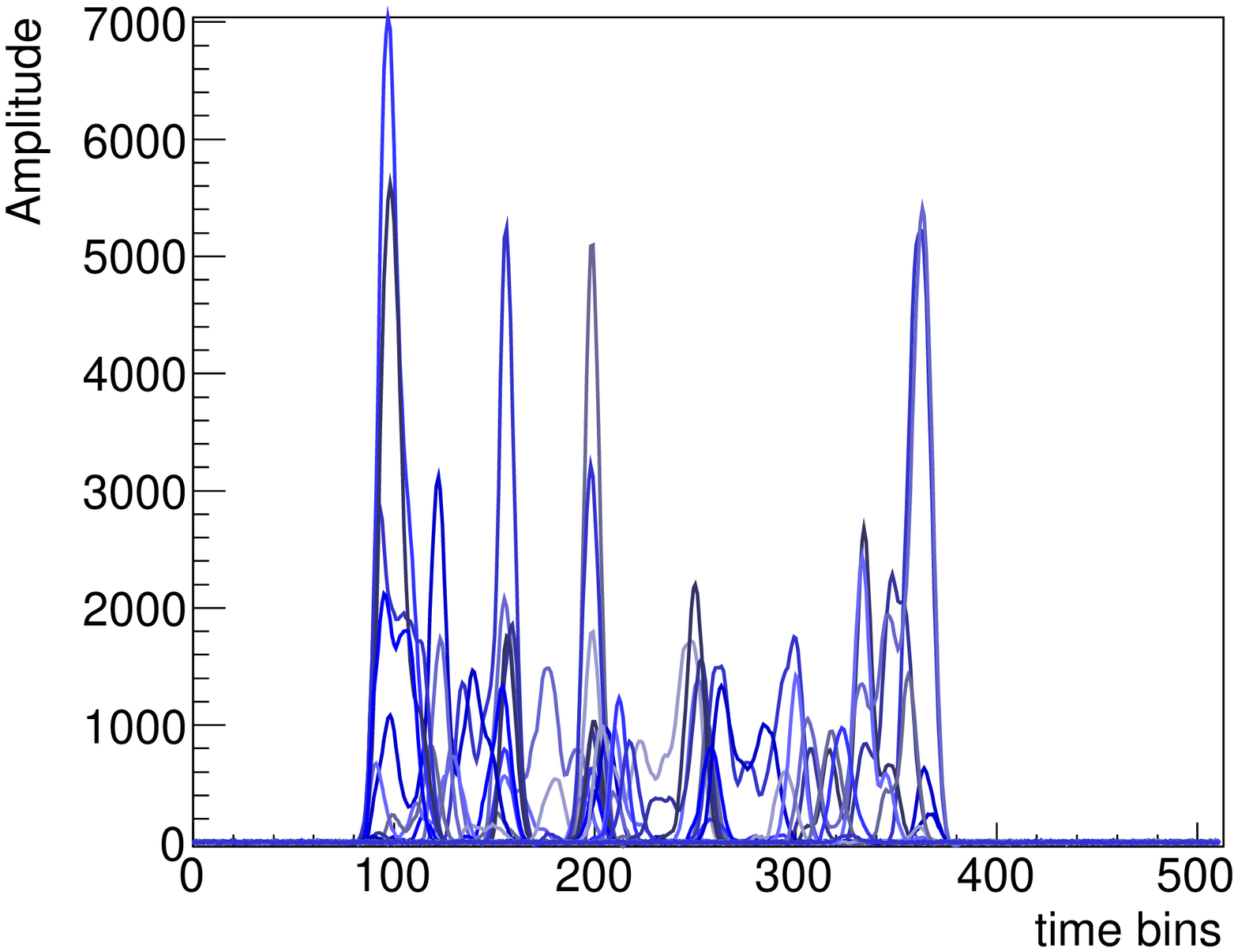} \\
(a) & (b) \\
\end{tabular}
\caption{A \emph{TRestRawSignalEvent} representation after applying the \emph{event processes} present on the signal conditioning data chain, including trigger definition, electronics shaping and electronic noise contributions. (a) The readout signals for a 10\,keV electron generated in the middle of our TPC volume, at z=500\,mm. (b) The readout signals for a \NLDBD signal event. }
    \label{fig:FullResponseEvts}
\end{figure}

In this extension of the processing chain we transform our \emph{event data} into a \emph{TRestRawSignalEvent} where we introduce different effects on the detector response, as a semi-Gaussian electronic signal shaper with 1\,$\mu$s peaking time and a Gaussian noise level, at each channel. In order to adjust the noise level we have simulated a small energetic deposition, produced by a 10\,keV electron, and we have chosen the noise level that approximately reproduces our signal-to-noise values when taking data in normal laboratory conditions. The unit values at this stage are completely arbitrary. Figure~\ref{fig:FullResponseEvts} shows the aspect of the final processed event after including all these effects.

These additional effects are included at the \emph{TRestRawSignalEvent} level to gain realism, but at the same time to exploit the REST analysis process, \emph{TRestRawSignalAnalysisProcess}, which is also used to analyze the real data of the detector. This analysis process is used right after \emph{TRestRawSignalAddNoiseProcess} to extract the diffusion parameter that we are interested in. In order to define our diffusion parameter, $\sigma_w$, we pre-select those channels exceeding a certain energy threshold measured in signal amplitude at the \emph{TRestRawSignalEvent}. The threshold value is defined as the 10\% of the maximum amplitude found inside the \emph{TRestRawSignalEvent}. Then, we obtain the FWHM of the maximum peak at each channel, and average the 9 channels which produce the lowest FWHM values\footnote{We describe in detail our particular definition of $\sigma_w$, although alternative definitions will certainly lead to similar results.}. The value of $\sigma_w$ is measured in samples, or bins, of 200\,ns width which is the sampling value we used in the detector response.

